\documentclass[11pt]{article}
\pdfoutput=1
\usepackage{graphicx}
\usepackage{amssymb,amsthm,amsmath,float}

 \renewcommand{\arraystretch}{1.2}

\newcommand{\Th}[1]{Th.~\ref{thm:#1}}
\newcommand{\Lem}[1]{Lem.~\ref{lem:#1}}

\newcommand{\Sec}[1]{\S \ref{sec:#1}}
\newcommand{\Fig}[1]{Fig.~\ref{fig:#1}}
\newcommand{\Tbl}[1]{Table~\ref{tbl:#1}}
\newcommand{\App}[1]{Appendix~\ref{app:#1}}

\newcommand{\InsertFig}[4]
{\begin{figure}[tbh!]
       \centerline{
         \includegraphics[width=#4]{./figures/#1}
       }
       \caption{{\footnotesize  #2}
       \label{fig:#3}}
\end{figure}}

\newcommand{\InsertFigTwo}[5] {
\begin{figure}[htb]
       \centerline{
         \includegraphics[width=#5]{./figures/#1}
         \hskip 0.5in
         \includegraphics[width=#5]{./figures/#2}
       }
       \caption{{\footnotesize  #3}
       \label{fig:#4}}
\end{figure}}

\newcommand{\InsertFigThree}[6] {
\begin{figure}[htb]
       \centerline{
         \includegraphics[width=#6]{./figures/#1}
         \hskip 0.5in
         \includegraphics[width=#6]{./figures/#2}
         \hskip 0.5in
         \includegraphics[width=#6]{./figures/#3}
       }
       \caption{{\footnotesize  #4}
       \label{fig:#5}}
\end{figure}}

\newcommand{\InsertFigFour}[7] {
\begin{figure}[ht]
       \centerline{
\renewcommand{\arraystretch}{0.01}
         \begin{tabular}{cc}
         \includegraphics[width=#7]{./figures/#1}&  \includegraphics[width=#7]{./figures/#2} \\
        \includegraphics[width=#7]{./figures/#3}  &  \includegraphics[width=#7]{./figures/#4}
        \end{tabular}
       }
       \caption{{\footnotesize  #5}
       \label{fig:#6}}
\end{figure}}

\newcommand{\R}{{\mathbb{ R}}}
\newcommand{\C}{{\mathbb{ C}}}
\newcommand{\T}{{\mathbb{ T}}}
\newcommand{\Z}{{\mathbb{ Z}}}
\newcommand{\N}{{\mathbb{ N}}}

\newcommand{\bS}{{\mathbb{ S}}}
\newcommand{\cB}{{\cal{B}}}
\newcommand{\cC}{{\cal{C}}}

\newcommand{\cL}{{\cal L}}
\newcommand{\cO}{{\cal O}}
\newcommand{\cR}{{\cal R}}
\newcommand{\eps}{\varepsilon}

\DeclareMathOperator{\sgn}{sgn}
\DeclareMathOperator{\tr}{tr}

\DeclareMathOperator{\diag}{diag}

\newcommand{\floor}[1]{{\lfloor{#1}\rfloor}}
\newcommand{\ceil}[1] {{\lceil{#1}\rceil}}

\newcommand{\Lmap}{\mathrm{L}}

\newtheorem{thm}{Theorem}

\newtheorem{lem}[thm]{Lemma}
\newtheorem{conjecture}{Conjecture}

\newcommand{\beq}[1]{\begin{equation}\label{eq:#1}}
\newcommand{\eeq}{\end{equation}}

\newcommand{\bsplit}[1]{\begin{equation}\label{eq:#1}\begin{split}}

\title{Quadratic Volume-Preserving Maps: \\Invariant Circles and Bifurcations}
\author{  
        H.~R. Dullin$^{1,2}$\thanks
      {
        HRD was supported in part by a Leverhulme Research Fellowship,
        and would like to thank the Department of Applied Mathematics in Boulder 
        for its hospitality. 
      } 
      and J.~D.~Meiss$^{2}$\thanks
      {
        JDM was supported in part by NSF grant DMS-0707659 and by the Mathematical
        Sciences Research Institute in Berkeley.
      }\\
$^{1}$School of Mathematics and Statistics\\
The University of Sydney\\
Sydney, NSW 2006, Australia\\
        {\tt hdullin@usyd.edu.au}\\
$^{2}$Department of Applied Mathematics\\
     University of Colorado \\
     Boulder, CO 80309-0526, USA\\
        {\tt James.Meiss@colorado.edu} 
}
\date{\today}
\begin{document}
\maketitle


\begin{abstract}
\vspace*{1ex}
\noindent

We study the dynamics of the five-parameter quadratic family of volume-preserving diffeomorphisms of $\R^3$. This family is the unfolded normal form for a bifurcation of a fixed point with a triple-one multiplier and also is the general form of a quadratic three-dimensional map with a quadratic inverse. Much of the nontrivial dynamics of this map occurs when its two fixed points are saddle-foci with intersecting two-dimensional stable and unstable manifolds that bound a spherical ``vortex-bubble''. We show that this occurs near a saddle-center-Neimark-Sacker (SCNS) bifurcation that also creates, at least in its normal form, an elliptic invariant circle. We develop a simple algorithm to accurately compute these elliptic invariant circles and their longitudinal and transverse rotation numbers and use it to study their bifurcations, classifying them by the resonances between the rotation numbers. In particular, rational values of the longitudinal rotation number are shown to give rise to a \textit{string of pearls} that creates multiple copies of the original spherical structure for an iterate of the map.    

\end{abstract}

\section{Introduction}\label{sec:Introduction}
The area-preserving H\'enon map \cite{Henon69} is the universal form for quadratic diffeomorphisms of the plane and provides perhaps the prototype for conservative systems with chaotic dynamics. As such it has deservedly been the subject of much study. We believe that the natural generalization of this map to three-dimensions is the volume and orientation-preserving diffeomorphism
\beq{StdMap}
	f(x,y,z)=
	\begin{pmatrix} 
     	x+y\\ y+z-\eps + \mu y +P(x,y) \\ z -\eps + \mu y + P(x,y)
	\end{pmatrix} \;.
\eeq
where $P$ is the quadratic form
\beq{quadForm}
  P(x,y) = a x^2 + b x y + c y^2 \;.
\eeq
Just as the H\'enon map arises as the normal form for a saddle-center bifurcation, the map $f$ (with $P$ a polynomial of any degree) is the normal form for the equivalent bifurcation in $\R^3$: a triple-one multiplier \cite{Dullin08a}. The H\'enon map and its generalizations to polynomials of arbitrary degree give a basis for the group of polynomial automorphisms of the plane \cite{Friedland89}. While such a decomposition is not known for higher dimensions, the map $f$ is the universal form of a quadratic diffeomorphism with quadratic inverse \cite{Lomeli98a}. More generally the normal form $f$ has an inverse with the same degree as the polynomial $P$ \cite{Dullin08a}. Finally, both the H\'enon map and $f$ arise as normal forms for certain homoclinic bifurcations \cite{Gonchenko06}, see \Sec{Contexts}.

In this paper we will study some of the dynamics of the map \eqref{eq:StdMap}. Numerical evidence presented in \Sec{BoundedOrbits} will show that the map has a nonzero measure of bounded orbits primarily near the simultaneous saddle-center and Neimark-Sacker (SCNS) bifurcation that occurs when $\eps = 0$ and $-4 < \mu < 0$. We will study the normal form for this bifurcation in \Sec{SNHBifurcation}. The bounded orbits of $f$ that appear in such a regime are built around the skeleton formed from its two saddle-focus fixed points, their stable and unstable manifolds, and the elliptic invariant circle that is created when this bifurcation is supercritical.

As we recall in  \Sec{SNHFlow}, the structure of this ``saddle-center-Hopf" bifurcation is well-known for the case of a volume-preserving flow \cite{Broer81, Holmes84}. For the supercritical case, the bifurcation creates a vortex-bubble structure analogous to the Hill's vortex of fluid mechanics \cite{Lamb45} or the spheromak configuration of plasma physics \cite{Goldenbaum80}. There are two saddle-focus equilibria whose two-dimensional stable and unstable manifolds coincide forming a sphere. The interior of the sphere is foliated by a family of two-tori enclosing an invariant circle that is normally elliptic. For the map, the normal form of the ``saddle-center-Neimark-Sacker" bifurcation is no longer integrable, but it still has two saddle-foci, a Cantor family of tori, and an elliptic invariant circle, see \Sec{SNHMap}. 

The quadratic map \eqref{eq:StdMap} is approximately described by this normal form for small $\eps$ and away from low-order resonances. We observe numerically that many of the invariant two-tori and the central invariant circle appear to persist for moderate values of $\eps$.  In this paper we will concentrate on the persistence and bifurcations of the elliptic invariant circle. We develop an algorithm to accurately compute this circle in \Sec{InvCircles}. When it is elliptic, the circle has two rotation numbers $\omega = (\omega_L, \omega_T)$, longitudinal and transverse, respectively. We compute these and compare the numerical results with the normal form calculations. 

Resonances of the form $m \cdot \omega = n$ lead to bifurcations that may give rise to new elliptic invariant circles and may result in the destruction or change of stability of the original circle. Interestingly, there are two types of doubling (or $m$-tupling) bifurcations, one in which the new invariant circle is a single circle that winds multiple times around the original circle (like in a flow), the other one in which $k$ new invariant circles appear that are mapped to each other. The latter case appears when $m_1$ and $m_2$ are not  coprime. 

Perhaps the most interesting of these bifurcations we call a \textit{string of pearls}. It occurs when $\omega_L$ becomes rational and typically results in a new SCNS bifurcation for some power of $f$, see \Sec{Resonances}. When this bifurcation is supercritical, the invariant circle is replaced by a set of small vortex-bubbles or \textit{pearls} linked by nearly coincident one-dimensional invariant manifolds of a pair of saddle-focus periodic orbits, the \textit{string}. Recently a similar bifurcation has been observed for dissipative $3D$ maps \cite{Broer08}. 

\section{Contexts}\label{sec:Contexts}
The quadratic map \eqref{eq:StdMap} arises naturally in at least three contexts. 

Suppose that $f:\R^3 \to \R^3$ is a smooth, volume and orientation-preserving map, $\det Df(\xi) = 1$, where $\xi = (x,y,z)$. If $\xi^*$ is a point on a period-$n$ orbit, $f^n(\xi^*)= \xi^*$, then the multipliers of this orbit must satisfy $\lambda_1\lambda_2\lambda_3 = 1$. Consequently, the characteristic polynomial of the linearization about the periodic orbit,
\beq{charPoly}
	\det (\lambda I - Df^n(\xi^*)) =\lambda^3 -\tau \lambda^2 + \sigma \lambda -1 \;,
\eeq
contains two parameters: the trace $\tau$ and second trace $\sigma$. There are eight stability regimes in the $(\tau,\sigma)$ plane, see \Fig{Stability}. The boundaries between these regimes contain two codimension-two points; one corresponds to a triple multiplier $\{1,1,1\}$ at $(\tau,\sigma)= (3,3)$ and the second to multipliers $\{-1,-1,1\}$ at $(\tau,\sigma) = (-1,-1)$. These two cases form organizing centers for the dynamics in all eight regimes.
Notice that in the interior of each of the eight regions, the fixed point is hyperbolic and  unstable.

\InsertFig{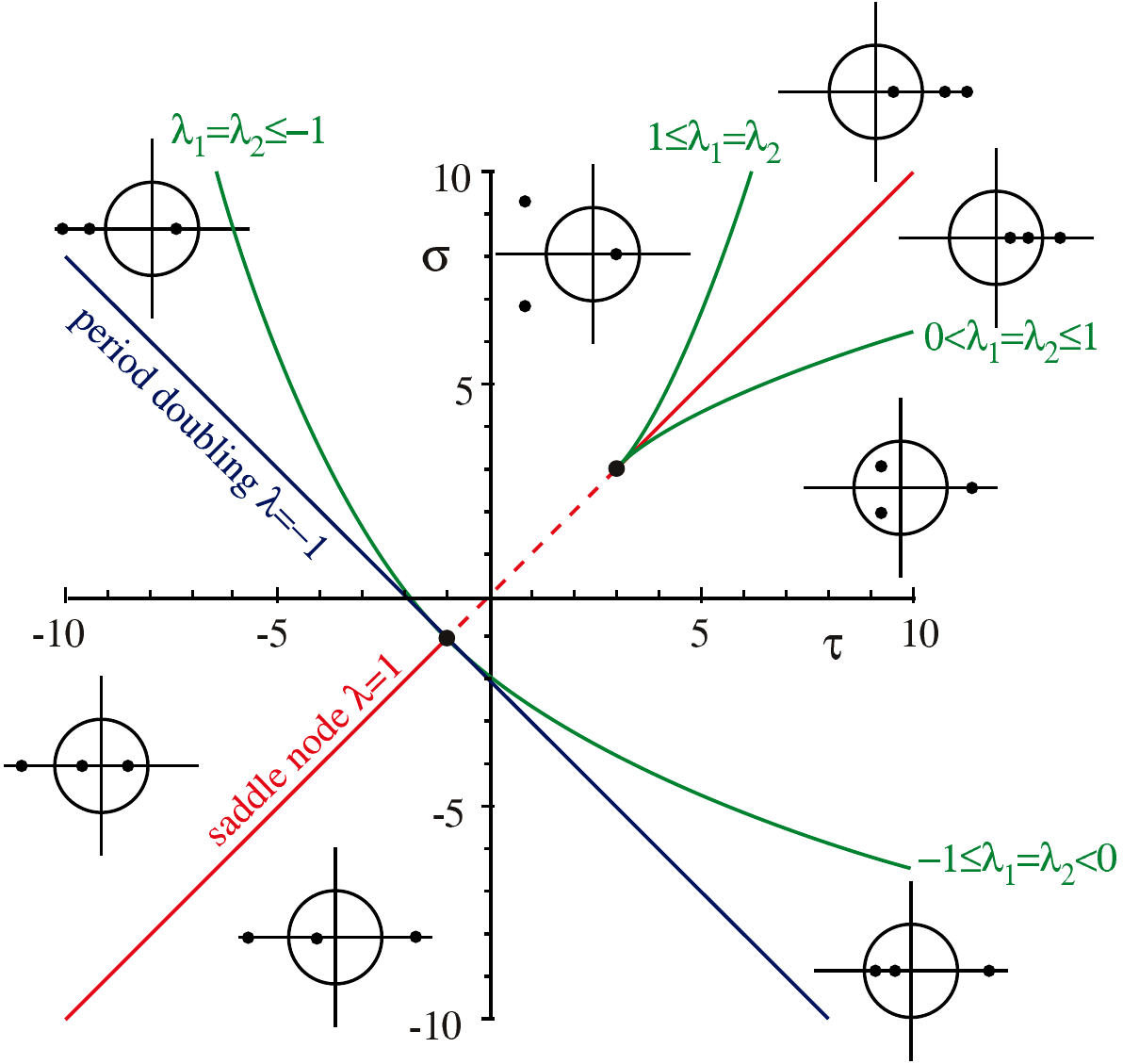}{Classification of the eigenvalues for a three-dimensional, 
volume-preserving map as a function of the trace $\tau$ and second trace $\sigma$.
The eight insets are the complex planes showing the multiplier configurations
relative to the unit circle.}{Stability}{3in}

If the Jacobian, $Df^n(\xi^*)$ has a multiplier $\lambda = 1$ with algebraic multiplicity three, then $\lambda$ generically has geometric multiplicity one; consequently, the Jacobian is similar to the Jordan block
\beq{111J}
  J = \begin{pmatrix} 1 & 1 & 0 \\ 0 & 1 & 1 \\ 0 & 0 & 1 \end{pmatrix} \;.
\eeq
We showed in \cite{Dullin08a} that near $\xi^*$, $f^n$ is formally conjugate to the normal form
\beq{111NF}\begin{split}
	(x',y',z')^T &=  J \left(x,y, z+p(x,y,\eps,\mu)\right)^T \;, \\
	p(x,y,\eps,\mu) &= -\eps + \mu_1 x +\mu_2 y + 
		a(\eps,\mu) x^2 + b(\eps,\mu)xy + c(\eps,\mu) y^2 + \ldots \;,
\end{split}\eeq
where the $\ldots$ indicates higher order terms in $x$ and $y$ only.
The normal form is ``formal" in the sense that if $f$ is expanded in a power series to any finite degree $d$ in the variables $\xi$ and a set of sufficiently general parameters, then this degree-$d$ map is conjugate to  \eqref{eq:111NF}. Generically,  one of the two parameters $(\mu_1, \mu_2)$ in \eqref{eq:111NF} can be eliminated, see \App{generic}, leaving two unfolding parameters; for example if $a \neq 0$ then $\mu_1$ can be set to zero, leaving the two parameters $(\eps, \mu = \mu_2)$.

The normal form provides a remarkable simplification of the full map since all of the nonlinearity is contained in the single polynomial $p$ that depends upon only two of the variables. To lowest nonlinear degree, the map \eqref{eq:111NF} is quadratic and if we view $(a,b,c)$ as independent parameters it becomes \eqref{eq:StdMap}.

The quadratic map \eqref{eq:StdMap} also arises in the study of polynomial diffeomorphisms: it was shown in \cite{Lomeli98a}, that any quadratic diffeomorphism of $\R^3$ that has a quadratic inverse and nontrivial dynamics is affinely conjugate to the \textit{shift-like} \cite{Bedford98a} map 
\[
  (x,y,z) \mapsto \left(-\eps + \tau x + \sigma y + z + Q(x, y), x , y \right) \;,
\]
where $Q(x,y)$ is a quadratic form. This map is linearly conjugate to the quadratic case of \eqref{eq:111NF} under the orientation reversing transformation $\xi \mapsto U \xi$ where
\[
   U = \left(\begin{array}{rrr}  
        0 & 1 & 0 \\
        1 & -1 & 0 \\
        1  & -2 & 1 
       \end{array}\right) \;,
\]
providing that $Q(x,y) = p(y,x-y)$, $\tau =\mu_2 +3$, and $\sigma = \mu_1-\mu_2-3$.
Thus the map \eqref{eq:111NF} is also a normal form in the sense of polynomial automorphisms. However, it is not known how to generalize this result to cubic or higher degree---unlike the planar case that was treated by Friedland and Milnor \cite{Friedland89}. Quadratic diffeomorphisms may have inverses of degree up to four; these were classified for $\C^3$ in \cite{Fornaess98, Maegawa01a}. Our map \eqref{eq:111NF} does have an inverse of the same degree for any polynomial $P$.

The map \eqref{eq:111NF} also appears in a third context as the normal form for certain homoclinic bifurcations \cite{Gonchenko06}. In particular consider a three-dimensional map with a saddle-focus fixed point such that the saddle value---the product of the multipliers of the fixed point---is $1$; this, of course, is true for the volume-preserving case. Suppose that the one-dimensional invariant manifold of the saddle-focus has a quadratic tangency with its two-dimensional (spiral) manifold. Unfolding this singularity leads to a map of the form \eqref{eq:111NF} that describes the dynamics of the return map in a neighborhood of the homoclinic point. There are two distinct cases: when the saddle-focus is of type $(2,1)$ (or \textit{type-A} \cite{Cowley73}), with a two-dimensional (spiral) stable manifold and a one-dimensional unstable manifold, then the homoclinic normal form has $P(x,y) = a(x+y)^2$, so that the quadratic form \eqref{eq:quadForm} is a perfect square. When the saddle-focus is of type $(1,2)$ (or \textit{type-B}), then $P(x,y) = ax^2$. 

An additional parameter can be introduced in \eqref{eq:StdMap} to make the Jacobian $\det{Df} = b \neq 1$. Maps similar to this often have strange attractors and have been studied in\cite{Arneodo83, Argoul84, Gonchenko05a, Du06}.

These considerations support our assertion that the normal form \eqref{eq:111NF} is an appropriate \textit{generalized H\'enon} map for three-dimensional dynamics.

\section{Fixed Points}
We begin the study of the dynamics of \eqref{eq:StdMap} looking at its fixed points; more details are given in \App{fixedPoints}.

When $\eps a < 0$ the quadratic map \eqref{eq:StdMap} has no fixed points; when $\eps a > 0$ it has two at 
\beq{fixedPoints}
 \xi_\pm = (x_\pm,0,0) \;, \quad x_\pm = \pm \sqrt{\frac{\eps}{a}} \;.
\eeq
Thus a pair of fixed points is created upon crossing the \textit{saddle-center} line $\eps = 0$. Stability of these fixed points is determined from the general stability diagram, \Fig{Stability}, by computing $(\tau, \sigma)$ for the fixed points, see \App{fixedPoints}. For fixed $(a \neq 0,b,c)$, the stability of each fixed point depends on the two parameters $(\eps, \mu)$. When $a > 0$ the stability diagram of $x_+$ is a diffeomorphic copy of the half plane $\tau > \sigma$; the corresponding diagram for $x_-$ covers the half-plane $\tau < \sigma$, see \Fig{QuadStability}. In this figure, the blue curves correspond to the fixed point $x_+$ and the red curves to $x_-$.  Note that over most of the parameter range, when one of the fixed points is type $(2,1)$ (two-dimensional unstable manifold), the other is type $(1,2)$ (two-dimensional stable manifold), see \App{fixedPoints}.
\InsertFig{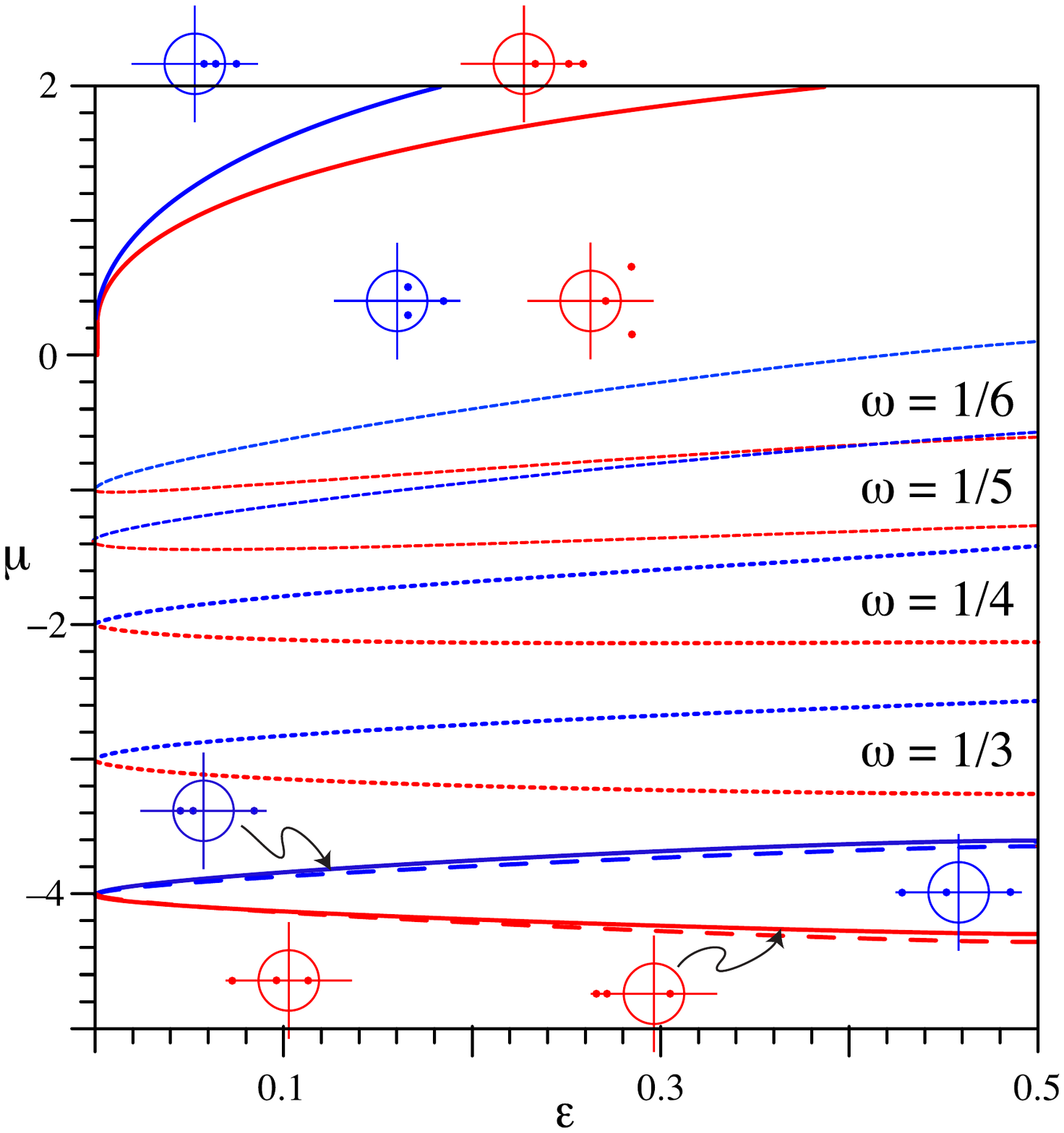}{Stability diagram for the map \eqref{eq:StdMap} with $a = 1$, $b= 0.5$ and $c$ arbitrary. The solid curves correspond to a double eigenvalue, the dashed curves to  period doubling, and the dotted curves to a complex conjugate pair at $re^{\pm2\pi i\omega}$ with $\omega$ as indicated. Bifurcation curves for the fixed point $x_+$ are blue and those for $x_-$ are red.}{QuadStability}{5in}

When $\eps = 0$ and  $-4 < \mu < 0$, the fixed points are created with a pair of multipliers $\lambda_\pm = e^{\pm 2\pi i \omega_0}$ on the unit circle with rotation number
\beq{fixedPtRotNum}
    \omega_0(\mu) = \frac{1}{\pi} \arcsin \sqrt{-\frac{\mu}{4}} \;.
\eeq
As we will see next, this range of $\mu$ seems also to correspond to the range for which there are other nontrivial bounded orbits for the map.

\section{Bounded Orbits}\label{sec:BoundedOrbits}
When the quadratic form $P$ of \eqref{eq:quadForm} is positive definite, it is not hard to show that all of the bounded orbits of \eqref{eq:StdMap} are contained in a cube, see \App{properties}. To find regions of parameter space that have nontrivial bounded orbits, we iterated a grid of initial conditions, declaring an orbit to be ``bounded" if it remains in a ball of radius $10$ for $N$ iterations. The resulting fraction of bounded orbits as a function of the parameters $(\eps, \mu)$ is shown in \Fig{VpBounded}. For this figure, as for most of the numerical studies reported below we fixed the parameters of $P$, choosing $a=1$, $b=c=\frac12$.

Our numerical studies indicate that the domain containing bounded orbits shrinks to zero as $\sqrt{\eps}$. To reflect this, we considered initial conditions on a three-dimensional grid in the box
\beq{box}
  \cB \equiv [-2x_+,2x_+]\times[-4x_+,4x_+]\times[-8x_+,8x_+] \;,
\eeq
where $x_+$ is the position of the fixed point \eqref{eq:fixedPoints}. In \Fig{VpBounded} we plot the fraction of bounded orbits with $N = 10^4$, choosing initial conditions on a cubical grid of $500^3$ points that covers $\cB$. It appears that there are bounded orbits only when $\eps \ge 0$ and $-4.1 \lesssim  \mu < 0$, though we have not exhaustively searched outside this parameter domain. We also observe that, for the parameters of \Fig{VpBounded}, there appear to be essentially no bounded orbits when $\eps \gtrsim 0.5$.

In the figure, the dark blue regions correspond to ``no" bounded orbits and the darkest red to the maximal fraction, $4.9\%$. There is considerable ``resonance tongue'' structure in the figure. Some of this is similiar to \Fig{QuadStability} for the multipliers of the fixed point. In particular the bounded fraction appears to be nearly zero at the doubling and tripling points $(\eps,\mu) = (0,-4)$ and $(0,-3)$, respectively, and is small near the quadrupling point $(\eps, \mu) = (0,-2)$. Indeed if we increase $N$ to $10^6$ and  double the grid resolution then we found no bounded orbits in $\cB$ whenever $\eps \ge 0.001$ for $\mu = -3$ and at most $0.043\%$ when $\mu = -4$.

\InsertFig{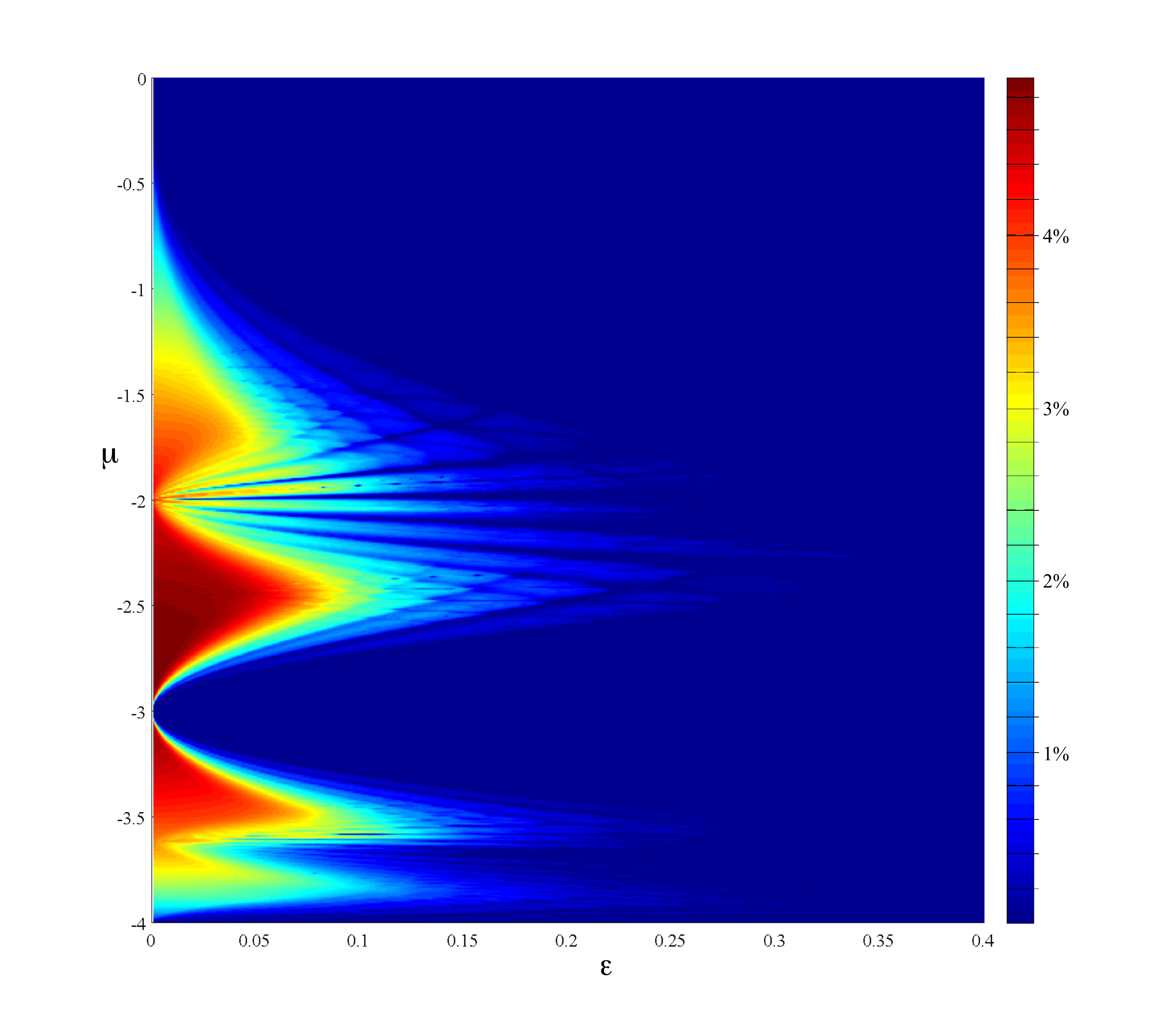}{Fraction of bounded orbits in $\cB$ for the map \eqref{eq:StdMap} with nonlinearity \eqref{eq:quadForm} and $a=1.0$ and $b = c = \frac12$. Orbits are ``bounded" if they stay inside the ball of radius $10$ for $N = 10^4$ iterates. For this calculation, the map was iterated a total of $1.69\times 10^{15}$ times over $500^3 \times 500^2$ initial conditions. The maximum number of trapped orbits was $6139040$, corresponding to $4.91\%$ of the initial conditions in $\cB$ when $(\eps,\mu) = (0.00078,-2.73)$. When $\eps = 0.4$  there are no bounded orbits except near $\mu = -2.26$ and $-3.95$ where less than $2\times 10^{-5}$ of the initial conditions are bounded. 
}{VpBounded}{5in}

Several constant-$\eps$ slices through the full dataset are shown in \Fig{relativeVolume}. Note that as $\eps \to 0^+$ the bounded fraction appears to converge to a smooth curve, shown as the dashed line, except for the resonances near $\mu = 0, -2$, and $-3$. This curve will be derived in \Sec{111SNH}.

Much of the structure of \Fig{VpBounded} and \Fig{relativeVolume} can be attributed to the orbits trapped by the two-dimensional manifolds of the fixed points, as we discuss below. 


\InsertFig{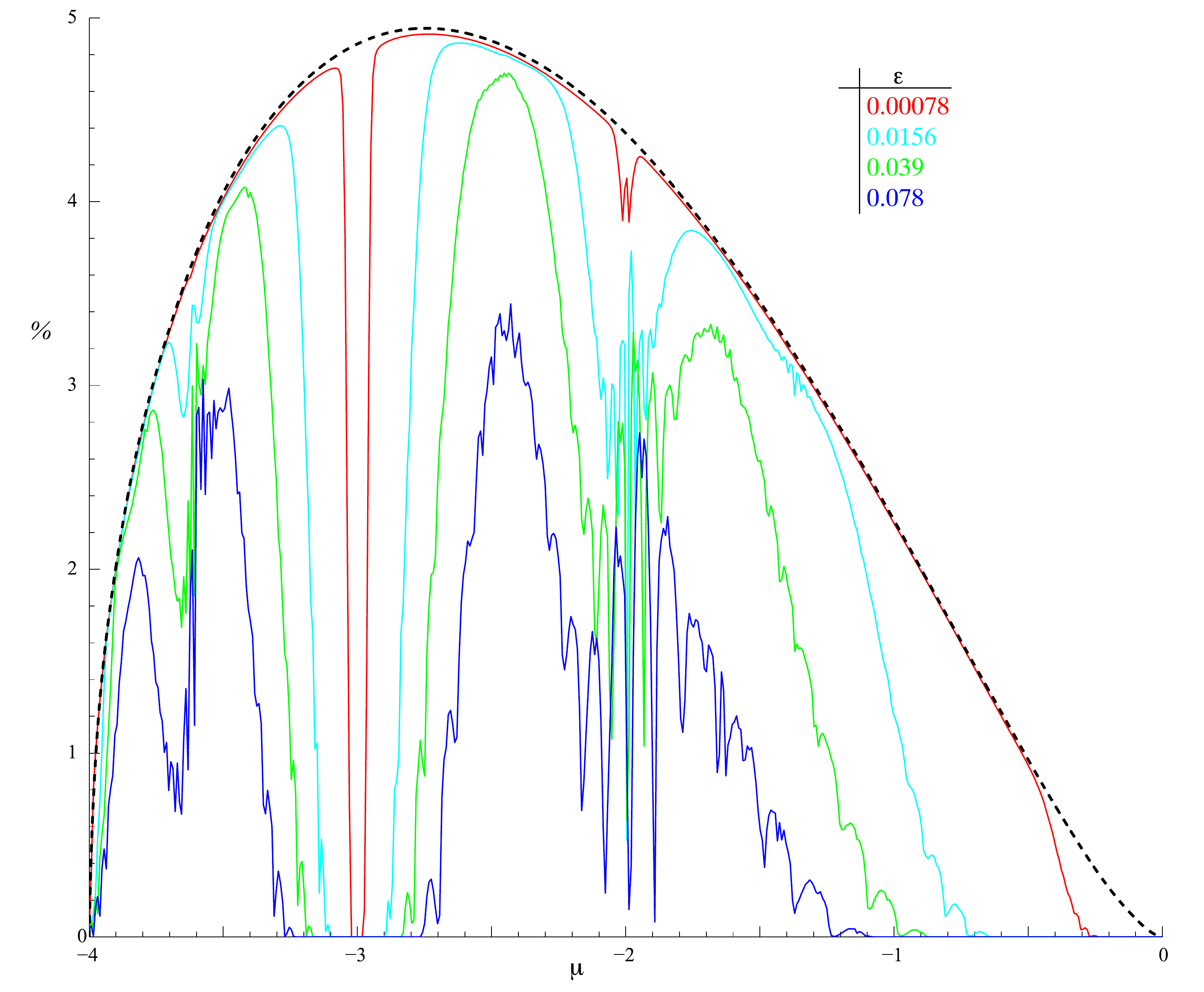}{Percentage of the box $\cB$ that is occupied by bounded orbits as a function of $\mu$ for several values of $\eps$. The dashed curve is the theoretical result \eqref{eq:relativeVol} that is valid as $\eps \to 0^+$, away from the major resonances.}{relativeVolume}{5in}

\section{Saddle-Center-Neimark-Sacker Bifurcation}\label{sec:SNHBifurcation}

In most of the regions with bounded orbits in \Fig{VpBounded}, the fixed points \eqref{eq:fixedPoints} are spiral foci with two-dimensional stable and unstable manifolds, respectively. These manifolds appear to intersect transversely for many parameter regimes and, roughly speaking, enclose a ball that appears to contain all of the bounded orbits \cite{Lomeli98a}. The structure is analogous to a vortex bubble in an incompressible fluid \cite{Lamb45, MacKay94}. 
Such a flow is generic for a volume-preserving vector field near a saddle-center-Hopf bifurcation, as we review in \Sec{SNHFlow}. The topology of the intersections changes as the parameters vary, but most often seems to include infinite spiral curves that connect the fixed points. This is necessarily true for a flow since every intersection point lies on a homoclinic orbit; however, for a map a number of different topological types of intersections can also occur \cite{Lomeli00b}.

The most prominent bounded orbits correspond to a family of invariant two-tori enclosing an invariant circle that lies approximately in the plane $x=0$, see \Fig{Torus}. If the rotation number \eqref{eq:fixedPtRotNum} of the fixed points is not close to a low-order rational number, then as $\eps$ tends to zero this structure limits on a ball that shrinks to zero as $\sqrt{\eps}$. The stable and unstable manifolds in this limit appear to nearly coincide and the family of invariant tori nearly completely fills the interior of the ball: the structure appears to be nearly ``integrable."

\InsertFig{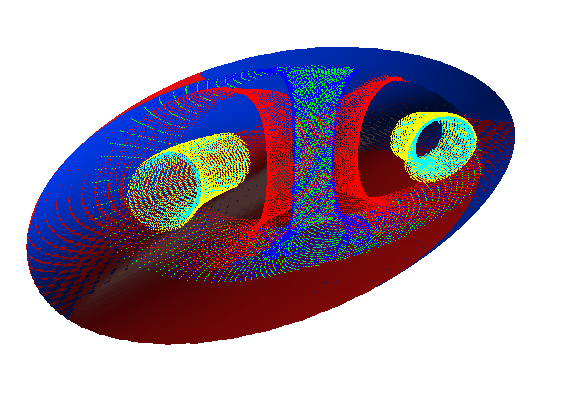}{Cutaway view of several orbits of the map \eqref{eq:StdMap} for $a=1.0$, $b=c=0.5$ and $(\eps,\mu) = (0.01, -2.4)$. Also shown are the two-dimensional stable (blue) and unstable (red) manifolds of the fixed points }{Torus}{3in}

Since the bounded orbits of the map \eqref{eq:StdMap} occur primarily along the saddle-center-Hopf line, we discuss in \Sec{SNHMap} the normal form for a map near such a bifurcation. We will show that both supercritical and subcritical bifurcations can occur. In the supercritical case, the normal form creates a pair of fixed points whose two-dimensional stable and unstable manifolds intersect and enclose a sphere that generically contains a Cantor-family of invariant two-tori. In \Sec{111SNH}, we will transform \eqref{eq:StdMap} into normal form, thereby obtaining the relationship between its parameters and those of the normal form. This will demonstrate that this bifurcation occurs on the saddle-center-Hopf segment of \Fig{QuadStability}.

Before proceeding to the discussion of the map, we review the standard results for volume preserving flow in \Sec{SNHFlow}. The time $\sqrt{\eps}$ flow of this vector field will be shown to approximate the map in \Sec{SNHMap}.

\subsection{Saddle-Center-Hopf Bifurcation}\label{sec:SNHFlow}

For a system of differential equations, the saddle-center-Hopf or Gavrilov-Guckenheimer bifurcation is the codimension-two bifurcation that occurs when an equilibrium has simultaneously one zero and one pair of imaginary eigenvalues. The three-dimensional, center manifold reduction to normal form is discussed in depth in \cite{Dumortier96, Dumortier01, Guckenheimer02, Kuznetsov04}. The key simplification is that the ``formal" normal form has cylindrical symmetry to all orders in  the power series expansion: it exhibits the symmetry of the linearized system. This symmetry is typically only formal and is broken by terms ``beyond-all-orders". For a system of divergence-free differential equations this bifurcation is codimension-one, and its unfolding is considerably simpler \cite{Broer81}.

In the neighborhood of an equilibrium point with eigenvalues $(2\pi i\omega, -2\pi i\omega,0)$, the normal form can be most easily obtained in the complex coordinates, $(u, v=\bar{u}, z)$, that diagonalize the linearization. The flow of the linear system then commutes with the rotation $u \mapsto u \exp( 2 \pi i \varphi)$. Consequently, in symplectic, cylindrical coordinates $(r,\theta,z)$ with
\beq{cylindrical}
	 u = \sqrt{2r} e^{2\pi i \theta} \;, \quad v = \sqrt{2r} e^{-2\pi i \theta} \;,
\eeq
the linearized vector field has the form
\[
  V = \omega \partial_\theta \;.
\]
The normal form has the same symmetry and since the rotation axis $r = 0$ is a fixed set of the symmetry, it is invariant under the dynamics. The normal form vector field therefore has the form
\[
   V =  r F(r,z) \partial_r + \Omega(r,z) \partial_\theta + Z(r,z) \partial_z \;,
\]
which is divergence free when $\partial_z Z + \partial_r (r F) = 0$.\footnote
{
	The divergence in the new coordinates looks Euclidean because the transformation
	to symplectic cylindrical coordinates is volume preserving .
}
The divergence-free condition implies that the two-dimensional projection of vector field onto $(r,z)$ is Hamiltonian with
\[
   H(r, z) = r G(r,z) \;,
\]
such that $F = -G_z$ and $Z = G + r G_r$.\footnote
{
	Indeed, any three-dimensional, volume-preserving flow with a Lie symmetry 
	has an invariant that gives rise to a Hamiltonian structure on the projection 
	of manifold by the group orbits \cite{Haller98}. Specifically
	suppose $V$ has the Lie symmetry $W$, i.e., $[V,W] = 0$. Then if
	both $V$ and $W$ preserve the volume form $\Omega$, we 
	have $dH = -i_V i_W \Omega$ and the symplectic 
	form is $\omega = -i_W \Omega$.
}
Thus the normal form is a skew-product system of the form
\beq{SNHflow}\begin{split}
	\dot r &= -\frac{\partial H}{\partial z} \;,\\
	\dot \theta &= \Omega(r,z) \;,\\
	\dot z &= \frac{\partial H}{\partial r} \,.
\end{split}
\eeq
Now consider the dynamics of \eqref{eq:SNHflow} near the origin. The first few terms in a power series expansion of the Hamiltonian are 
\[
	H(r,z) = r( A_{0,0} + A_{1,0}r + A_{0,1}z + A_{0,2} z^2 +\ldots) \;.
\]
When $A_{0,2} \neq 0$, the implicit function theorem implies that the coefficient $A_{0,1}$ can be eliminated by an affine shift in $z$. In the new coordinates, we replace $A_{0,0}$ by $-\delta$; this represents the unfolding parameter. The shape of the contours of $H$ near the origin for small $\delta$ are determined by the privileged scaling $r = \cO(\delta)$ and $z = \cO(\sqrt{\delta})$. In this case, as $\delta \to 0$ $H$ is equivalent to
\beq{unscaledHam}
	H(r,z) = r(-\delta + \beta r  + \alpha z^2) + \cO(\delta^{5/2}) \;.
\eeq

When $\alpha\beta\delta \neq 0$, the shape of the contours of $H$ depends only upon the two signs
\beq{signs}
    s_1 = \sgn (\delta \alpha)\;, \quad s_2 = \sgn(\alpha \beta) \;. 
\eeq
Indeed, the transformation 
\beq{scaleTrans}
	\rho = 2\left|\frac{\beta}{\delta}\right| r \;, \quad
	\zeta = \left| \frac{\alpha}{\delta}\right|^{\frac12} z \;,\quad
	\tau = h t \;, \quad
	h \equiv \sqrt{|\delta\alpha|} \sgn{\alpha} \;,
\eeq
leads to the scaled Hamiltonian
\beq{saddleNodeHam}
    \tilde{H} (\rho,\zeta) =  -s_1\rho + \frac{s_2}2\rho^2 + \rho \zeta^2 + 
    \cO(\sqrt{\delta}) \;,
\eeq
which has no continuous parameters. The implication is that there are two types of saddle-center-Hopf bifurcation, see \Fig{Vp_sn}. When $s_2 = 1$, three equilibria are created as $\delta\alpha$ changes from negative to positive. Two, at $(\rho,\zeta) = (0, \pm 1)$, are saddle-foci of the three-dimensional flow when $\Omega$ is nonzero. The third equilibrium is a center at $(1,0)$; this corresponds to an elliptic invariant circle in $\R^3$.

By contrast, when $s_2 = -1$, the bifurcation is subcritical: the saddle-foci exist when $s_1 < 0$; they annihilate at $\delta = 0$; and a hyperbolic invariant circle is created for $s_1 > 0$.

The separatrix of stable and unstable manifolds of the saddles is the contour $H=0$; it is the $\zeta$-axis together with the parabola
$
     \rho = 2 s_2 \left( s_1 - \zeta^2 \right) 
$.
When $s_1 = s_2 = 1$ this separatrix encloses a region of closed trajectories that become, for the three-dimensional flow, a family of two-dimensional invariant tori that surround the invariant circle. In the original, unscaled variables, the volume of the vortex-bubble as delimited by the invariant manifolds of the fixed points on the symmetry axis can be easily computed to be
\beq{volume}
  V_H = \int_{H<0} dr \wedge d\theta \wedge dz = \frac{8\pi \delta}{3\beta} \sqrt\frac{\delta}{\alpha}  \,.
\eeq
However if $s_1 = s_2 = -1$, then the stable and unstable manifolds of the saddles are unbounded and there is no heteroclinic connection apart from the $\zeta$-axis.

\InsertFigFour{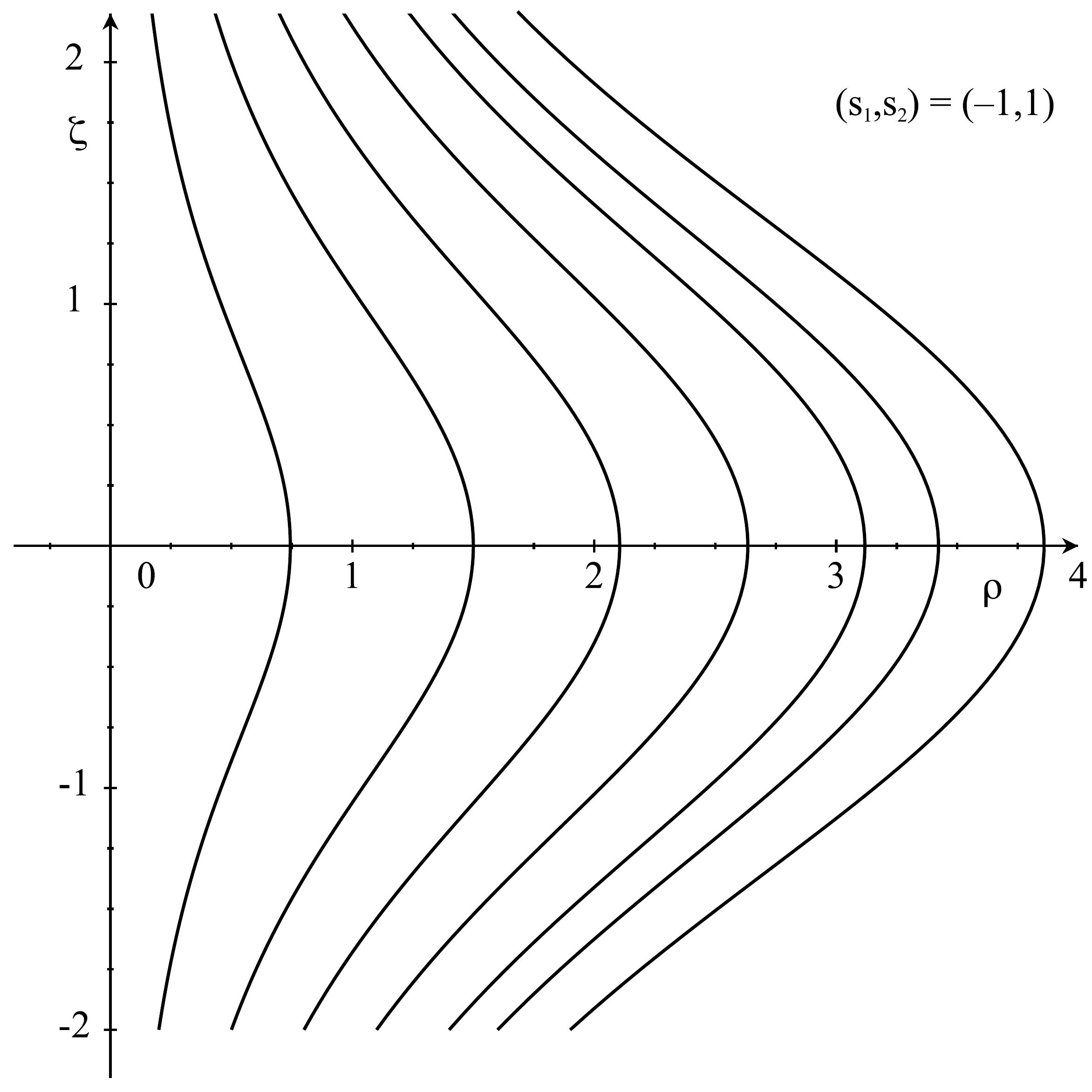}{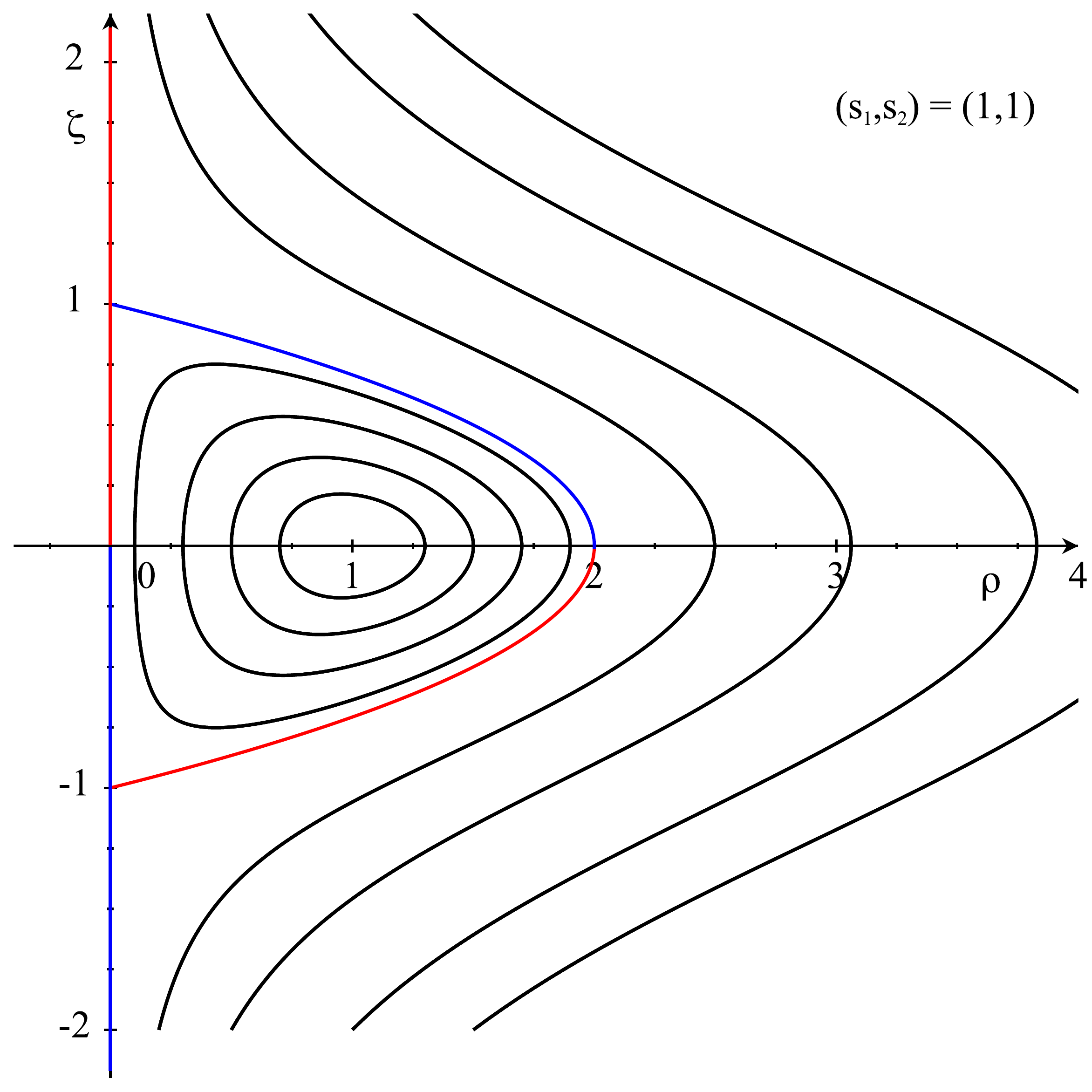}{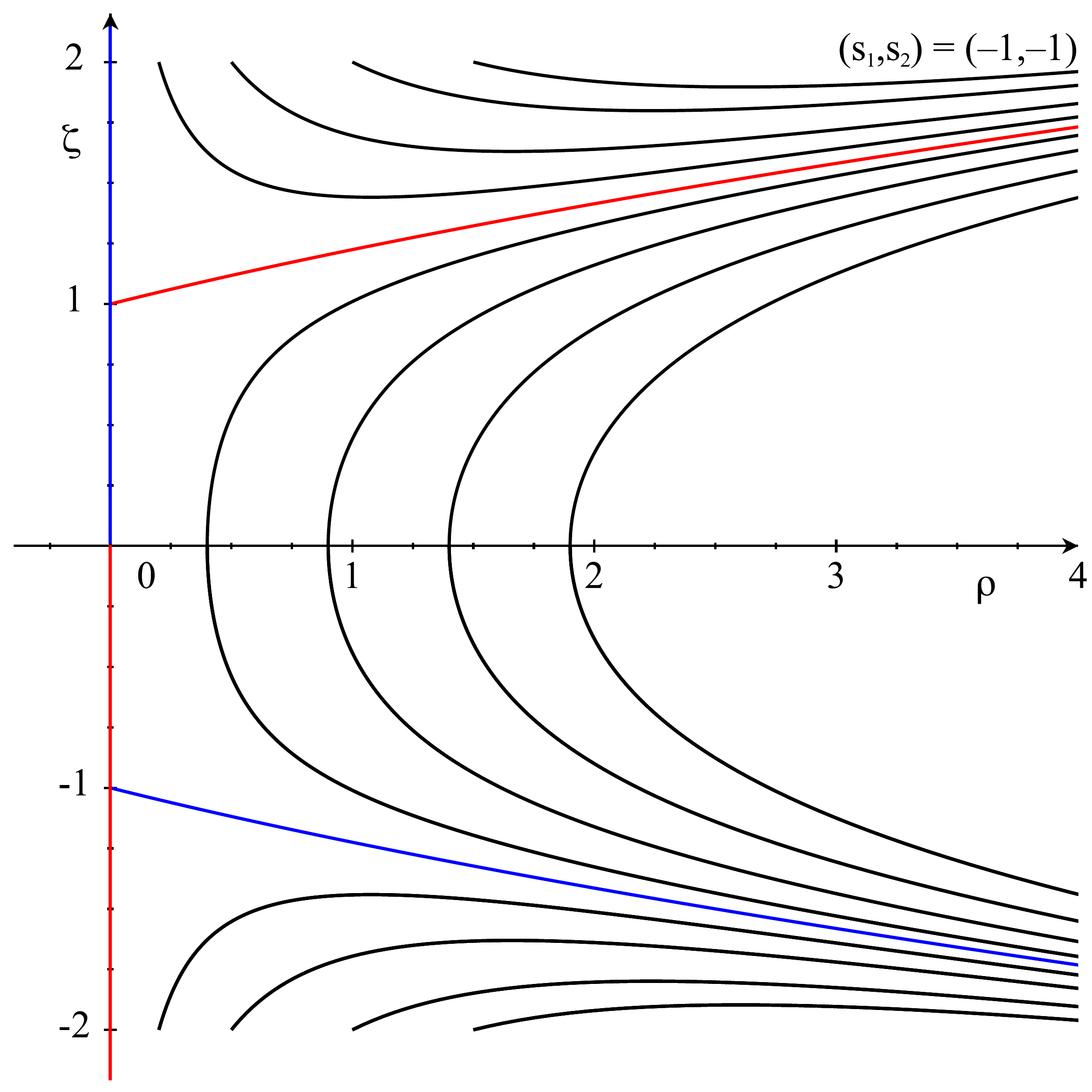}{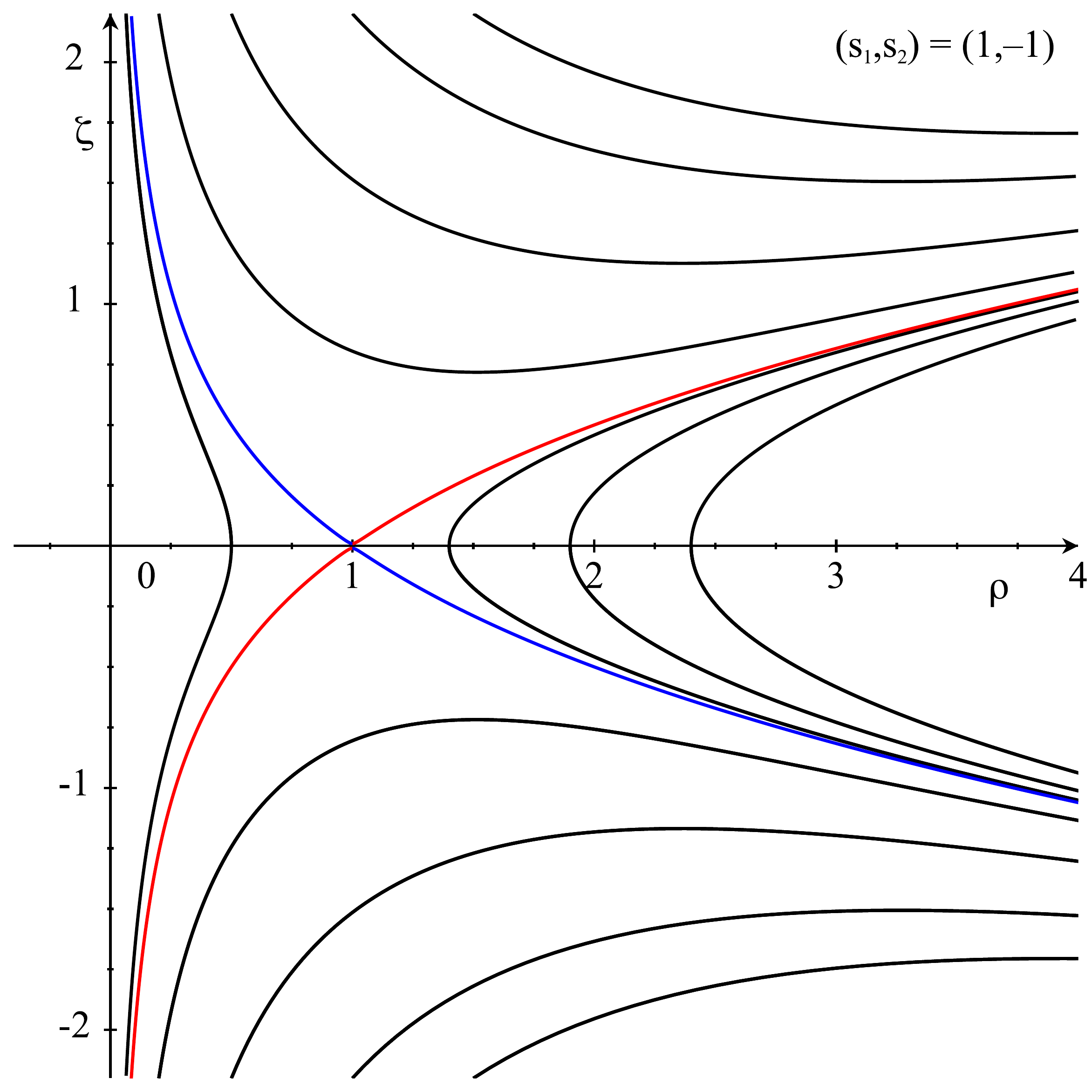}{Contours of the Hamiltonian \eqref{eq:saddleNodeHam} for the four choices of the signs $s_1 = \sgn{\delta\alpha}$ and $s_2 = \sgn{\alpha\beta}$. The upper two panes illustrate the supercritical case, $s_2 = 1$, and the lower two the subcritical case $s_2 = -1$. Stable and unstable manifolds of the saddle equilibria are shown in blue and red, respectively.}{Vp_sn}{2.5in}

On any invariant torus, the dynamics is conjugate to a rigid rotation. Since the flow of $H$ is integrable, there exist angle-action coordinates $(\phi,I)$ that are valid inside the separatrix. In these coordinates the normal form vector field would become
\[
     V = \nu(I) \partial_\phi  + \Omega(\phi,I) \partial_\theta \;,
\]
where $\nu(I) = \partial H / \partial I$ is the Hamiltonian frequency.
Thus the dynamics of the cylindrical angle can be trivially solved
to obtain
\[
   \theta(t) = \theta(0) + \frac{1}{\nu(I)} \int_{\phi(0)}^{\phi(t)} \Omega(\phi,I) d \phi \;.
\]
Consequently, the winding number on the torus is
\[
	\frac{1}{\nu(I)} \left< \Omega(\phi,I) \right>_\phi \;.
\]

As we will see below, the flow is a good model of the map for $\delta \ll 1$. Consequently, the structure that we observed in \Fig{Torus} corresponds to the creation of a bubble of bounded orbits in a supercritical saddle-center-Hopf bifurcation.

\subsection{Saddle-Center-Neimark-Sacker Bifurcation}\label{sec:SNHMap}

Consider a three-dimensional volume-preserving map with a fixed point whose multipliers are
\beq{HopfEigen}
   \lambda_1 = \bar{\lambda}_2 = \lambda \equiv e^{2\pi i \omega} \;,\; \lambda_3 =1 \;,
\eeq
where, without loss of generality, $\omega \in (0,\frac12)$. As for the flow case, it is convenient to use complex coordinates to diagonalize the linearization: let $\zeta =(u,v,z)$ where $u$ is the complex eigen-coordinate for $\lambda_1$, $v = \bar{u}$ for $\lambda_2$, and $z$ is the real eigen-coordinate for $\lambda_3$.

The dynamics in the neighborhood of this fixed point can be studied by a standard normal form analysis \cite{Murdock03,BridgesCushman93}.  We summarize the aspects that are new for the volume-preserving case in \App{SNHNormalForm}.

As for the flow case, the normal form, \eqref{eq:SNH2}, commutes to all orders with the phase shift $u \rightarrow u e^{i\psi}$. Consequently, the dynamics of $|u|$ and of $z$ are independent of those of the argument of $u$. To make this explicit, we use the symplectic, cylindrical coordinates \eqref{eq:cylindrical} so that if $u = x + i y$, the volume form obeys 
\beq{volumeForm}
	\Omega = dx \wedge dy\wedge dz = \frac{i}{2} du \wedge d\bar{u} \wedge dz 
	    = dr \wedge d\theta \wedge dz \;. 
\eeq
To cubic degree, the volume-preserving normal form obtained in \App{SNHNormalForm}, \eqref{eq:SNH2} is
\beq{SNHFinal}\begin{split}
     r'  &= r \left(1-2\alpha z -(\gamma+2\alpha\beta) r +(4\alpha^2-3\kappa)z^2 \right) \;, \\
     \theta' &= \theta + \Omega(r,z)   \;,\\   
     z'  &= -\delta + z + \alpha z^2 + 2\beta r + 2\gamma rz + \kappa z^3\;,     
\end{split}\eeq
where
\[
	2\pi \Omega(r,z) = 2\pi \omega + 
		 A_i z +2C_i r + (\alpha A_i+B_i) z^2 + (2\alpha C_i - 2A_i C_r)rz \;.
\]
Here $4 C_r \equiv -\gamma -2 \alpha\beta$, but the remaining eight parameters $(\alpha, \beta, \gamma,\delta, \kappa, A_i, B_i, C_i)$ are independent.
We summarize this result as the theorem:

\begin{thm}\label{thm:SCNS}
Suppose $f_p$ is a family of $C^3$ volume-preserving maps of $\R^3$ and that $f_0$ has a fixed point $x^*$ for which
\begin{itemize}
	\item $Df_0(x^*)$ has eigenvalues $(e^{ 2\pi i \omega},e^{-2\pi i \omega},1) $;
	\item $\omega \notin \{0, \frac14, \frac13, \frac12\}$ ;
	\item the vector $\frac{\partial f_p(x^*)}{\partial p}|_{p = 0}$ is
	not in the range of $Df_0(x^*)-I$.
\end{itemize}
Then for $p$ and $x-x^*$ sufficiently small, there is a coordinate change so that $f_p$ reduces to \eqref{eq:SNHFinal} through cubic order.
\end{thm}

The $(r,z)$ components of the map \eqref{eq:SNHFinal} are independent of $\theta$; consequently, the $(r,z)$ map is area preserving to $\cO(3)$. A typical phase portrait for this two-dimensional map is shown in \Fig{SNHPhasePortrait}.

\InsertFig{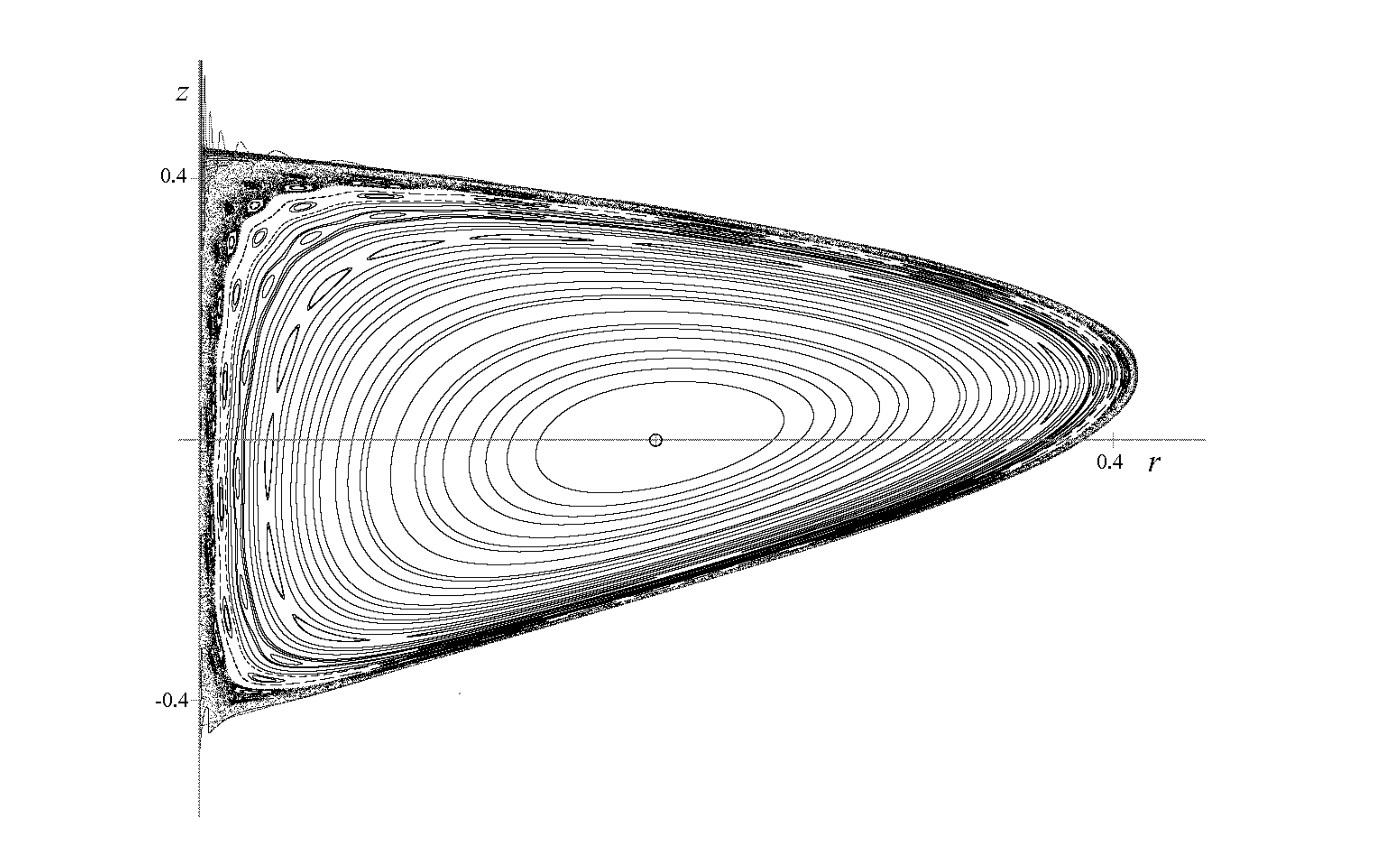}{Phase portrait of the map \eqref{eq:SNHFinal} for $\delta = 0.2$,
$\alpha = \beta = 1$, and $C_r=\gamma=\kappa=0$ in the $(r,z)$ plane. The unstable manifold of the lower saddle is shown in red and the stable manifold of the upper saddle in blue. Since
$C_r$ has been set to zero, the invariant circle is artificially at $z=0$.}{SNHPhasePortrait}{4in}

As $\delta \to 0$, the $(r,z)$ projection of \eqref{eq:SNHFinal} is approximately equivalent to the time $h$ map of the flow of the Hamiltonian \eqref{eq:saddleNodeHam}. To see this, introduce the scaled coordinates \eqref{eq:scaleTrans} to the map to obtain
\begin{align*}
	\rho' &= \rho -2 h  \rho \zeta + \cO(h^2) \;,\\
	\zeta' &= \zeta + h( -s_1 + \zeta^2 +s_2\rho) + \cO(h^2) \;.
\end{align*}
This, to $\cO(h)$, is the flow of \eqref{eq:saddleNodeHam};
consequently, the analysis of \Sec{SNHFlow} implies that the phase portrait of the map will asymptotically approach the pictures shown in \Fig{Vp_sn} as $\delta \to 0$.

To obtain the portrait in more detail, we study the map directly.
Since we have expanded in a power series, the valid fixed points of \eqref{eq:SNHFinal} emerge from the origin when $\delta = 0$. Assuming that $\alpha\beta \neq 0$, there are three such fixed points:
\beq{SNHfixedPoints}\begin{split}
 	(0,z_\pm) &= \left(0, \pm \sqrt{\frac{\delta}{\alpha}} \right)  
		+ \cO(\delta^{3/2}) \;,\\
	(r_c, z_c) &=  \left( \frac{\delta}{2\beta},\frac{C_r\delta}{\alpha\beta}  \right) 
		+ \cO(\delta^2) \;,
\end{split}\eeq
where $C_r$ is given by \eqref{eq:vpCoeffs}.

The stability of the fixed points can be classified by computing the ``residue"
\[
   R \equiv \frac14 (2-\tr Df ) = \alpha\beta r -\alpha^2 z^2 + \cO(3) \;.
\]
Recall that fixed points are hyperbolic saddles when $R < 0$, elliptic when $0 < R < 1$ and 
hyperbolic reflection-saddles when $R > 1$.

The fixed points on the $z$-axis exist when $\delta\alpha > 0$ and have residue
\[
	R_\pm =  -\delta\alpha +\cO(\delta^{3/2})\;.
\]
Thus these points are always hyperbolic saddles when they exist. The $z$-axis corresponds to the one-dimensional stable/unstable manifolds of these two fixed points with the corresponding multiplier $\lambda_\pm = 1+2\alpha z_\pm +\cO(\delta^{3/2})$.  Thus
when $\alpha > 0$ the $z$-axis is the unstable manifold of the upper fixed point and the stable manifold of the lower fixed point. For the three-dimensional map, the other two multipliers form a complex conjugate pair,
\[
	\lambda_\pm = (1-2\alpha z_\pm) e^{2\pi i \omega_\pm} +\cO(\delta^{3/2})\;.
\]
provided that $\omega_\pm = \Omega(0,z_\pm) \neq 0$.
The graph of two-dimensional invariant manifolds for these fixed points has the form\footnote{
	To make sense of this expansion, we should use the scaled variables 
	\eqref{eq:scaleTrans}. This scales the bubble to fixed size as $\delta \to 0$. 
	In this case each of the terms in the series begins with $\cO(\delta^0)$ terms.
}
\[
	 z = W(r) = z_\pm - \frac{\beta}{2\alpha z_\pm} r - \frac{\beta^2}{8 \alpha^2 z_\pm^3}r^2 + \cO(r^3) \;.
\]
Note that the manifold of the upper (lower) fixed point has negative (positive) slope when $\alpha\beta > 0$; these manifolds are thus inclined towards each other and, as we observed in \Fig{SNHPhasePortrait}, they generally intersect, enclosing, roughly speaking, a topological sphere. When $\alpha\beta <0$ the manifolds are inclined away from each other and, as $\delta \to 0$, there is no local intersection of the manifolds.

The third fixed point corresponds to an invariant circle $\cC$ of the three-dimensional map \eqref{eq:SNHFinal}. Since $r \ge 0$, the circle exists only when $\delta\beta > 0$. The dynamics restricted to the circle is a rigid rotation with rotation number
\beq{omegaL}
	\omega_L = \Omega(r_c,z_c) = \omega + \frac{1}{2\pi \alpha\beta} 
		\left(\alpha C_i +C_r A_i \right) \delta + \cO(\delta^2) \;.
\eeq
We call this the \textit{longitudinal rotation number} of $\cC$.

The invariant circle corresponds to a fixed point in the $(r,z)$ plane with residue
\[
	R_c = \frac{\alpha \delta}{2} + \cO(\delta^2) \;.
\]
Consequently if $\alpha\beta < 0$, the invariant circle is transversely hyperbolic. Alternatively, if $\alpha\beta >0$ the invariant circle is elliptic when $\alpha \delta< 2$. Nearby circles in the $(r,z)$ plane have a rotation number defined through $R = \sin^2(\pi \omega_T)$,  
\beq{omegaT}
	\omega_T = \frac{\sqrt{2\alpha\delta}}{2\pi} + \cO(\delta^{3/2}) \;.
\eeq
This is the \textit{transverse rotation number} of the invariant circle $\cC$.

\subsection{Quadratic Map near the Saddle-Center-Hopf Line}\label{sec:111SNH}
When $\eps$ is small and $-4 < \mu < 0$, the normal form \eqref{eq:StdMap} can be transformed to the saddle-center-Hopf form, \eqref{eq:SNHFinal}. As this transformation is somewhat tedious, we use computer algebra to perform the manipulations.

The fixed point at the origin for $\eps = 0$ and $-4 < \mu < 0$ has the linearization
\[
	Df = \begin{pmatrix} 1 & 1 & 0\\
						0 & 1+\mu & 1 \\
						0 & \mu & 1
		\end{pmatrix} \;.
\]
To apply the results of \Sec{SNHMap} this matrix is transformed to the diagonal form $M$, \eqref{eq:SNHstart}, using
\[
	T = \begin{pmatrix}\frac{1}{\mu} & \frac{1}{\mu} & 1 \\ 
	                   \frac{\lambda}{\lambda-1} & -\frac{1}{\lambda-1} & 0 \\
	                   1 & 1 & 0 
	    \end{pmatrix} \;.
\]
Applying this transformation to \eqref{eq:StdMap} gives the new map $T^{-1} \circ f \circ T$. 
Note that the Jacobian of this transformation is $\det{T} = i \cot{\pi\omega}$, so the volume in the transformed coordinate system must be scaled by this factor. The transformation $\psi$ to the new coordinates is constructed by Lie series and is obtained first to $\cO(\eps^0)$ through second order in $\xi$ and then to first order in $\eps$. As discussed in \App{SNHNormalForm}, this gives a map of the form \eqref{eq:SNH3}. A final affine transformation on $z$ restores the magnitude of the multipliers to $1$, giving a map of the form \eqref{eq:SNHFinal} with parameters
\beq{111Coefficients}
\begin{split}
	\alpha & = -\frac{a}{\mu} \;, \quad  
     \beta = -\frac{2a+(b-2c)\mu}{\mu^3} \;,\quad
	\gamma = \frac{2(a-b)(3+\mu)}{\mu(4+\mu)} \beta \;,\quad
    \delta = -\frac{\eps}{\mu} \;, \\
	A_i &= \frac{a-b}{\sqrt{-\mu(4+\mu)}} \;,\\
	B_i &= -\frac{ \left(4(\mu+2)(\mu+3) a^2 + \mu(4(\mu+4)c-(5\mu +16)b)a + b^2 \mu(\mu +2)\right)}{2\mu(-\mu(\mu +4))^{3/2}} \;,\\
	C_i &= \frac{-2 (\mu +1) a^2+\mu  (2 c (5 \mu+13)-b (3 \mu +5)) a-\mu  
	 \left(4 \mu (2 \mu +5) c^2-2 b \mu  (3 \mu +7)c+b^2 (\mu +1)^2\right)}
	 {2 \mu ^4 (\mu+3) \sqrt{-\mu  (\mu +4)}} \;,\\
  \omega &= \omega_0(\mu) + \eps \frac{b^2-4ac}{4\pi a \mu \sqrt{-\mu(4+\mu)}} + \cO(\eps^2) \;,\\
\end{split}
\eeq
where $\omega_0(\mu)$ is given in \eqref{eq:fixedPtRotNum}.
Note that these coefficients have singularities when $\mu \in \{-4,-3,0\}$ corresponding to the resonant frequencies $\omega \in \{\frac12, \frac13, 0\}$, respectively.\footnote
{
	The value $(b-2c)\mu +2a = 0$ is also special since at this point 
	both $\beta$ and $\gamma$ vanish. This corresponds to the boundary 
	between supercritical and subcritical saddle-center-Hopf bifurcations. 
	To study the dynamics at this point would require computing higher order terms.
}

Thus the dynamics of \eqref{eq:StdMap} near the saddle-center-Hopf line are approximated by the normal form \eqref{eq:SNHFinal} away from the low-order resonances. The signs  \eqref{eq:signs} that determine the character of the bifurcation now become
\begin{align*}
	s_1 &= \sgn(\delta \alpha) = \sgn(\eps a) \;,\\
	s_2 &= \sgn(\alpha \beta) = \sgn(a) \sgn (2a+(b-2c)\mu)) \;.
\end{align*}
Recall from \Sec{SNHFlow} that the bifurcation is super(sub)-critical for $s_2 =1$ ($=-1$), see \Fig{Vp_sn}. For the parameter values $(a,b,c) = (1,\frac12,\frac12)$ that we have primarily studied, $s_2 = 1$ for all $\mu < 0$, thus the bifurcation is always supercritical. Consequently as $\eps$ increases through zero, an elliptic invariant circle and a pair of saddle-focus fixed points whose two-dimensional manifolds intersect are created. This is confirmed by numerical computations; for example in \Fig{Torus} where $\eps = 0.01$, the map is nearly integrable and the ball enclosed by the two-dimensional manifolds is predominantly filled with invariant tori. As $\eps$ increases the intersection angle between the two-dimensional stable and unstable manifolds of the saddle-foci grows, many of the tori are destroyed, and orbits near these manifolds become unbounded, see \Fig{OrbitsMu-24}. Along the way there are many torus bifurcations that create new families helically wound around the original invariant circle. We postpone a discussion of these until \Sec{Resonances}.

Meanwhile, these results permit comparison with \Fig{relativeVolume} near $\eps = 0$. Taking into account the scaling of the volume due to the transformation $T$ and that due to the complex transformation \eqref{eq:volumeForm}, the fraction of bounded orbits in the box $\cB$, \eqref{eq:box} is
\beq{relativeVol}
  V_{\mbox{rel}} = 2i\frac{V_H}{{\mbox{vol}(\cB)} \det{T}} = 
   \frac{\pi a}{96\beta} \sqrt{\frac{4+\mu}{-\mu^3}},
\eeq
where $V_H$ is the volume \eqref{eq:volume}. This formula gives the dashed curve shown in \Fig{relativeVolume}. Since $\beta = O(\mu^{-3})$, \eqref{eq:relativeVol} implies that the bounded fraction is $O(\mu^{3/2})$ when $\mu \to 0^-$. The actual results in \Fig{relativeVolume} deviate from this result when $|\mu| < 0.05$ due to the $\omega = 0$ resonance. Moreover, the resonances near $\mu = -2$ and $-3$ are also not captured by this lowest-order formula. However, \eqref{eq:relativeVol} agrees well with the numerical results near the period-doubling point $\mu = -4$.

\InsertFig{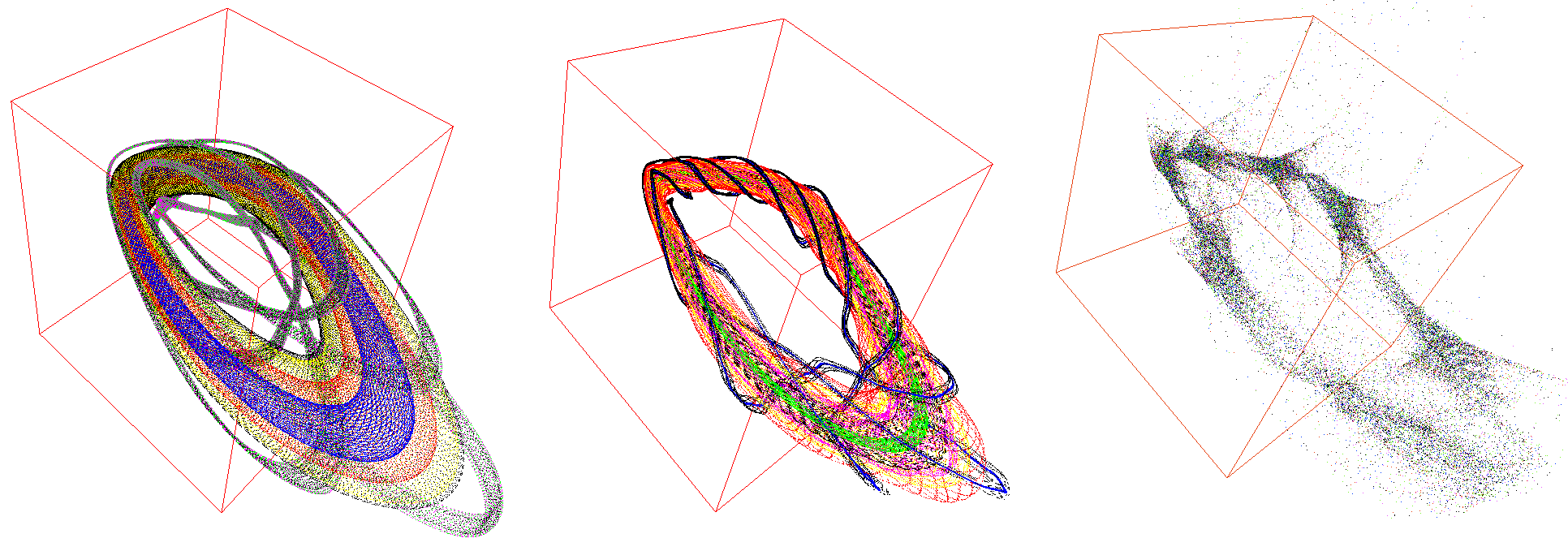}{Orbits of the map \eqref{eq:StdMap} for $\mu = -2.4$, $a=1$, $b=c=\frac12$, and  $\eps = 0.1$ (left), $0.2$ (center), and $0.32$ (right). In the last case there appear to be no bounded orbits, however there is probably a pair of period $7$ saddles. The
orbits that are shown lie near the stable and unstable manifolds of these orbits. The red cubes are centered on the origin and have sides with length $\frac{10}{3}\sqrt{\eps}$.}{OrbitsMu-24}{6in}

From \eqref{eq:111Coefficients} we can compute the position of the fixed points and invariant circle \eqref{eq:SNHfixedPoints}. These expressions are rather complicated and not especially useful. 
However, the rotation numbers of the invariant circle will be used extensively in the following sections.
%
For the values $(a,b,c) = (1,\frac12, \frac12)$ these become
\beq{RotNormForm}\begin{split}
	\omega_L &= \omega_0(\mu) +\frac{5 \mu ^3+19 \mu^2-28 \mu-64} 
		{16 \pi \mu(\mu+3)(\mu-4)(-\mu(\mu+4))^{3/2}}\epsilon + \cO(\eps^2) \;,\\
	\omega_T &=  -\frac{1}{\pi\mu}\sqrt{\frac{\eps}{2}}\left(1 -
		g(\mu) \eps + \cO(\eps^2) \right) \;, \\
	g(\mu) &\equiv \frac{58\mu^7+648\mu^6-377\mu^5-23526\mu^4-59027\mu^3+139032\mu^2+
	689616 \mu +695808}{96 \mu^3 (\mu-4)^2 (\mu +3)^2 (\mu+4)^2} \;.
\end{split}\eeq

As a first step in understanding the invariant tori, we will study the evolution of the invariant circle. As we will see, resonances between the rotation numbers 
$(\omega_L, \omega_T)$ will be key to this discussion.

\section{Invariant Circles}\label{sec:InvCircles}

In a volume-preserving flow, the supercritical, saddle-center-Hopf bifurcation creates a pair of saddle-focus fixed points, one of type-$(2,1)$ and one of type-$(1,2)$ separated by $\cO(\sqrt{\delta})$, recall \Sec{SNHFlow}. The two-dimensional unstable and stable manifolds of these fixed points coincide in the formal normal form and bound a ball that is foliated by a family of two-dimensional invariant tori. These tori enclose an invariant circle that has a diameter of size $\cO(\sqrt{\delta})$.

As we showed in \Sec{SNHMap}, the analogous bifurcation for a volume-preserving map has a normal form \eqref{eq:SNHFinal} that is approximated by this flow when $\delta \ll 1$. However, the cylindrical symmetry of both normal forms is only formal: even when $\delta$ is very small, the invariant circles presumably exist only for a Cantor set of parameter values when their rotation numbers satisfy appropriate Diophantine conditions \cite{Cheng90b, Xia96}.

Our goal in this section is to study numerically the invariant circles for the quadratic map \eqref{eq:StdMap} that are created in the saddle-center-Neimark Sacker bifurcation. In order to do this, we must first develop an algorithm to accurately compute elliptic invariant circles.

\subsection{Ellipsoid Algorithm} \label{sec:Algo}

A number of algorithms have been proposed for finding invariant circles and tori; some of the history is discussed in \cite{Schilder05, Haro06b}. Many of the techniques assume that the dynamics on the invariant circle is conjugate to a rigid rotation with a given rotation number $\omega_L$. In this case, the conjugacy can be expanded in a Fourier series and a Newton method employed to compute the Fourier coefficients. Castello and Jorba have used this idea for Hamiltonian systems that have a Cantor set of circles whose frequencies vary  \cite{Castella00}. If the gaps in the Cantor set are ``small" they can fix a rotation number and search for the corresponding circle. Haro and de la Llave \cite{Haro06a, Haro06b, Haro07} also use spectral methods to find tori in quasiperiodically forced maps. The important simplification for the forced case is that the rotation vector and the conjugacy to rigid rotation are automatically known. A very different method was developed by Edoh and Lorenz \cite{Edoh03} to find nonsmooth, but attracting invariant circles; here the asymptotic stability of the invariant circles is key. A related method is to define a variational method, using the distance between the circle and its image, and obtain the circle by following the gradient flow \cite{Lan06}.

An iteration-based method was developed by Simo \cite{Simo98}. Letting $T = \frac{1}{\omega_L}$ be the ``irrational period", Simo defined the $T$th iterate of the map by interpolation between the points $f^{\floor T}$ and $f^{\ceil T}$.\footnote
{
	Note that these points may be far apart and so the interpolation may not be accurate.
}
An invariant circle corresponds, roughly speaking, to a fixed point of $f^T$.
An advantage of this method is that it does not depend upon the existence of a conjugacy. It can also be used to determine $\omega_L$ by fixing a section and looking for approximate returns to define $T$ \cite{Castella00}.

Our method is also iteration-based, but in contrast to Simo's idea, uses a slice to localize iterates and a number of points to locate the invariant circle. Suppose $\cC$ is an elliptic invariant circle with a local cross section $\Sigma$; for \eqref{eq:StdMap}, the section
\[
	\Sigma = \{(x,y,0): y > 0\}
\]
works in most cases. Let $\Sigma_\Delta = \{ |z| < \Delta, y > 0 \}$ be a thin slice enclosing $\Sigma$. Assume that there is a neighborhood of $\cC$ in which the local dynamics predominantly lies on two-dimensional invariant tori that enclose $\cC$. Thus the intersection of a typical orbit near $\cC$ with $\Sigma_\Delta$ is a slightly curved, elliptic cylinder, see \Fig{Ellipsoid}.

\InsertFig{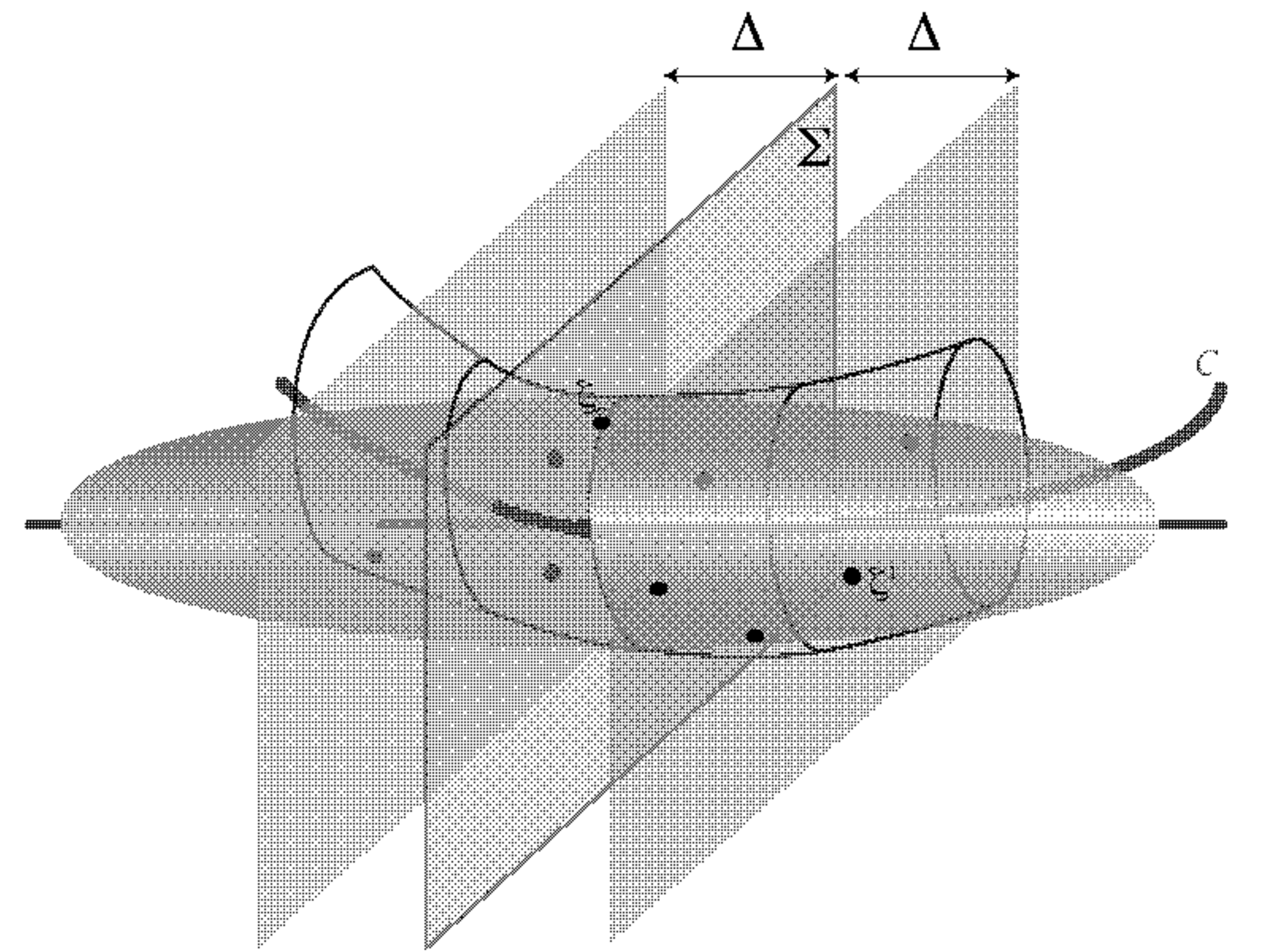}{Finding an invariant circle by fitting a section of a nearby torus to an ellipsoid.}{Ellipsoid}{4in}

Begin with a point $\zeta_0 \in \Sigma $ in the neighborhood of $\cC$ and find a set of $N$ returns of its orbit to $\Sigma_\Delta$: 
\[
	\zeta^{i} = f^{t_i}(\zeta_0) \in \Sigma_\Delta\;, \quad 
	0 < t_1 < t_2 < \ldots <t_N \;, \quad  
	i = 1,\ldots , N \;.
\] 
By assumption, these points lie on an invariant torus; the goal is to find the ``axis" of this torus. Instead of fitting the points to an elliptic cylinder, it is easier to fit them to a general ellipsoid, defined as level set of a quadratic form
\[
  e(x,y,z) = Ax^2 + By^2 + Cz^2 + Dxy + Exz + Fyz + Gx + Hy + Jz \;.
\]
Since the returning points are, presumably, all close to the circle and since they lie on the ellipsoid by hypothesis, consider the $N$ differences $\delta \zeta^i = \zeta^i -\zeta_0$. In this coordinate system the ellipsoid goes through the origin, so $e(\delta \zeta^i) = 0$. 
Since all the $\delta \zeta^i$ approach zero when the method converges we normalize by $l =\frac{1}{N} \sum_{i=1}^N | \delta \zeta^i |$ and introduce $\delta \xi^i = \delta \zeta^i/l$.
This removes the singular behavior from the algorithm that originates from the fact that
the fitted ellipsoid becomes thinner, and it also helps avoid round-off errors. 

Since the equations $e(\delta \xi^i) = 0$ are homogeneous in the coefficients, an additional equation is needed to fix the solution; we chose to normalize the sum of the squared coefficients of $e$, giving $N+1$ linear equations in the coefficients:
\begin{align*}
	e(\delta \xi^i) &= 0 \;, \quad i = 1,\ldots, N \;, \\
	A+B+C &= 1 \;.
\end{align*}
We must choose at least $N = 8$ to fix the $9$ coefficients of $e$. It seems most convenient to use exactly $8$ returns, since the number of iterates to find a return can be quite large, especially when the slice thickness, $\Delta$, is small.\footnote
{
This could be a problem if the points do not lie in a general position; to get around this one could use more points and solve the system using least squares.
}

Once the ellipsoid is determined, the center of the ellipse in the section $\Sigma$ is used as the next guess for a point on $\cC$; for the section $\Sigma = \{z = 0\}$, this gives the iterative step
\[
   \zeta_0 \mapsto \zeta_0 + l \frac{1}{4AC-B^2} ( BH-2CG,BG-2AH,0) \;.
\]
The whole process is then repeated, finding a new set of returns to the slice $\Sigma_\Delta$ and fitting a new ellipsoid. The error is estimated as the change in distance between the new point and the previous one. The iteration stops either when the error decreases below an error tolerance (we typically chose $10^{-10}$), or after a fixed number of iterations (we chose $10$). If the given tolerance is not obtained for a fixed $\Delta$, we decrease $\Delta$ by a factor of $100$ and then restart the algorithm with the best previous value as the initial point $\zeta_0$ (the results typically converged when $\Delta = 10^{-6}$). 

This process appears to be quadratically convergent, see \Fig{CircleError}. Indeed, it is not hard to show that this is the case for a two-dimensional version of this algorithm for finding an elliptic fixed point: since the linearized map has invariant ellipses the linearization of our iteration at a solution is superstable.

We used our algorithm to compute invariant circles for the normal form \eqref{eq:StdMap} for a grid of parameters near the saddle-center-Hopf line with the standard values $a=1, b=c=\frac12$.  Typically we started at a fixed value of $\mu$ with $\eps = 0.01$. If we are able to find the invariant circle, $\eps$ is incremented using the previous solution as an initial guess for the new parameters.

\InsertFig{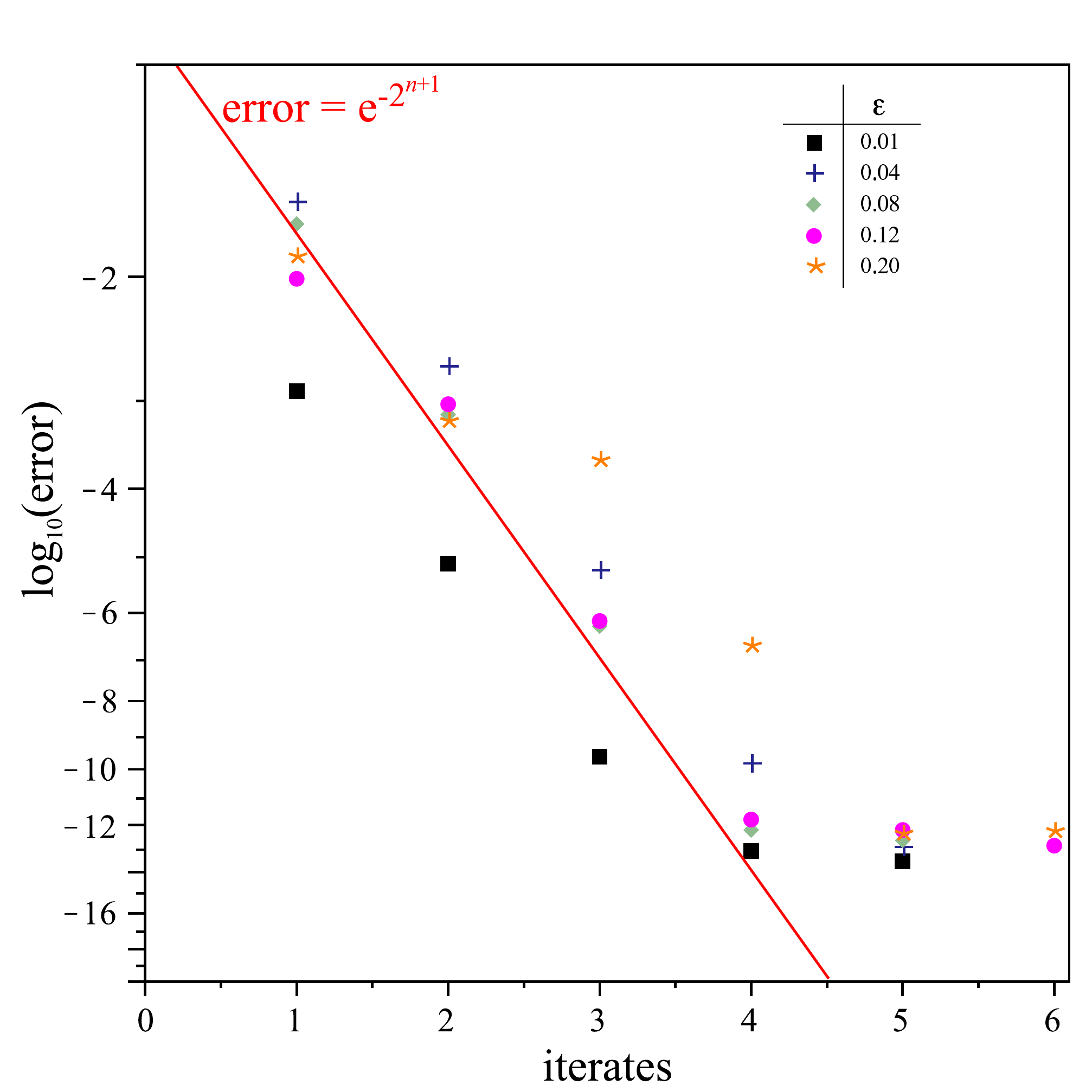}{Error on a log(log) scale for successive iterates of the invariant circle position for $\mu = -2.4$, with $\Delta = 10^{-7}$ and the values of $\eps$ indicated. The line represents an error of $e^{-2^{n+1}}$ which would mean quadratic convergence. With this value of $\Delta$, the minimal error seems to be about $10^{-13}$, and is achieved in $4-6$ iterations.}{CircleError}{3.5in}

An example of the output of this algorithm with $\mu = -2.4$ is given in \Tbl{circles}. For this value of $\mu$ the algorithm finds an invariant circle when $\eps < 0.3129$. Beyond this value it no longer converges; moreover, when $\eps > 0.326$ the computation of \Sec{BoundedOrbits} (with a $500^3$ resolution and $10^4$ iterates) also gives no bounded orbits. Consequently we believe that the circle is either destroyed or becomes unstable in this range of $\eps$. Several of the computed invariant circles are overlaid in \Fig{CirclePlot}. This figure also shows a chaotic orbit for $\eps = 0.315$ near where the invariant circle used to be. 

\begin{table}
\centering
\begin{tabular}{c|cc|c|cc}
$\eps$  	 & 		$x$			  &   $y$		 	 &	  error 	   &  $\omega_L$		  &	  $\omega_T$	 \\ \hline
 0.01    &	 -0.000077100956  &	 0.110334966059   &	1.6e-11    &	   0.282171317669   &	  0.00939482\\ 
 0.02    &	 -0.000218667148  &	 0.156587932592   &	2.7e-11    &	   0.282294227518   &	  0.01331187\\ 
 0.03    &	 -0.000401253015  &	 0.192348589289   &	7.8e-12    &	   0.282415802529   &	  0.01633514\\ 
 0.04    &	 -0.000606226180  &	 0.222722556427   &	1.5e-11    &	   0.282536006132   &	  0.01889745\\ 
 0.05    &	 -0.000263469053  &	 0.250886913814   &	1.1e-11    &	   0.282654815793   &	  0.02117252\\ 
 0.06    &	 -0.001281607227  &	 0.273786477131   &	1.4e-13    &   0.282772154554   &	  0.02324300\\   
 0.07    &	 -0.001578623424  &	 0.296421783954   &	8.4e-11    &	   0.282888017858   &	  0.02516009\\ 
 0.08    &	 -0.001928385257  &	 0.317549729412   &	9.6e-12    &	   0.283002357933   &	  0.02695743\\ 
 0.09    &	 -0.002313763896  &	 0.337472296148   &	3.3e-12    &	   0.283115135201   &	  0.02865947\\ 
 0.10    &	 -0.002732253247  &	 0.356389819268   &	7.2e-13    &	   0.283226310896   &	  0.03028362\\ 
 0.11    &	 -0.003191438702  &	 0.374443365374   &	1.1e-11    &	   0.283335847312   &	  0.03184166\\ 
 0.12    &	 -0.003632379081  &	 0.391797224416   &	8.9e-11    &	   0.283443708609   &	  0.03309808\\ 
 0.13    &	 -0.004150372053  &	 0.408459677586   &	3.4e-12    &	   0.283549859693   &	  0.03480205\\ 
 0.14    &	 -0.004689339128  &	 0.424551322589   &	2.5e-12    &	   0.283654271356   &	  0.03622014\\ 
 0.15    &	 -0.005256682085  &	 0.440125047284   &	2.9e-11    &	   0.283756917316   &	  0.03760506\\ 
 0.16    &	 -0.005618175619  &	 0.455286683823   &	5.7e-11    &	   0.283857779176   &	  0.03896231\\ 
 0.17    &	 -0.006487124526  &	 0.469903756471   &	5.7e-12    &	   0.283956842148   &	  0.04029660\\ 
 0.18    &	 -0.007146262564  &	 0.484186136060   &	4.7e-12    &	   0.284054109448   &	  0.04161036\\ 
 0.19    &	 -0.007838687257  &	 0.498105595397   &	6.6e-12    &	   0.284149593962   &	  0.04291542\\ 
 0.20    &	 -0.008559583750  &	 0.511685650160   &	4.9e-11    &	   0.284243328335   &	  0.04420893\\ 
 0.21    &	 -0.009315347590  &	 0.524951884960   &	6.6e-11    &	   0.284335369737   &	  0.04549702\\ 
 0.22    &	 -0.010104271994  &	 0.537923761008   &	2.9e-12    &	   0.284425807465   &	  0.04678551\\ 
 0.23    &	 -0.010926038044  &	 0.550616855400   &	5.2e-11    &	   0.284514773462   &	  0.04807963\\ 
 0.24    &	 -0.011781835115  &	 0.563047279372   &	1.4e-11    &	   0.284602457095   &	  0.04938499\\ 
 0.25    &	 -0.012671957974  &	 0.575230413835   &	3.6e-12    &	   0.284689126532   &	  0.05070952\\ 
 0.26    &	 -0.013597339065  &	 0.587176221467   &	9.2e-11    &	   0.284775161033   &	  0.05206063\\ 
 0.27    &	 -0.014558693626  &	 0.598897558338   &	2.4e-11    &	   0.284861102396   &	  0.05345099\\ 
 0.28    &	 -0.015555864549  &	 0.610403755970   &	2.8e-13    &	   0.284947742217   &	  0.05489500\\ 
 0.29    &	 -0.016590212710  &	 0.621704492198   &	5.2e-12    &	   0.285036283041   &	  0.05642285\\ 
 0.30    &	 -0.017662158192  &	 0.632808106481   &	3.0e-11    &	   0.285128671076   &	  0.05807352\\ 
 0.31    &	 -0.018772488310  &	 0.643722521310   &	8.1e-12    &	   0.285228402490   &	  0.06007135\\
 \end{tabular}
\caption{Position $(x,y,0)$ on $\Sigma$ and rotation numbers of the elliptic invariant circle of \eqref{eq:StdMap} for $\mu = -2.4$, and $(a,b,c) =(1,\frac12,\frac12)$ as function of $\eps$. The error is the distance on $\Sigma$ between the last and the penultimate iterates. We attempted
to find a circle with an error $< 10^{-10}$ using a slice thickness $\Delta \le 10^{-6}$.}
\label{tbl:circles}
\end{table}

\InsertFig{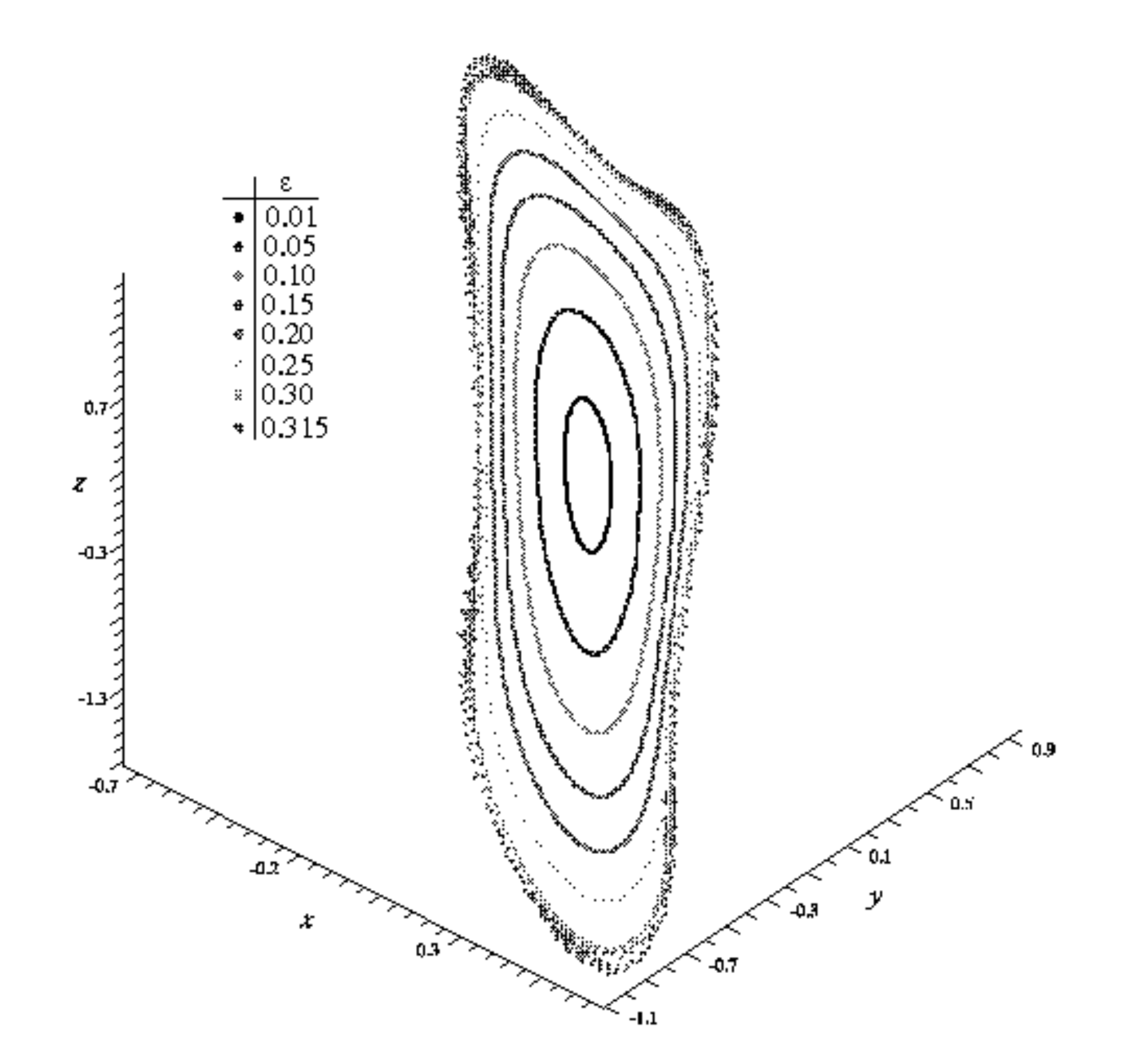}{Several of the invariant circles for $\mu = -2.4$ and $\eps$ values as shown. Each circle is displayed with $1000$ iterates. For the outermost set of points, at $\eps = 0.315$, there is apparently no invariant circle.}{CirclePlot}{4in}

The ellipsoid algorithm can fail in two ways. One the one hand if the invariant circle has a very small stable neighborhood, the initial guess $\zeta_0$ may lie on an unbounded trajectory even when there is an elliptic circle. This typicially happens as the parameters approach a bifurcation where $\cC$ becomes hyperbolic. To approach the stability border, we can simply take smaller steps in $\eps$. This is apparently what happens for $\eps> 0.31$ when $\mu = -2.4$. A second failure mode occurs when the orbits remain bounded, but the given error tolerance cannot be achieved; this indicates that there is no elliptic invariant circle. For example, when $\mu =-2.4$, the algorithm does not converge for $ 0.3018 \le \eps \le 0.3044$; in this range, the circle appears to be unstable and to have undergone a doubling bifurcation, see \Sec{Resonances}. 

In our computations for other values of $\mu$, we observe that the invariant circle often undergoes a complex sequence of bifurcations as $\eps$ changes, sometimes loosing stability or simply vanishing for one $\eps$ and then regaining stability or reforming for a larger value of $\eps$; more details are discussed in \Sec{Resonances} below. In every case there appears to be a maximal value of $\eps$ beyond which the circle ceases to exist or never regain stability. To make more sense of these bifurcations, we compute the longitudinal and transverse rotation numbers.

\subsection{Longitudinal Rotation Number}
If there is an invariant circle, then the restriction of the map to the circle, $f|_\cC$, is a homeomorphism. Recall that the rotation number of a circle homeomorphism always exists and is independent of the initial point. This is the longitudinal rotation number of the circle, $\omega_L$.

The longitudinal number can be easily computed under the assumption that the invariant circle projects to the plane $x=0$ as a Jordan curve that encloses the the origin,\footnote
{
	If this is not the case, we could also use the \textit{self-rotation} number
	\cite{Dullin00}.
}
and that the angle advance between iterates ``avoids an angle'' \cite{Dullin00}.
The avoided angle is chosen as the location for the branch cut 
of the arctan function. Typically the avoided angle can be chosen as the 
angle diametrically opposite to the average rotation angle.
Then we simply sum the polar angular increments between images, defining
\[
	\Theta(N) = \frac{1}{2\pi} \sum_{t=1}^{N} 
			 \mbox{atan2}(y_t y_{t-1} + z_t z_{t-1}, y_t z_{t-1} - z_t y_{t-1})
\]
where $\mbox{atan2}(y,x) = \mbox{arg}(x + i y)$. An approximate rotation number is then simply $\Theta(N)/N + \cO(N^{-1})$. 

However as was first suggested by H\'enon \cite{Efstathiou01}, a more accurate value for $\omega_L$ is easily obtained by carefully choosing $N$. To do this, compute the continued fraction expansion of $\omega_L$ by constructing the sequence, $\frac{p_j}{q_j}$, of its continued fraction convergents. Each convergent is defined to be a rational number closer to $\omega_L$ than any others with denominators $q < q_j$. Since each orbit on $\cC$ must be ordered as a rigid rotation with rotation number $\omega_L$, we can obtain the convergents by finding the closest returns to the initial point. Let $q_j$ be the time of the next closest return to the initial point, $d_{j} = |\zeta_{q_j} - \zeta_0| < d_{j-1}$, and $p_j = \floor{\Theta(q_j) + \frac12}$ be the nearest integral number of rotations. To start the process we arbitrarily chose $d_{j_0} = 0.01$; this means that the first computed convergent will not be the leading convergent of the continued fraction for $\omega_L$; however, it is easy to find the earlier convergents from the continued fraction expansion for $p_1/q_1$.
For example with $(\eps,\mu) = (0.2, -2.4)$, the computed convergents are
\begin{align*}
	q_j &= \{971, 3205, 7381, 150825, 158206, 309031, 776268 \} \;, \\
	p_j &= \{276, 911, 2098,  42871, 44969, 87840, 220649 \} \;,
\end{align*}
which implies that the continued fraction for $\omega_L$ is
\[
	\omega_L \approx \frac{220649}{776268} = [0,3,1,1,13,3,3,3,2,20,1,1,2] \;.
\]
Note that this rotation number is relatively close to the rational $\frac27 = [0,3,1,1]$---see \Sec{Resonances}.
Recall that the rotation number is close to its convergents:
\[
	\left|\omega_L  - \frac{p_j}{q_j}\right| <  \frac{1}{q_j^{2}}\;.
\]
Indeed, this is the error observed in practice: \Fig{FreqError} shows the difference between the value at the $j$th step and the final value.
As discussed in \cite{Efstathiou01}, a value correct to $\cO(q_j^{-4})$ can be obtained by adding to $\omega_L$ the average angular deviation from exact $q_j$-periodicity over the next $q_j$ iterations; however, the estimate above is sufficient for our purposes.  

\InsertFig{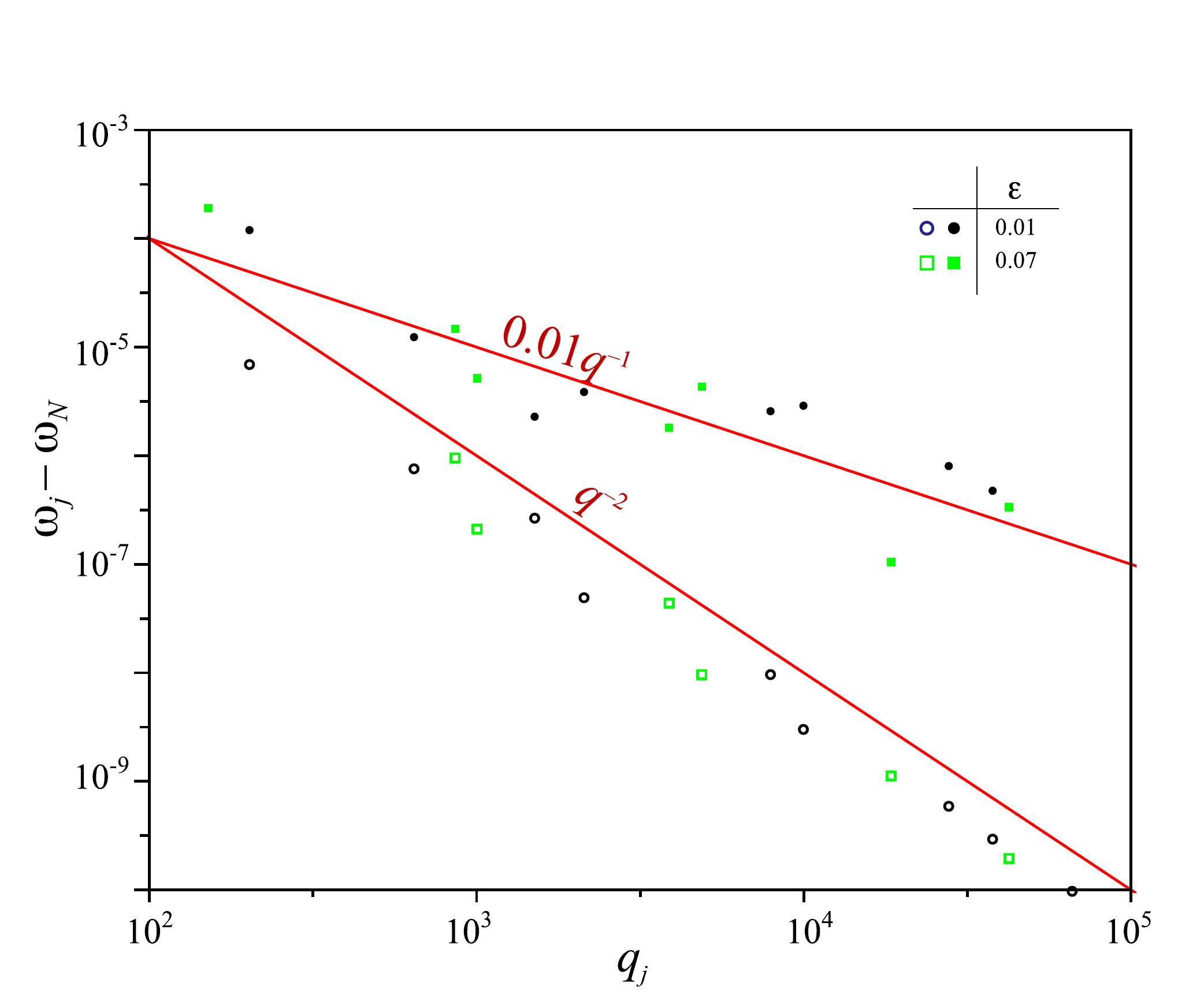}{Error in the computation of the rotation numbers of an invariant circle at $\mu = -2.4$, for two different $\eps$ values as shown. The open points correspond to $\omega_L$ and the solid to $\omega_T$. The error in the longitudinal rotation number is $\cO(q^{-2})$, but that in the transverse rotation number is only $\cO(q^{-1})$.}{FreqError}{3.5in}

The values in \Tbl{circles} were computed using the sequence of closest returns for $t \le 10^6$, giving a result accurate to $\cO(10^{-12})$. We can also use this computation to check that the orbit is properly ordered on the circle. Indeed, the sequence $q_j$ must obey the recursion
\[
	q_{j+1} = a_{j} q_{j} + q_{j-1} \;,
\]
where $q_{-1} = 0$, and $q_0 = 1$, and $a_j \in \N$ are the continued fraction elements. Thus a necessary condition for a valid sequence of convergents is that
\[
	(q_{j+1} -q_{j-1}) \mod {q_j} = 0 \;.
\]
Usually this criterion only fails when the error bound in the circle algorithm is also not achieved.

The computations of $\omega_L$ are compared with the normal form results 
\eqref{eq:RotNormForm} in \Fig{OmegaL}. The dominant behavior is the zeroth order rotation number $\omega_0(\mu)$ in \eqref{eq:fixedPtRotNum}. For the figure we subtracted this value from the computed results and then compare with the theoretical $\cO(\eps)$ term. The curves in the figure show the results for fixed values of $\eps$ as a function of $\mu$. The agreement between the numerical results and the theory is nearly perfect for $\eps < 0.15$ away from the resonances where the normal form is not valid. The computations show that an ellipitical invariant circle does not even exist near the main resonances $\mu = -4, -3, -2$ and $0$, see \Sec{Resonances}. The numerical results indicate an additional singularity in $\omega_L$ near $\mu = -2$ that is not present in the normal form to $\cO(\eps)$. To find this singularity in the normal form we would have to keep quartic terms.

\InsertFig{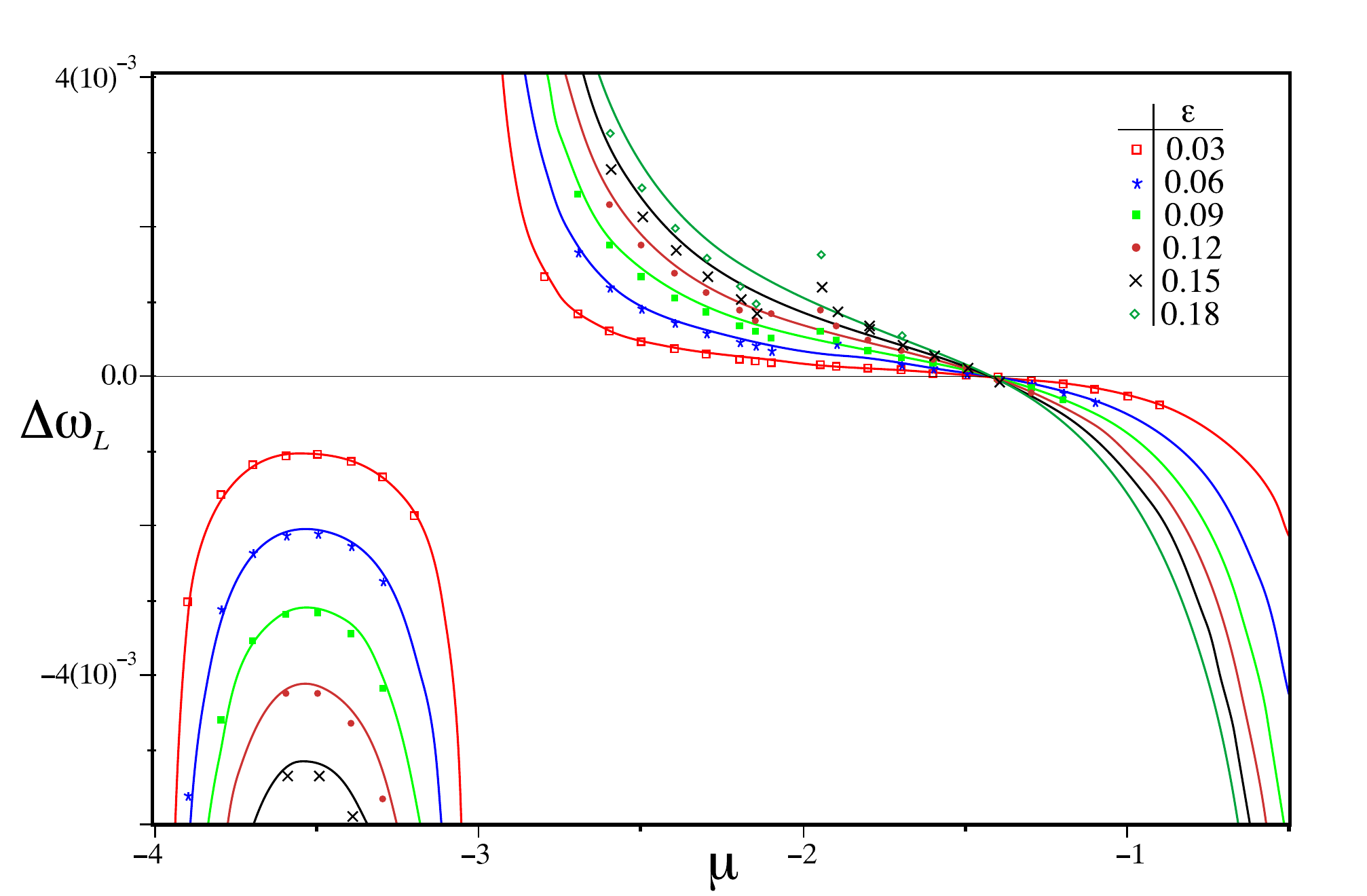}{Comparison of the computed $\Delta \omega_L = \omega_L - \omega_0(\mu)$ (dots) with the normal form \eqref{eq:RotNormForm} (curves) as a function of $\mu$ for $(a,b,c) = (1,\frac12,\frac12)$ and the values of $\eps$ indicated.}{OmegaL}{5in}

\subsection{Transverse Rotation Number}
Assume that $f$ does have an invariant circle $\cC: \{\zeta(\theta_L): \theta_L \in \bS^1 \}$ that is $C^1$ and on which the dynamics is conjugate to rigid rotation with irrational rotation number $\omega_L$,
\[
    \zeta(\theta_L+\omega_L) = f(\zeta(\theta_L)) \;.
\]
The linearization of $f$ then gives rise to a quasiperiodic skew-product on $\R^3 \times \bS^1$
\beq{skewProduct}\begin{split}
	 \xi' &= A(\theta_L)  \xi \;,\\
	 \theta' &= \theta_L + \omega_L \mod 1 \;,
\end{split}\eeq
where $A(\theta_L) = Df(\zeta(\theta_L))$ is a periodic matrix. 

The transverse rotation number of $\cC$ is the average rotation rate of a transverse vector $\xi$ ``around" $\cC$, if that average exists.
Indeed, Herman proved that the ``fiberwise" rotation number for a quasiperiodically forced circle map always exists \cite{Herman83b}. The map \eqref{eq:skewProduct} reduces to this case if the $\xi$ dynamics are projected onto a family of circles transverse to $\cC$. To do this, 
define the transverse angle $\varphi$ relative to a ribbon $\cR$ attached to the invariant circle, see \Fig{TransverseOmega}. The transverse dynamics then induce a circle map $\varphi \mapsto g(\varphi,\theta_L)$. Consequently, under the assumption that $\cC$ exists and its dynamics are conjugate to a rigid rotation, the rotation number of $g$ exists and is independent of the initial $\theta$ and $\varphi$. 


\InsertFig{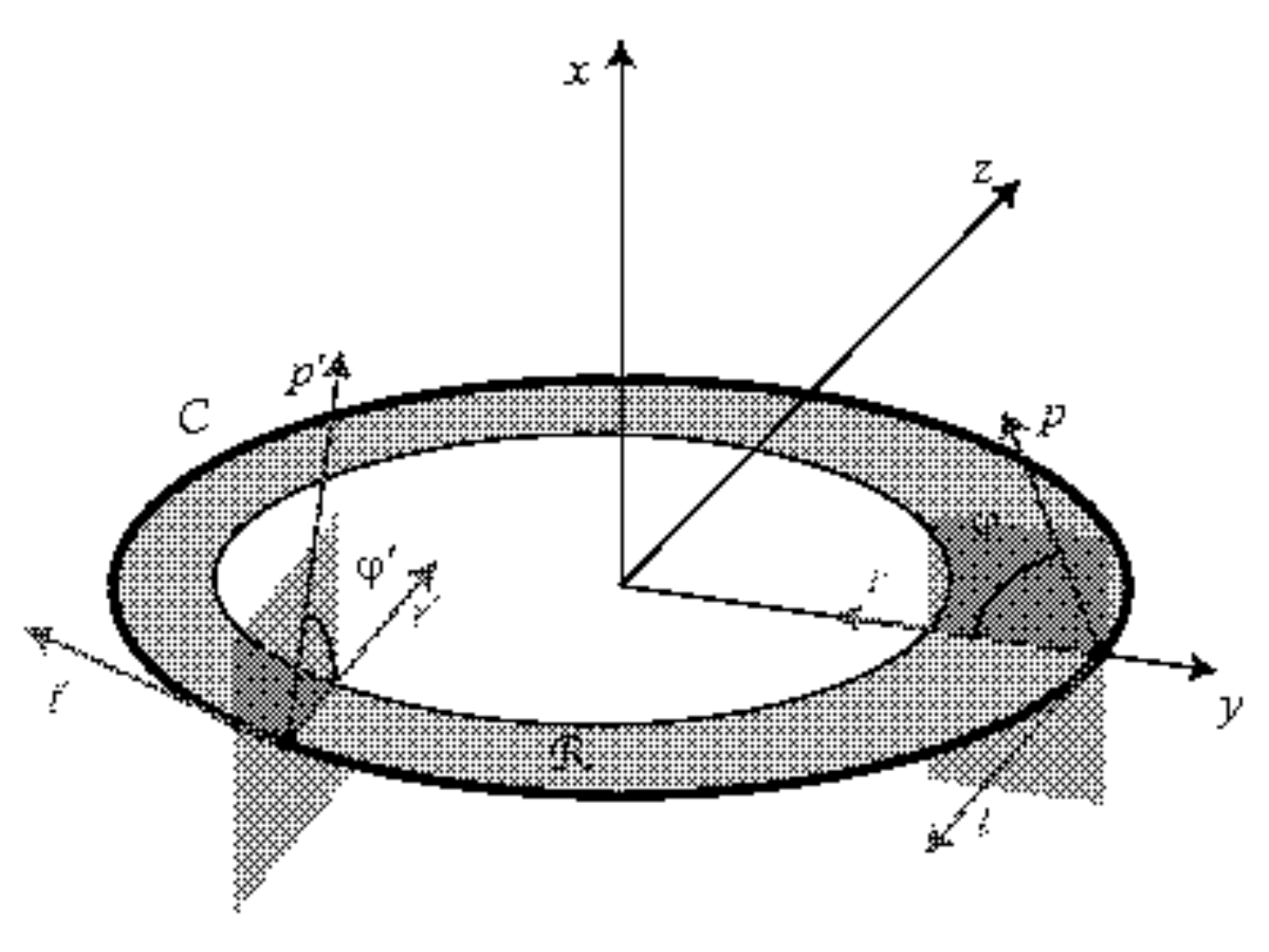}{Computing the transverse rotation number for an invariant circle $\cC$ relative to a ribbon $\cR$.}{TransverseOmega}{3in}

Let $\hat{t}(\theta_L)$ be the unit tangent vector to $\cC$ at $\zeta(\theta_L)$; it can be approximated using the closest approaches from the computation of $\omega_L$.  Since the invariant circle of \eqref{eq:StdMap} appears to be almost always everywhere transverse to the $x$-axis,\footnote{
	An exception is shown in \Fig{OmegaLTmu-15-19} for $(\eps, \mu) = (0.068, -1.9)$, where the invariant circle develops ``curlicues". Here the algorithm computes $\omega_T$ incorrectly because the projection of the circle on the $x=0$ plane is not one-to-one.
}
define the ribbon by the vector $r(\theta_L) = \hat{t}(\theta_L) \times \hat{e}_1$. Beginning with an arbitrary vector $v_0$ attached to the point $\zeta(0) \in \cC$, we iterate to obtain the sequence $v_{j+1} = A(j\omega_L) v_j$. These vectors are then projected onto a plane orthogonal to the local tangent vector $\hat{t}(j\omega_L)$; for convenience, we also rescale, defining
\[
	p = |r| \left( v - (\hat{t}\cdot v) \hat{t}\right) \;.
\] 
The projected vector is effectively two-dimensional with components 
$p = (p^\parallel, p^\perp)$ defined relative to the ribbon direction
\begin{align*}
	p^\parallel &= p \cdot \hat{r} = \hat{e}_1 \cdot \hat{t} \times v \;,\\
	p^\perp     &= \hat{t} \cdot p \times \hat{r} = \hat{e}_1 \cdot (v- (\hat{t}\cdot v) \hat{t}) \;.
\end{align*}
Thus the angle of $p$ is $\varphi = \arctan(p^\perp/p^\parallel)$.
As before, we compute the change in rotation angle at the $j$th iterate with the two-argument arc-tangent:
\[
	\Delta \varphi_j 
	      = \mbox{atan2}(
				p^\parallel_{j-1} p^\parallel_j+p^\perp_{j-1} p^\perp_j, 
				p^\parallel_{j-1} p^\perp_j- p^\perp_{j-1} p^\parallel_j) \;.
\]
The transverse rotation number is the average change along the orbit
\[
	\omega_T \approx \frac{1}{2\pi q} \sum_{j=1}^{q} \Delta \varphi_t \;.
\]
It is not clear how to optimize the error in this computation as we did for the longitudinal rotation number. We simply use the time of closest approach that we computed for the longitudinal rotation number; however, as can be seen in \Fig{FreqError}, the accuracy for this computation is only $\cO(q^{-1})$.

We compare the computations of $\omega_T$ to the normal form results \eqref{eq:RotNormForm} in \Fig{OmegaT}. Again, the agreement between the two results is extremely good for small $\eps$ and away from the resonant values. The $\cO(\eps^{1/2})$ normal form result is finite at the resonances at $\mu = -4, -3$, and $-2$, but, as before, the numerical computations indicate that the actual rotation number diverges there. Such divergences are found in the normal form at $\cO(\eps^{3/2})$; however, since these corrections to the rotation number are normally very small, we do not show in them in the figure.

\InsertFig{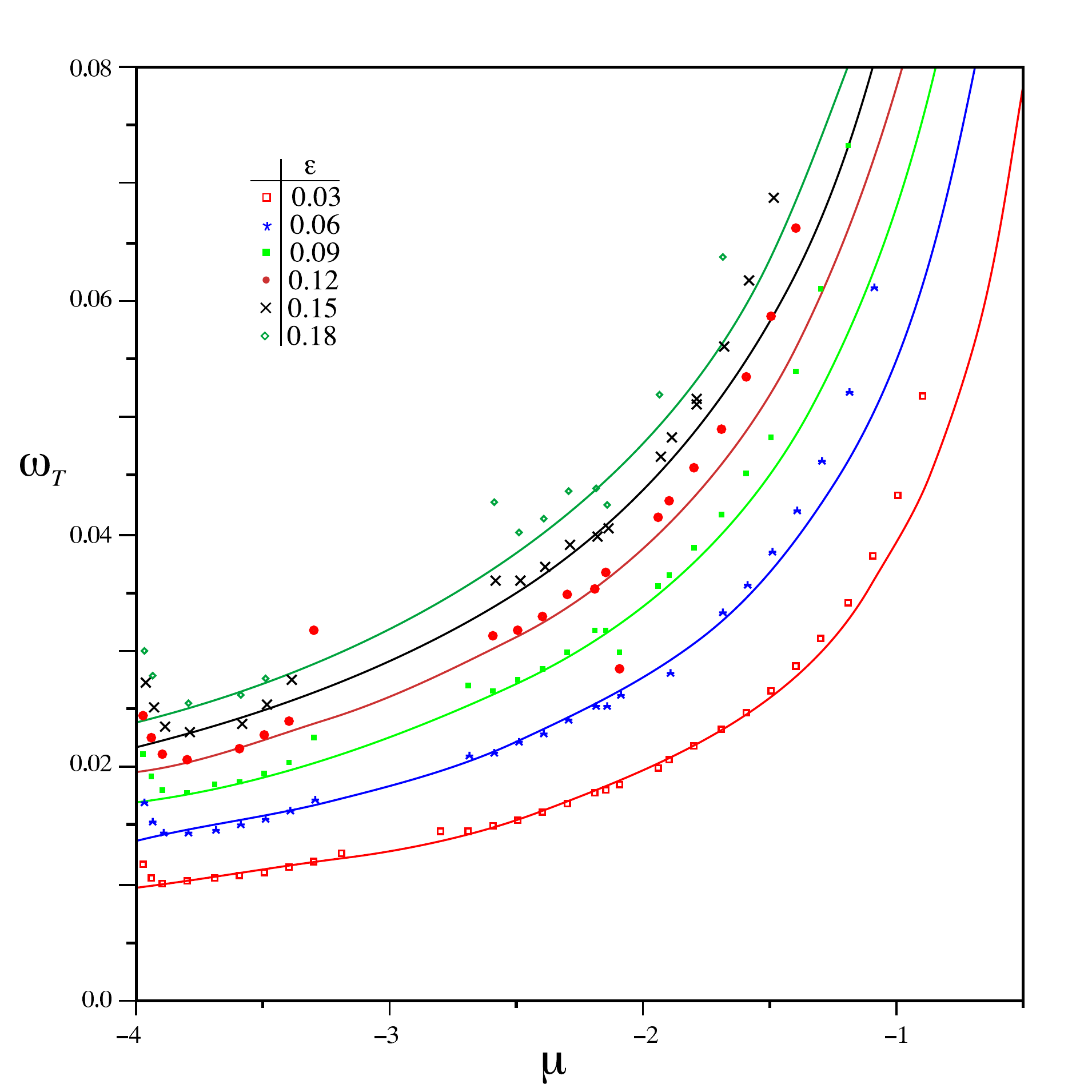}{Comparison of the the measured transverse rotation number (dots) with the theory \eqref{eq:RotNormForm} as a function of $\mu$ for $(a,b,c) = (1,\frac12,\frac12)$ and the values of $\eps$ indicated.}{OmegaT}{5in}

\subsection{Frequency Map}\label{sec:FreqMap}

If $\mu$ and $(a,b,c)$ are held fixed, the rotation numbers $(\omega_L, \omega_T)$ of the circle vary along a curve as $\eps$ changes; an example is shown in \Fig{OmegaLTmu-24} for $\mu = -2.4$. For small $\eps$ this curve lies close to the parabola \eqref{eq:RotNormForm} defined by the normal form results. Though we expect the curve to be defined only for a Cantor set of parameter values, it appears to be continuous for most values of $\eps$; there are, however, several small intervals visible in which the ellipsoid algorithm does not converge. The algorithm can find an invariant circle up to $\eps = 0.312$, at this point the circle is apparently destroyed by a resonant bifurcation, see \Sec{Resonances}.
The invariant circles shown in \Fig{OmegaLTmu-24} are also discussed in more detail in  \Sec{Resonances}.

\InsertFig{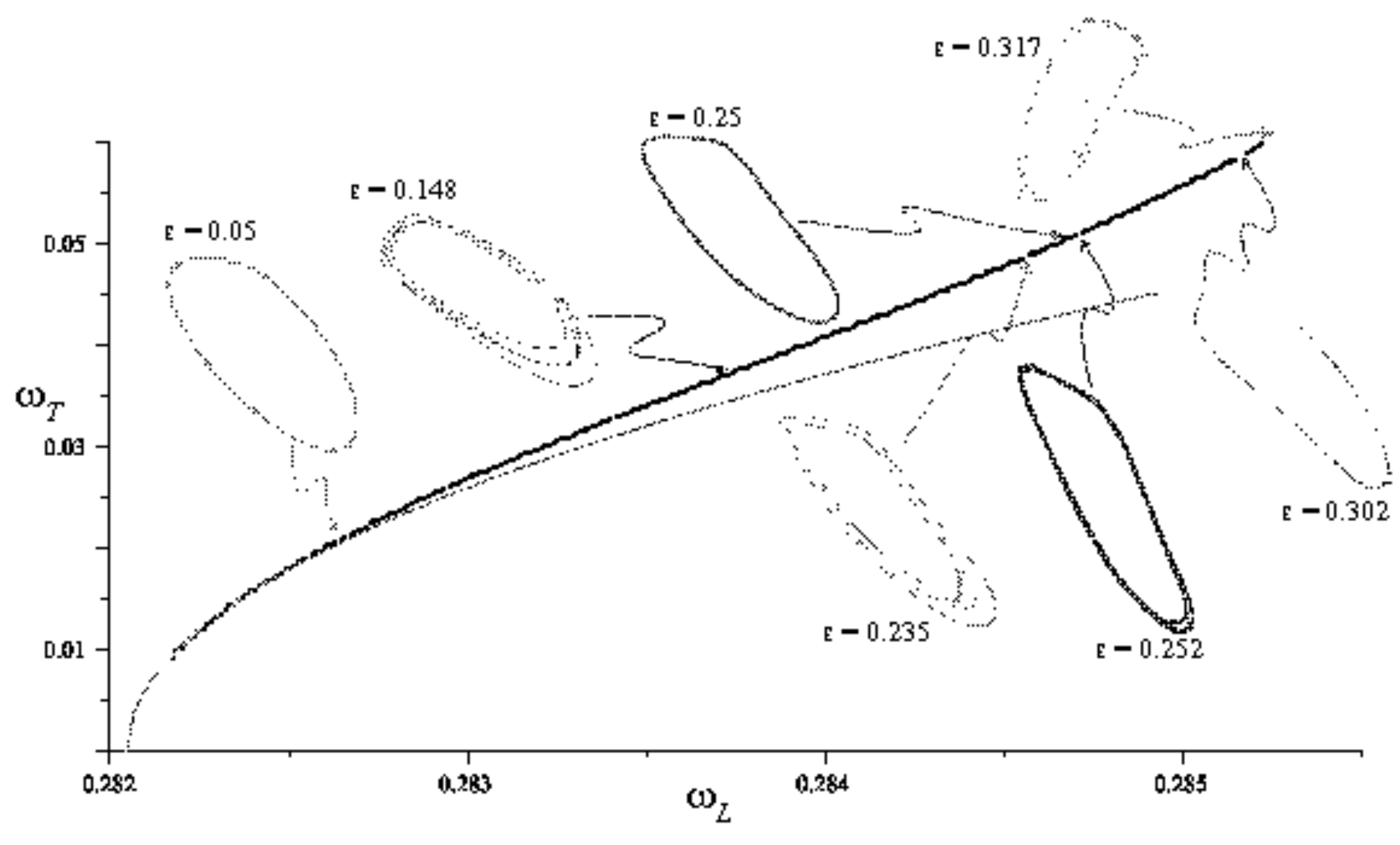}{Numerically computed frequency map for the invariant circle with $\mu = -2.4$ and the standard values $(a,b,c) =(1,\frac12, \frac12)$ as $\eps$ varies from $0.01$ to $0.312$ (black dots). The curve (brown) represents the normal form of \eqref{eq:RotNormForm}. Insets show several phase portraits of the circle and/or nearby orbits. The illustrated resonances in the order of increasing $\epsilon$ are
$(7,1,2)$, $(3,4,1)$, $(3,3,1)$, $(10,3,3)$, $(46,-2,13)$, $(7,0,2)$.}{OmegaLTmu-24}{6in}

For some values of $\mu$, the frequency map exhibits singularities as $\eps$ varies; two examples are shown in \Fig{OmegaLTmu-15-19}. Note that the normal form curve (brown) still fits the results for small $\eps$. The singularities are in $\omega_L$; those shown occur near 
$(\eps,\mu) = (0.1,-1.5)$  and $(\eps,\mu) = (0.07,-1.9)$. Near the singularities, the original circle (blue) becomes highly distorted and as $\eps$ increases it can no longer be found. The singularities appear to be associated with a pair of circle-saddle-center bifurcations:
the old circle is destroyed in a saddle-center bifurcation below the resonance line,
while a new elliptic invariant circle is born above (red), see \Sec{SaddleNode}. 
Remarkably, the new circle persists to a much larger value of $\eps$ and its rotation numbers approximately follow those of the normal form. In each case the old circle eventually appears to lose smoothness and is apparently destroyed.

\InsertFig{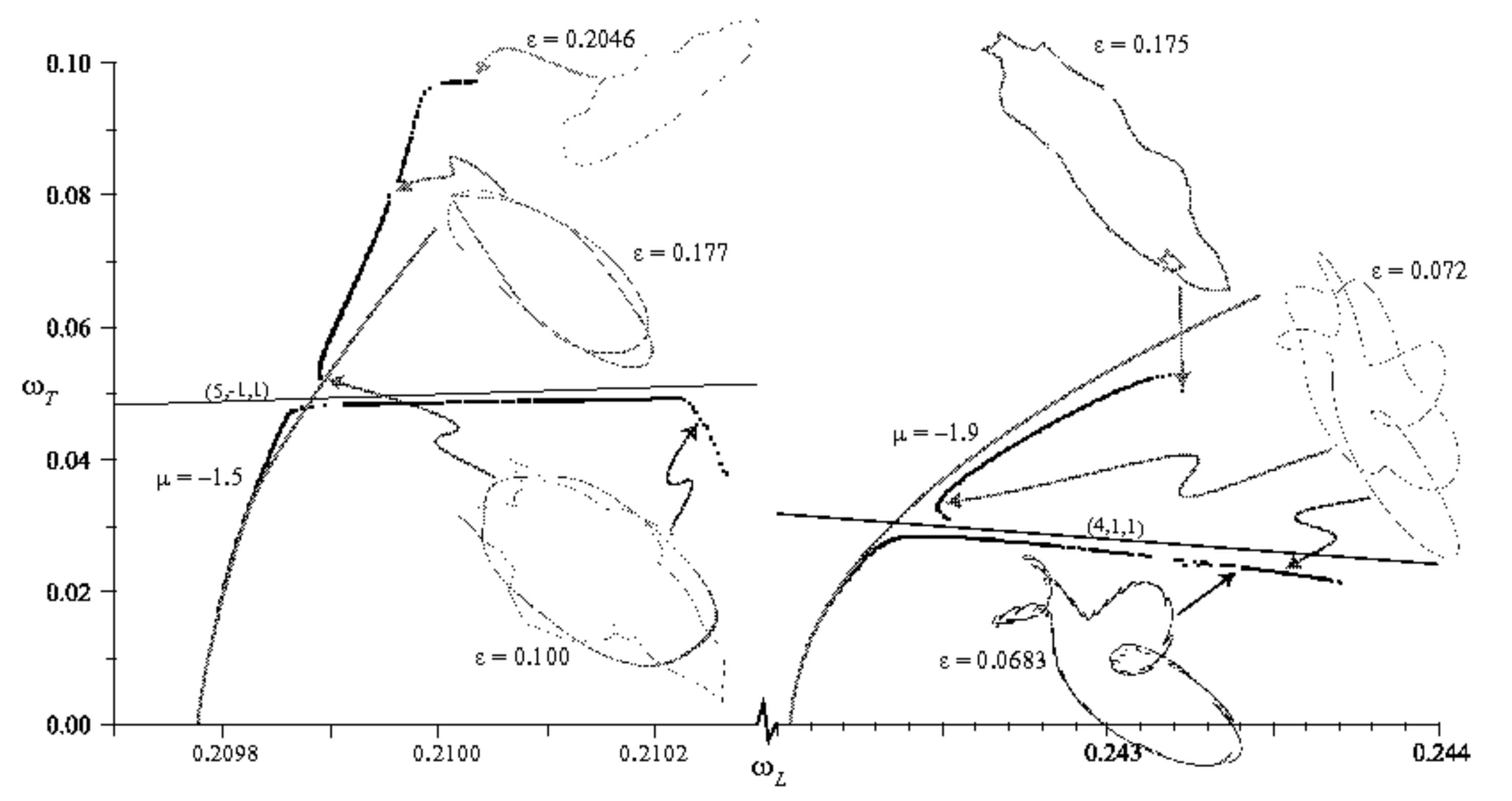}{Numerically computed frequency map for $\mu = -1.5$ and $-1.9$ (dots) and the corresponding normal form results (brown curves). Also shown are the computed invariant circles for several values of $\eps$.  The singularities in the frequency maps
are due to $(5,-1,1)$ and $(4,1,1)$ resonances that result in pairs of circle-saddle-center bifurcations, see \Sec{SaddleNode}.}{OmegaLTmu-15-19}{7in}

The image of the frequency map for $\mu = -3.9$ is shown in
\Fig{OmegaLTmu-39}. For this $\mu$ the longitudinal rotation number decreases with $\eps$ as predicted by \eqref{eq:RotNormForm}. The computed curve is fairly smooth, but there are bifurcations along the way, e.g., for $\eps = 0.0767$, but the resolution is too rough to resolve them. The gap between $0.0173 < \eps < 0.0177$ is caused by a period doubling and the doubled circle is shown in green.

\InsertFig{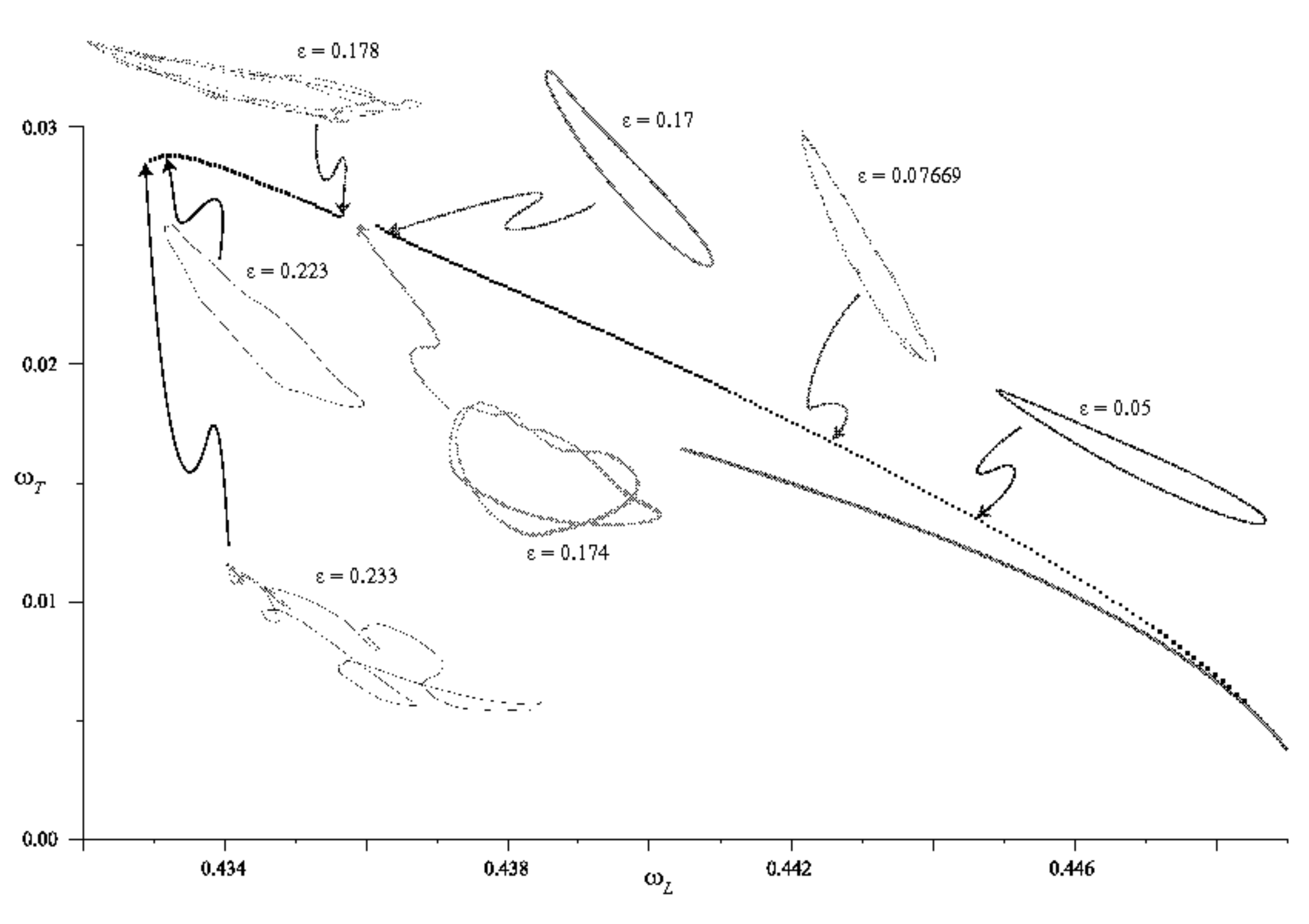}{Frequency map for $\mu = -3.9$ as $\eps$ ranges from $0.01$ to $0.173$. Insets show invariant circles for three values of $\eps$. The doubled circle in the gap near $\eps = 0.174$ is caused by a $(7,-2,3)$ resonance.}{OmegaLTmu-39}{7in}

A summary of our computations is given by the two-parameter frequency map 
\[
	\Omega: (\eps,\mu) \mapsto (\omega_L, \omega_T) \;,
\]
shown in \Fig{OmegaTvsL}. Here we computed the invariant circle for $\mu \in [-4,0]$ in steps of $0.1$  and for $\eps\ge 0.01$. We were able to find invariant circles for some range of $\eps$ for each $\mu$ except for the values $\{0, -2,-3, -4\}$ where there are strong resonances. When $ -0.6 < \mu < 0$, an invariant circle is stable only for $\eps < 0.01$, so these points do not appear in the figure  

There are many cases in which the algorithm fails to find a circle at one value of $\eps$, but then succeeds for a slightly larger value. These gaps are typically very small, but can be seen in several of the fixed $\mu$ curves; in most cases they correspond to resonant bifurcations that either cause the circle to lose transverse stability or to be destroyed. It is to the study of these bifurcations that we turn next.

\InsertFig{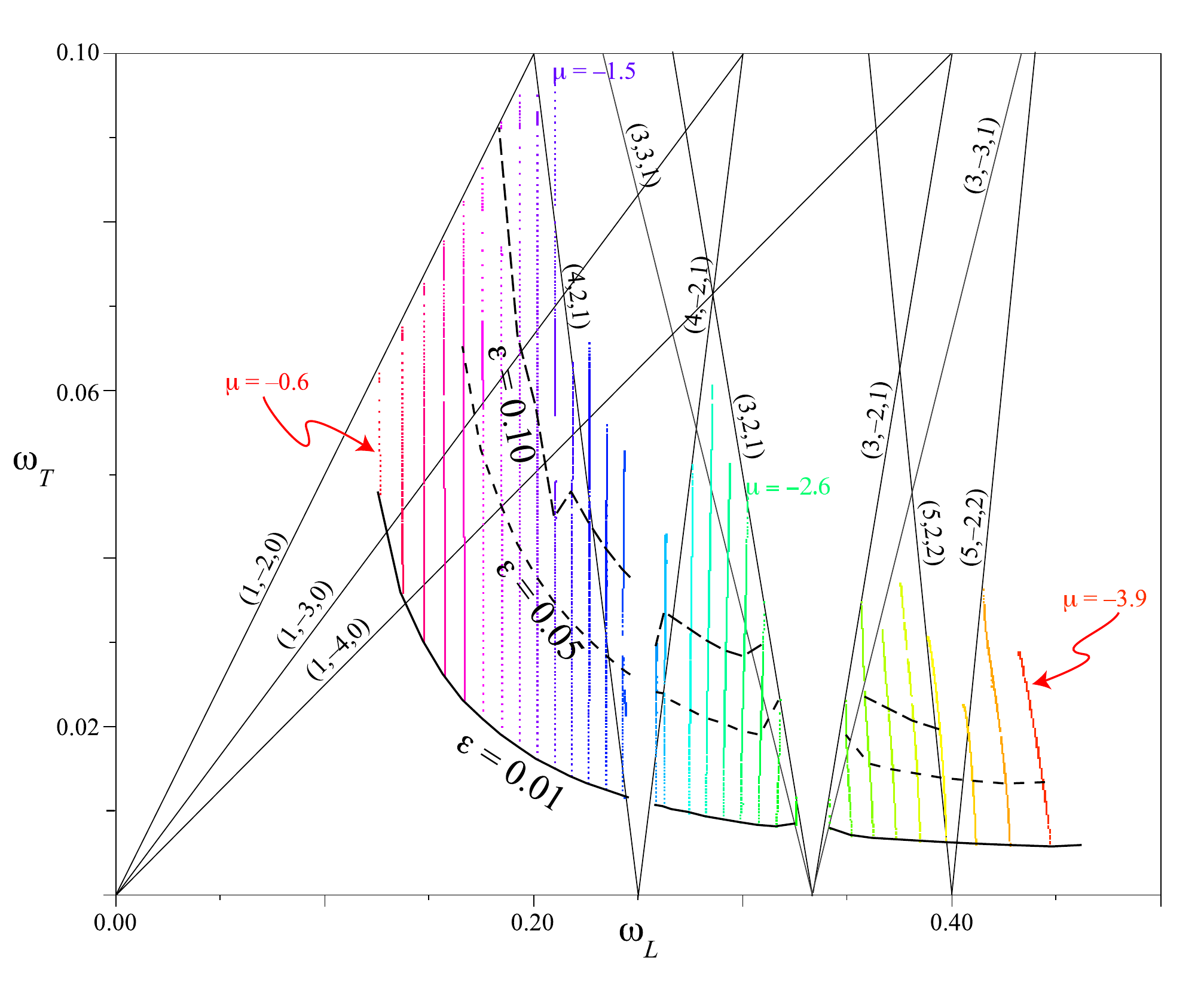}{Frequency map $(\omega_L,\omega_T)$ for $\mu \in [-3.8,-0.6]$ with step $0.1$ and a grid of $\eps$ values from $0.01$ up to the largest $\eps$ for which the ellipsoid method converged to an invariant circle. Also show are several of the important resonance lines defined by \eqref{eq:rCond}.}{OmegaTvsL}{5in}

\section{Resonant Bifurcations of Invariant Circles}\label{sec:Resonances}

The rotation vector $\omega = (\omega_L, \omega_T)$ is resonant if there are integers $(m_1,m_2,n)$, not all zero, such that
\beq{rCond}
	m \cdot \omega = n \;.
\eeq
If $\omega$ does not satisfy such a relation then it is ``nonresonant."

Since \eqref{eq:rCond} is homogeneous, if $(m,n)$ is a solution then so is $(lm,ln)$, for $l \in \Z$; consequently, the set of integers that satisfy \eqref{eq:rCond} form a sublattice of $\Z^3$,
\beq{module}
	\cL(\omega) = \{ (m,n) \in \Z^3 : m \cdot \omega = n \} \;, 
\eeq
called the \emph{resonance module}. The dimension of $\cL$ is the number of independent resonance conditions. Using homogeneity, we will assume that $n$ is nonnegative and $\gcd(m_1,m_2,n) = 1$. 

In $\omega$-space, \eqref{eq:rCond} defines a line for each $(m,n)$; a few such lines are shown in \Fig{OmegaTvsL}. Using the normal form results \eqref{eq:RotNormForm}, the resonance curves for the invariant circle can also be obtained in parameter space. 
The fine structure in the bounded volume \Fig{VpBounded} is caused by these resonances.

For example, resonances with $|m_2| = 2$ are often observed to result in the destruction or loss of stability of the invariant circle $\cC$, as well as a dramatic drop in the volume of bounded orbits. Figure~\ref{fig:OmegaTvsL} shows that $\cC$ appears to lose stability at the $(1,-2,0)$ resonance for $-1.4 < \mu < 0$ and at the $(3, \pm2, 1)$ resonances for $-3.3 < \mu < -2.5$ . In each case, a new elliptic, doubled circle is created when the original circle crosses the resonance; phase space portraits are shown in the top and bottom panels of \Fig{doubling}. Since $\omega_T/\omega_L = \frac12$ for the $(1,-2,0)$ resonance, tori near $\cC$ just below the bifurcation approximate a  M\"obius strip with a half-twist, and the new circle undergoes one-half turn about the original circle for each longitudinal period. For the second case, the $(3, \pm 2, 1)$ resonance, the new circle rotates by $3$ turns transversally in $2$ longitudinal rotations.
In both cases the new elliptic invariant circle is a double covering of the original one, 
as is familiar for period doubling of periodic orbits of flows. 

Interestingly, there is also another type of period doubling bifurcation that cannot occur
in flows. This happens, for example, at the $(4,2,1)$ resonance; a phase space portrait
is shown in the middle panels of \Fig{doubling}. Instead of one invariant circle that is a double cover of the original, here there are two disjoint invariant circles, each
covering the original circle once. Though the new invariant circles are geometrically disjoint, they are mapped onto one another by the dynamics. The reason for the different 
behavior is that $\gcd(m_1, m_2) \neq 1$ in this case. This is the simplest example of a general scheme that is easily understood by studying an integrable map of the two-torus with resonant rotation vector $\omega$. 

\InsertFig{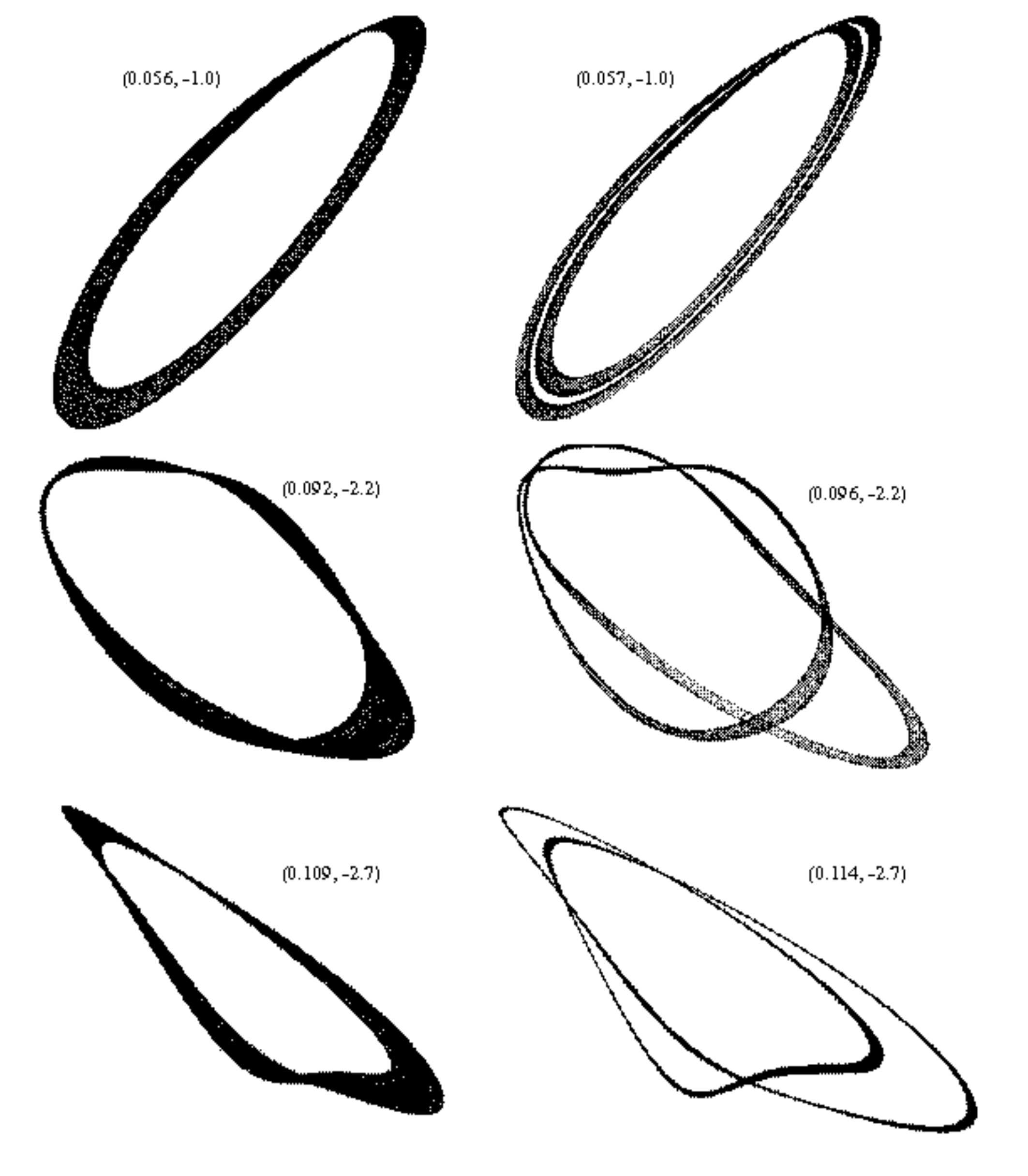}{Circle doubling bifurcations. Shown are some of the orbits in the neighborhood of an ellipitical invariant circle (left panels) just before the doubling bifurcations, and orbits in the neighborhood of the doubled circle (right panels) just after. 
The corresponding resonances are $(1,-2,0)$, $(4,2,1)$, and $(3,-2,1)$ from top to bottom.
The values of $(\eps,\mu)$ are shown.}{doubling}{4in}

\begin{lem}\label{lem:torusMap}
  The orbits of the torus map
	\beq{torusMap}
	   x' = x + \omega \mod 1 \;,
	\eeq
	for $x, \omega \in \T^2$, are either
  \begin{enumerate}
  \item dense on $\T^2$ if $\omega$ is nonresonant ($\dim \cL(\omega) = 0$);
  \item dense on $k$ circles if $\omega$ satisfies one resonant relation and
     $\gcd(m_1,m_2) = k$ ($\dim \cL(\omega) = 1$); or
  \item periodic if $\omega$ satisfies more than one resonance 
     relation ($\dim \cL(\omega) = 2$).
\end{enumerate}
\end{lem}

\proof
If $\omega$ is nonresonant, then the result is standard \cite{Cornfeld82}.
Suppose now that there is a single resonance relation ($\dim \cL = 1$), 
let $k = \gcd(m_1, m_2)$ be the common divisor, and define
\beq{linknumber}
     (m_1, m_2)  = k(p,q) \;, 
\eeq
where $p,q$ are coprime.\footnote
{
	Note that if one frequency is rational, then $m_1m_2 = 0$ and hence $pq = 0$. 
	This case can be formally included here with $k = m_1 + m_2$ and $p$ or $q$ equal to $1$.
}
As is shown in \cite{HardyWright79} there is a solution $(\hat{p}, \hat q) \in \Z^2$ to the equation
\[
    p \hat{q} - q \hat{p} = 1 \;,
\]
if and only if $p,q$ are coprime. Taking any such solution, define
the unimodular transformation
\[
    y = \begin{pmatrix} p & q \\ \hat{p} &\hat{q} \end{pmatrix} x \;.
\]
The variables $y$ are simply a new set of coordinates on $\T^2$. The map in these new coordinates takes the form
\beq{torusMapII}\begin{split}
	y_1' &= y_1 + \frac{n}{k} \;, \\
	y_2' &= y_2 + \hat{p}\omega_1+ \hat{q}\omega_2 \;.
\end{split}\eeq
By assumption $k$ and $n$ are coprime integers, so the orbit of $y_1$ is periodic with period $k$ (in particular if $n=0$, then $k = 1$). Moreover, since $\omega$ satisfies exactly one resonance relation, the combination $(\hat{p},\hat{q})\cdot\omega$ is irrational. Thus the orbit of $y_2$ is dense
in $\T^1$. Hence the combined orbit on $\T^2$ is dense on $k$ circles. This still holds true in the original coordinate system, but the circles will wrap around the torus in both directions.

Finally, if $\omega$ satisfies two independent resonance relations ($\dim \cL = 2$), 
$m \cdot \omega = n$ and $\hat{m} \cdot \omega = \hat{n}$, then the components 
of $\omega$ must each be rational because then 
\[
	\begin{pmatrix} m^T \\ \hat{m}^T \end{pmatrix} \omega = 
	\begin{pmatrix} n \\ \hat{n} \end{pmatrix} \;,
\]
and the matrix on the left is nonsingular since the vectors $m$ and $\hat{m}$ are not parallel.
Thus both frequencies are rational, and all orbits are periodic.
\qed

Structures similar to the second case of \Lem{torusMap} are also found in quasiperiodically forced circle maps \cite{Jager06a, Jager06b} where they have been called $(k,q)$-invariant graphs.

We now embed these dynamics into three dimensions as invariant sets of a family of ``integrable'' volume-preserving maps $f_0$ on the solid torus, $\T^2 \times \R^+$,
\beq{integrable}
	f_0(\theta, r) = (\theta + \omega(r,\mu), r) \;.
\eeq
Here $r$ represents the radius, so that $r = 0$ corresponds to the toroidal or longitudinal axis, $\theta_L$ is the toroidal angle, and $\theta_T$ is the poloidal or meridional angle.  We assume there is a parameter $\mu$ to unfold the frequency map (just as for our model \eqref{eq:StdMap}). The phase space of \eqref{eq:integrable} is foliated into invariant tori $r = r_0$, and on each such torus $f_0$ becomes \eqref{eq:torusMap}.

Now suppose that at $(r_0,\mu_0)$, the rotation vector of \eqref{eq:integrable} satisfies a single resonance condition, that is $\omega(r_0,\mu_0) \cdot m = n$ where $m$ satisfies \eqref{eq:linknumber}. Each orbit on this resonant torus densely covers a circle ($k=1$) or set of circles ($k>1$) on this torus. These curve(s) correspond to a torus knot (torus link) \cite{Adams04}; several examples are shown in \Fig{TorusLinks}.\footnote
{
	Famous knots and links include the trefoil knot $(3,2)$, 
	the Hopf link $(2,2)$, and Solomon's knot (which is a link) $(4,2)$.
	Note that Borromean rings are not included in the class of torus links.
}
A torus knot of type $(p,q)$ coprime corresponds to the closed loop $\{(qt, -pt, r_0) : t \in \R\}$ that has $p$ poloidal wraps for each $q$ toroidal circuits. A torus link of type $(m_1,m_2)$ is the collection of $k$ loops given by 
$\{(q t, -p t  + i/m_1, r_0): i  = 0,\, \dots, k-1,\, t \in \R\}$.

For example, the $(3,4,1)$ resonance leads to a single $(3,4)$ torus knot,
the $(4,4,1)$ resonance leads to four $(1,1)$ torus knots (each of which is not knotted), 
which together give the $(4,4)$ torus link. Similarly, the $(10,4,3)$ resonance leads to two $(5,2)$ torus knots, together forming a torus link of type $(10,4)$. 

Note that the number $n$ does not enter into the geometric specification of the knot or link. 
However, by \eqref{eq:torusMapII}, $n \bmod k$ does determine the dynamics on the torus link.

\InsertFigThree{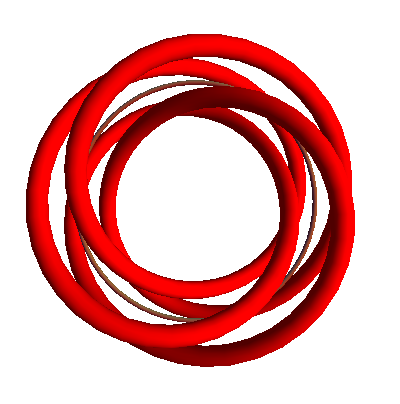}{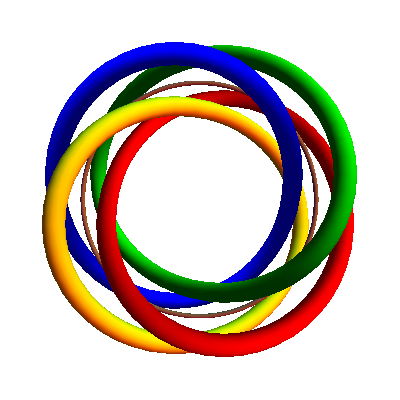}{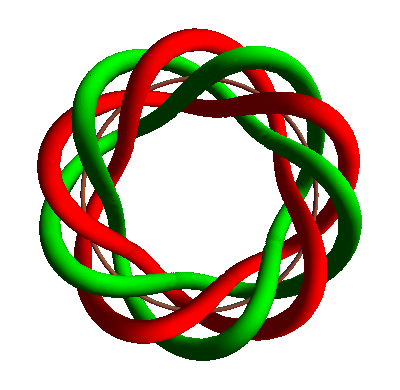}{Schematic pictures of torus links of type $(m_1, m_2) = (3,4)$, $(4,4)$ and $(10,4)$. Different colors distinguish the 
$k = \gcd(m_1, m_2)$ disjoint circles of the link. The central invariant circle (not part of the torus link) is also shown.}{TorusLinks}{2in}

Here we are specifically interested in local bifurcations at the elliptic invariant circle $\cC = \{ r = 0\}$ of \eqref{eq:integrable}. Some of these are analogous to the generic bifurcations of an elliptic fixed point of an area-preserving map in which periodic orbits are created or destroyed when the multiplier of the linearization passes through a root of unity.  For the invariant circle, this would correspond to the transverse rotation number $\omega_T$ passing through a rational, $n/m_2$: an $(0,m_2,n)$ resonance.
More generally, suppose that $\cC$ is resonant, i.e., that $\omega(0,0)$ satisfies a single resonance condition \eqref{eq:rCond}, and that $\partial_\mu \omega(0,0) \neq 0$. Then as $\mu$
passes through zero, a resonant torus will be created in the neighborhood of $\cC$. As above, this resonant torus is foliated into a one-parameter family of torus knots or links. If we now perturb \eqref{eq:integrable}, $f_\eps = f_0 + O(\eps)$, then by analogy with the Poincar\'e-Birkhoff theorem for area-preserving maps we expect that all but a finite number of these circles will be destroyed. Indeed, a generalization of this theorem to the volume-preserving map case has been obtained by Cheng and Sun \cite{Cheng90a}.
Thus the bifurcation should create a finite number of invariant circles in the neighborhood of $\cC$.

Though the torus map \eqref{eq:torusMap} does not distinguish between its two angle variables, the embedding of \eqref{eq:integrable} into $\R^3$ assigns different roles to the longitudinal ($\theta_L$) and transverse ($\theta_T$) angles. The two-dimensional analogy then implies that for the volume-preserving case, $m_2$ may play a more important role than $m_1$. In particular, recall that an elliptic fixed point of an area-preserving map is ``strongly'' resonant if the denominator of $\omega_T$ is small, i.e., $m_2 \le 4$. An analogous classification will pertain 
to the volume-preserving case.

From these qualitative considerations and our numerical observations, we propose the following conjecture:

\begin{conjecture}
An elliptic invariant circle of a volume-preserving map with frequencies 
$\omega = (\omega_L,\omega_T)$ that satisfy the single resonance condition
$\omega_L m_1 + \omega_T m_2 = n$ ($\gcd(m_1, m_2 ,n) = 1$, $\gcd(m_1, m_2)=k$)
generically undergoes one of the following bifurcations:
\begin{itemize}
\item $m_2  = 0$  (String of pearls bifurcation): 
The circle is destroyed and a pair of saddle, period-$m_1$ orbits are born in a SCNS bifurcation, recall \Sec{SNHMap}. The one-dimensional invariant manifolds of the saddles nearly coincide along the location of the destroyed circle. If the bifurcation is supercritical, it also creates a family of $m_1$, almost-invariant balls (pearls) bounded by the manifolds of neighboring points on the orbits and containing elliptic circles.

\item $|m_2| = 1$ (Saddle-center bifurcation):
The transverse multiplier of the invariant circle becomes $1$ and the circle is destroyed in a saddle-center bifurcation as the parameters cross the resonance line.

\item $|m_2| \ge 2$ (Torus-link bifurcation): The invariant circle persists but may loose stability. In its neighborhood, $k$ invariant circles are born that form an $(m_1,m_2)$ torus link.
\end{itemize}
\end{conjecture}

We believe that this classification of different types of $m_2$-tupling bifurcations of invariant circles is new. In the following subsections, we will separately treat each of the three cases of the conjecture.

\subsection{Torus-Link Bifurcations, $|m_2| \ge 2$}

In the beginning of this section, we discussed several examples of period 
doubling with $|m_2| = 2$, showing that a torus knot (link) is created if $m_1$ is odd (even), recall \Fig{doubling}. We will now present more examples with larger $|m_2|$,
and then give an explanation of the observed bifurcations.


Several of such bifurcations for $\mu = -2.4$ are illustrated using Poincare\'e \textit{slices} in \Fig{SectionsMu-24}. A slice is analogous to a Poincar{\'e} section of a three-dimensional flow; however, since an orbit of a three-dimensional map would almost never intersect a two-dimensional section, we must instead consider a slice of some nonzero thickness. In the figure, we use the same slice, $\Sigma_\Delta = \{(x,y,z): y > 0, |z| < \Delta\}$, that was used to find the elliptic circles in \Sec{Algo}.

For example, at $(\eps,\mu) \approx (0.121, -2.4)$, the elliptic circle crosses the $(4,-4,1)$ resonance. Since $(m_1,m_2)$ have a common factor, this bifurcation creates an elliptic torus link of type $(4,-4)$ which consists of four torus knots of type $(1,-1)$, recall the schematic \Fig{TorusLinks}. The new circles move away from the central circle as $\eps$ grows. In the Poincar{\'e} slice---the top row of \Fig{SectionsMu-24}---this bifurcation looks just like a generic quadrupling bifurcation of an elliptic fixed point in an area-preserving map. The slice reveals that both an elliptic and a hyperbolic $(4,-4)$ torus link are created, and that the central circle appears to persist and remain stable through the bifurcation. We observe qualitatively similar Poincar{\'e} slices for the  
$(3,4,1)$ resonance (at $(\eps,\mu) \approx (0.147, -2.4)$, recall \Fig{OmegaLTmu-24}), and  $(10,4,3)$ (at $(0.169, -2.4)$); however, the corresponding torus-links are quite different, recall \Fig{TorusLinks}.

It is easy to observe many such resonances; a list of the lower order resonances encountered for $\mu = -2.4$ is given in \Tbl{resonances} in \App{tables}.
It is interesting to note that many resonances appear in sequences: $(3,4,1)$ and $(10,4,3)$
are the two first elements of the sequence $(3 + 7n, 4, 1 + 2n)$.
These resonances lines have one conjugate point, $(\omega_L, \omega_T) = (\frac27, \frac{1}{28})$, in common. Since this limit point is close to the image of the frequency map for $\mu = -2.4$ many of these resonances are encountered in the family. 
However, since the slope of these resonance lines is proportional to $n$ and the last observable, stable invariant circle has $\omega_L \approx 0.2853 = [0,3,1,1,44,\ldots]$, which is close to the rational $\frac27$, the frequency curve will eventually miss the resonance lines. Any low-order, doubly-rational point in the image of the frequency map can similarly serve as a limit point for many resonance sequences.  In the present case we observe, for example, the sequences $(7n, 1, 2n)$, $(7n, 2, 2n)$, $(3 + 7n, 3, 1 + 2n)$, and $(4 + 7n, -3, 1+2n)$. 
As a consequence, the invariant circle will repeatedly undergo 
bifurcations that have the same $m_2$ but that correspond to different associated torus links.
%
%

Resonances with larger $m_2$ have similar structure: for example a $(1,-5,0)$ resonance at  $(\eps,\mu) \approx (0.026, -1.1)$ creates a new elliptic and a new hyperbolic circle that, as $\eps$ grows, give rise to a five-island structure when viewed in a Poincar\'e slice.

Tripling bifurcations ($|m_2| = 3$) may result in a brief loss of stability of the invariant circle. For example, there is a gap in the $\mu = -2.4$ frequency map of \Fig{OmegaLTmu-24} near $\eps = 0.235$ where the circle crosses the $(3,3,1)$ resonance. This bifurcation again resembles the two-dimensional case. Indeed the new hyperbolic and elliptic $(3,3)$ torus links are created in a circle-saddle-center bifurcation before the ``tripling'', near $\eps = 0.234$. Moreover, the central circle appears to lose stability in a collision with the hyperbolic period-three family near $\eps = 0.235$; cross sections are shown in \Fig{SectionsMu-24}. This and the similar $(3,-3,1)$ resonance occur for a range of $\mu$, as can be seen in \Fig{OmegaTvsL}. 
Another tripling for $\mu = -2.4$ occurs in a gap around $\eps = 0.252$; this is a $(10,3,3)$ resonance. For this case $(m_1,m_2)$ are coprime and the bifurcation results in the formation of a single, elliptic torus-knot with three longitudinal wraps and ten meridional wraps (the dark blue circle in \Fig{OmegaLTmu-24}). Similar tripling bifurcations occur whenever $\mu > -1.6$ as the invariant circle crosses the $(1,-3,0)$ resonance. However, for these cases the tripled circle is a $(1,-3)$ torus knot.

\InsertFig{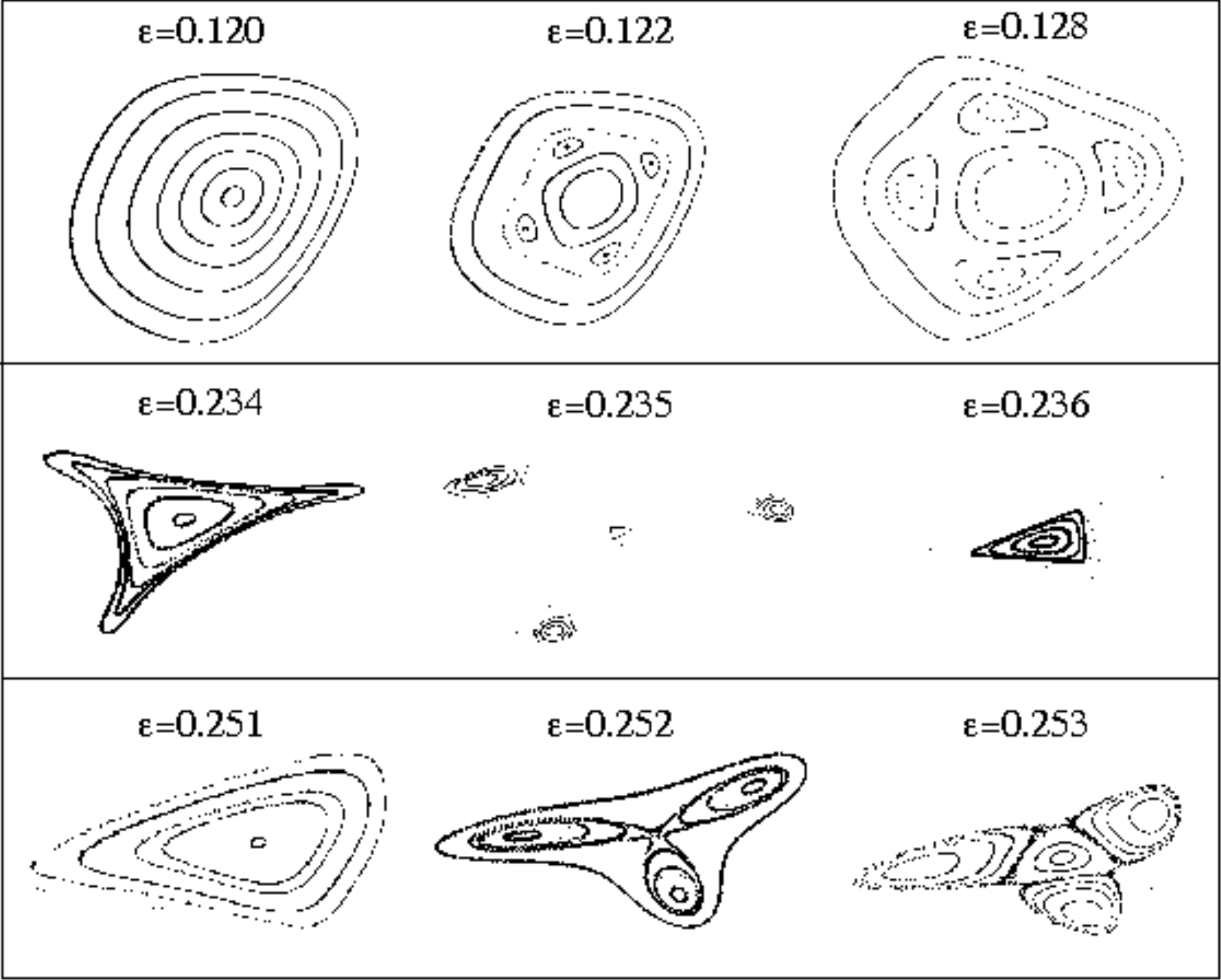}{Poincar\'e slices $\Sigma_\Delta$ for $\mu = -2.4$ and the $\eps$ values shown. The upper sequence crosses the $(4,-4,1)$ resonance, the middle the $(3,3,1)$, and the lower the $(10,3,3)$ resonances.}{SectionsMu-24}{5in}


From the examples discussed so far it seems as if the bifurcations of invariant 
circles are simply those of fixed points of area-preserving maps multiplied by a (twisted) circle. However, this appears to be false. We found many portraits in the Poincar\'e slice 
that are not generic in area-preserving maps. To discuss this, we first recall the generic bifurcations of a fixed point of an area-preserving map. When the multiplier of the fixed point
is $\exp( \pm 2\pi i n/m)$, $n, m$ coprime, then there are 5 cases (assuming the generic higher order transversality conditions are satisfied) \cite{MeyerHall92}:
\begin{itemize}
	\item $m= 1$ (saddle-center): collision of saddle and center 
	  fixed points (double multiplier 1, not semisimple),
	\item $m=2$ (doubling): a periodic orbit of period two is created 
	  and the fixed point changes stability (double multiplier $-1$, not semisimple).
	\item $m=3$ (tripling): a saddle-center bifurcation of creates a 
	  pair of period three orbits away from the fixed point, and the 
	  saddle passes through the fixed point, causing a momentary loss of stability. 
	\item $m = 4$ (quadrupling): this bifurcation can behave either like 
	  $m=3$ (the hyperbolic orbit colliding with the fixed point with loss of stability) 
	  or like $m>4$ (elliptic and hyperbolic pair being created without loss of 
	  stability), depending on the relative strengths of higher order terms.
	\item $m>4$ ($m$-tupling): a pair of elliptic and hyperbolic periodic 
	  orbits of period $m$ is created at the origin without loss
	  of stability. These orbits form an island chain with $m$ islands.
\end{itemize}
As a result resonances are usually grouped into ``low order" or ``weak'' ($m\le4$)
and ``high order" or ``strong'' ($m\ge4$) cases. High-order resonances do not cause 
loss of stability of the fixed point at the bifurcation.

For $m \ge 3$ these resonances are described by an area-preserving map of the form
\[
	z' = e^{2\pi i n/m} z + O(2) \;.
\]
The resonant terms correspond to the monomials $z^{k_1} \bar z^{k_2}$ with $k_1, k_2 \in \N$ and $|k_1 - k_2|$ divisible by $m$. Upon setting $z = \sqrt{2I} \exp{i \phi}$, for angle-action variables $(\phi,I)$, these bifurcations can be described by an approximate Hamiltonian
\begin{equation} \label{HamGenBif}
	H(\phi,I) = \mu \alpha I + \beta I^2 + \dots + (\gamma
	 + \delta I + \ldots)  I^{m/2}\cos m \phi \,,
\end{equation}
where the bifurcation parameter $\mu$ detunes the resonance. When $m = 3$ the resonant term $\gamma$ is the leading-order nonlinear term, and thus the behavior is different than for $m > 4$. When $m = 4$ the leading-order resonant and nonresonant terms have the same order and
the behavior depends on whether $|\beta| \gtrless |\gamma|$. 
For $m \ge 5$ the resonant term is not of leading order and the usual island chain is born.

Resonant bifurcations of invariant circles of volume-preserving maps follow a similar scheme, but there are more possibilities. This is for two reasons. Firstly, instead of the single period $m$ there is now a vector $(m_1, m_2)$. Roughly speaking, an area-preserving map describes a slice of the dynamics transverse to the invariant circle with ``period'' $m_2$. In the simplest situation, these $2D$-dynamics rotate as the slice moves along the invariant circle; this is described by $m_1$. Secondly and less obviously, the distinction between low-order and high-order resonances is not so clear cut: indeed all the ``low-order" resonances can also appear without loss of stability. Our observations can be summarized in three cases:
\begin{itemize}
\item $m_2 = 2$: the central circle looses stability and an elliptic doubled circle appears.
Depending of $k = \gcd(m_1, m_2)$ there is either one circle that is a double cover 
of the original one ($k=1$), or there are two circles each covering the original circle once 
that are mapped into each other ($k=2$). These two circles form an $(m_1, m_2)$ torus link.
However, unlike generic bifurcations in Hamiltonian systems this bifurcation may be ``weak'', creating a pair of elliptic and hyperbolic circles. This happens, e.g., for the $(7,2,2)$ resonance at $(0.0167, -2.4)$.
\item $m_2 = 3$: the tripling may lead to a momentary loss of stability of the central circle. 
When $m_1$ is divisible by 3, then 3 circles are created, and $n \bmod 3$ determines 
how they are mapped to each other. Unlike the generic area-preserving case, the tripling bifurcation can be ``weak''. This occurs, e.g., for the $(11,-3,3)$ resonance at $(0.1791, -2.4)$.
\item $m_2\ge4$: here a pair of elliptic/hyperbolic torus links of type $(m_1, m_2)$ is created. If $m_1$ and $m_2$ are coprime, there is only a single circle, while if $\gcd(m_1, m_2) = k$ then there are $k$ disjoint circles. This difference is invisible in a slice, and these bifurcations do look like generic bifurcations of area-preserving maps. The order in which the $k$ circles are mapped into each other is determined by $n \bmod k$.
\end{itemize}

It is possible that the additional cases when $m_2 < 4$ could be described as non-generic bifurcations of area-preserving maps with $m = m_2$ but in which the lowest order resonant term vanishes. The approximate Hamiltonian would still be given by \eqref{HamGenBif}, but now $\gamma = 0$.
For $m = 2$ there are now two cases, depending on whether $|\beta| \gtrless |\delta|$. Thus $m = 2$ becomes similar in some sense to $m = 4$ for the area-preserving case.
When $|\beta| > |\delta|$ an elliptic/hyperbolic pair is born 
like the generic case for $m \ge 5$. However, when $|\beta| < |\delta|$, the bifurcation behaves like the generic $m = 3$ case: i.e., a hyperbolic orbit passes through 
the center with loss of stability.
We have not yet observed this case, but we expect that is is also generic
for bifurcations of invariant circles in volume-preserving maps.
If $\gamma = 0$ for $m = 3$, an elliptic/hyperbolic pair of orbits are born: this case behaves like the generic case $m\ge 5$. The mechanism by which the lowest order resonant terms vanishes in these cases presumably is discrete symmetry. We hope to return to this question in a future work.

Besides the local bifurcations discussed so far there are also some interesting sequences of bifurcations associated with period doubling. We have already discussed several doubling bifurcations ($|m_2| = 2$) that result in the permanent loss of stability  of the invariant circle, and thus indicate the loss of most of the bounded orbits. There are also doublings for which $\cC$ is unstable only for a small range of $\eps$. For example the frequency map curves shown in \Fig{OmegaTvsL} sometimes pass through the $(4,\pm2,1)$ and $(5,\pm2,1)$ resonance lines---with a short interruption---indicating that the original circle regains stability after creating the doubled circle. Even when $|m_1|$ is large, a doubling bifurcation can result in the temporary loss of stability of $\cC$. For example, the gap in the frequency map of \Fig{OmegaLTmu-24} near $\eps = 0.302$ corresponds to a $(46,-2,13)$ resonance.  A pair of stable, circles is created that each wrap around the longitudinal direction once, while encircling the original circle in the transverse direction $23$ times, an enlargement is shown in \Fig{CircleOverlaymu-24}.

\InsertFig{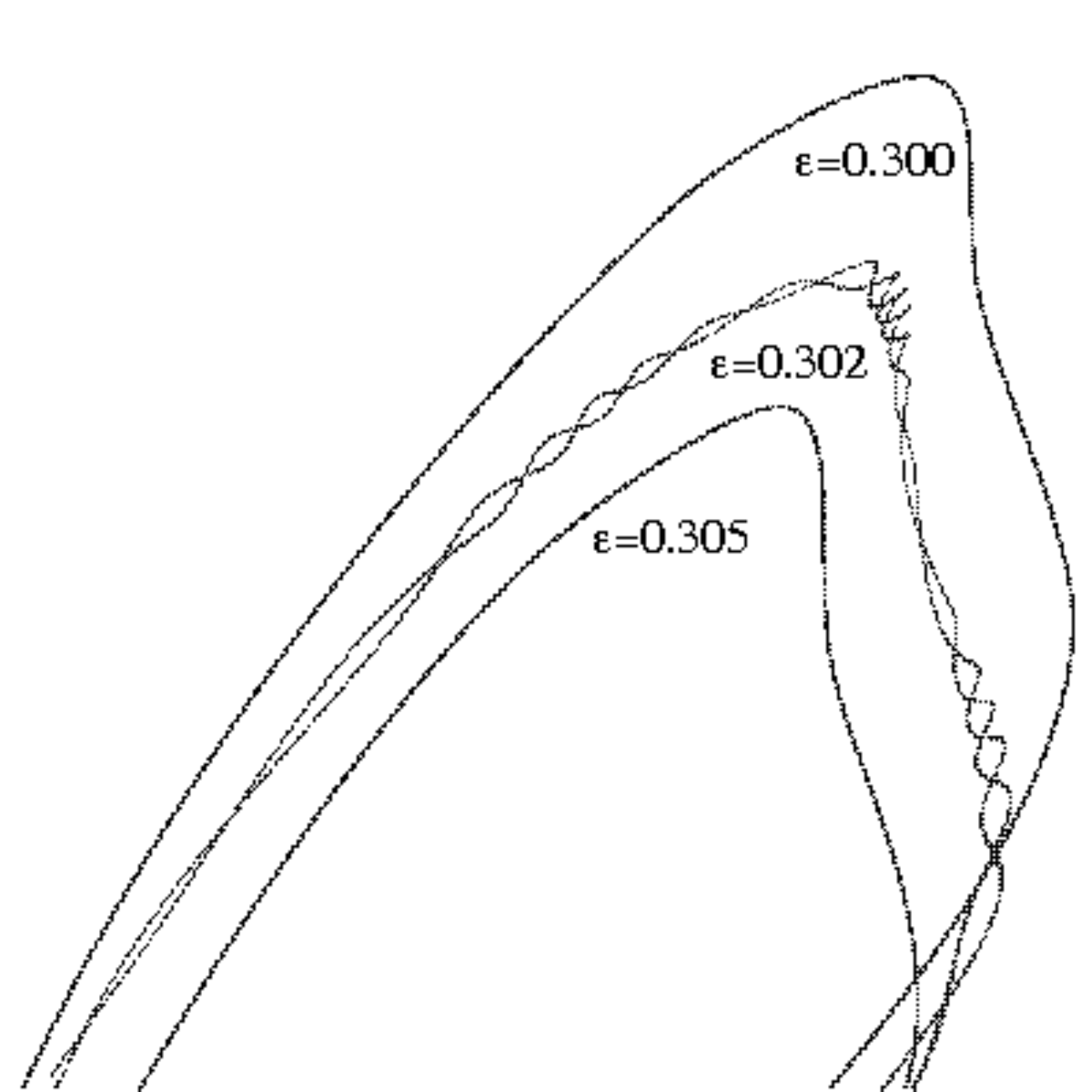}{Portions of the elliptic invariant circles near the $(46,-2,13)$ resonance when $\mu = -2.4$. In the gap $0.3017<\eps<0.3045$ the central circle is unstable. The circles are shifted vertically for clarity.}{CircleOverlaymu-24}{3.in}

The class of torus knot (or torus link) bifurcations of invariant circles of volume-preserving maps of $\R^3$ shows a rich bifurcation structure. While there are similarities to area-preserving maps, there is also non-trivial three-dimensional structure and slices reveal some bifurcations that would be considered non-generic in the area preserving category.
A more detailed investigation of these bifurcations would be very interesting.

\subsection{\textit{String of Pearls} Bifurcations}

Resonances of an invariant circle with $m_2=0$ correspond to rational values of the longitudinal rotation number. We will argue below that this bifurcation corresponds to a SCNS bifurcation for the map $f^{m_1}$, i.e., the bifurcation discussed in \Sec{SNHMap}. In the supercritical case the invariant circle typically degenerates into a pair of period-$m_1$ saddle-foci whose two-dimensional stable and unstable manifolds enclose a periodic family of elliptic invariant circles. The result is a structure that looks like a \textit{string of pearls} on a necklace. This is a common phenomenon in our quadratic family, and we expect this to be a generic bifurcation of invariant circles of volume-preserving maps 
of $\R^3$. Moreover, a similar bifurcation occurs in dissipative three-dimensional maps \cite{Broer08}; they studied in detail an example with $\omega_L = \frac15$.

For example when when $\eps = 0$ and $\mu = -2.0$ for the map \eqref{eq:StdMap}, $\omega_L = \frac14$. As $\eps$ is increased from zero, there is a period-four \textit{string of pearls} created as shown in \Fig{PeriodFourPearls}. In the figure the new period-four family of invariant two-tori and several orbits that spiral around the one-dimensional manifolds of the period-four saddle-foci are clearly visible.

\InsertFigTwo{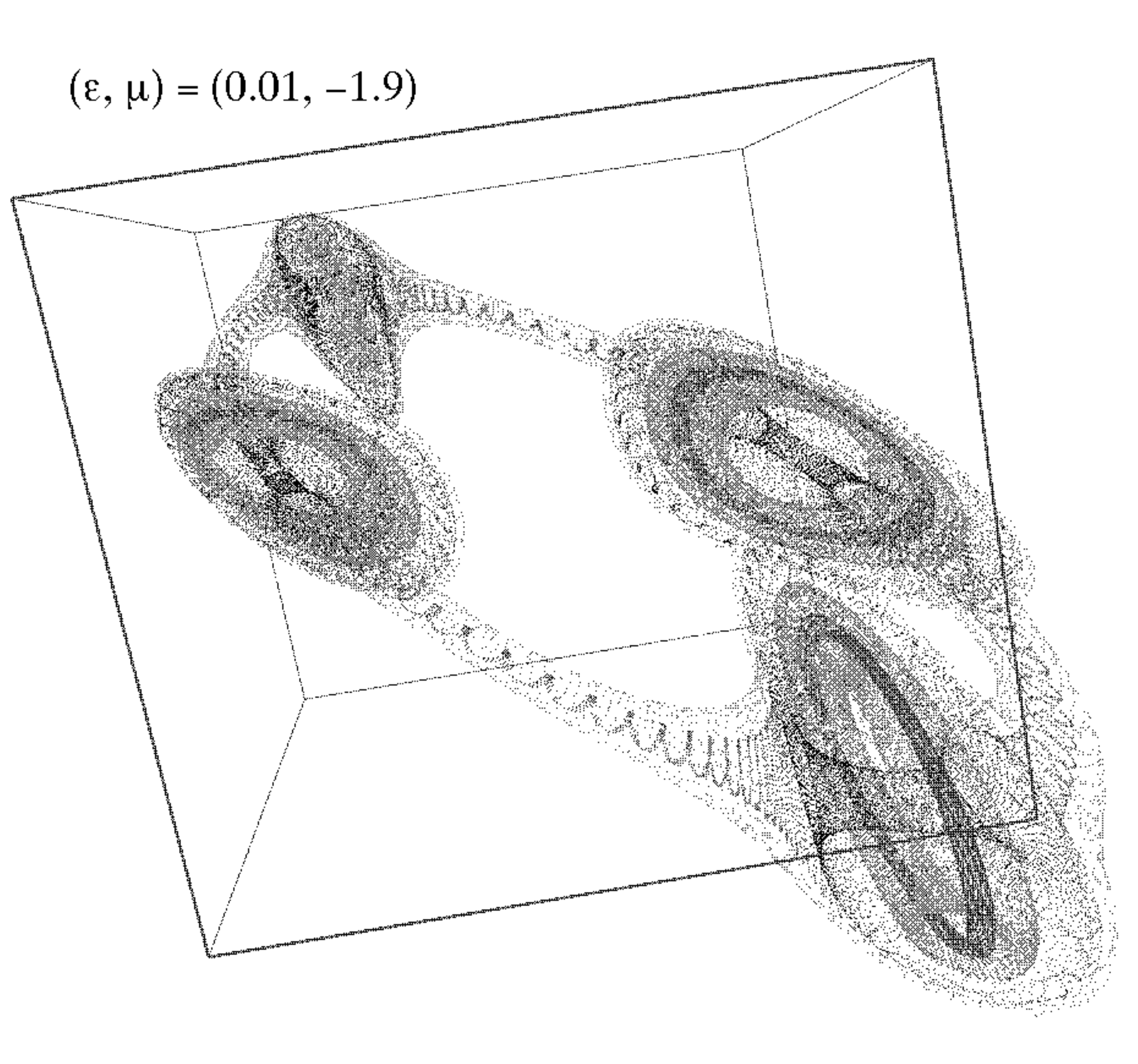}{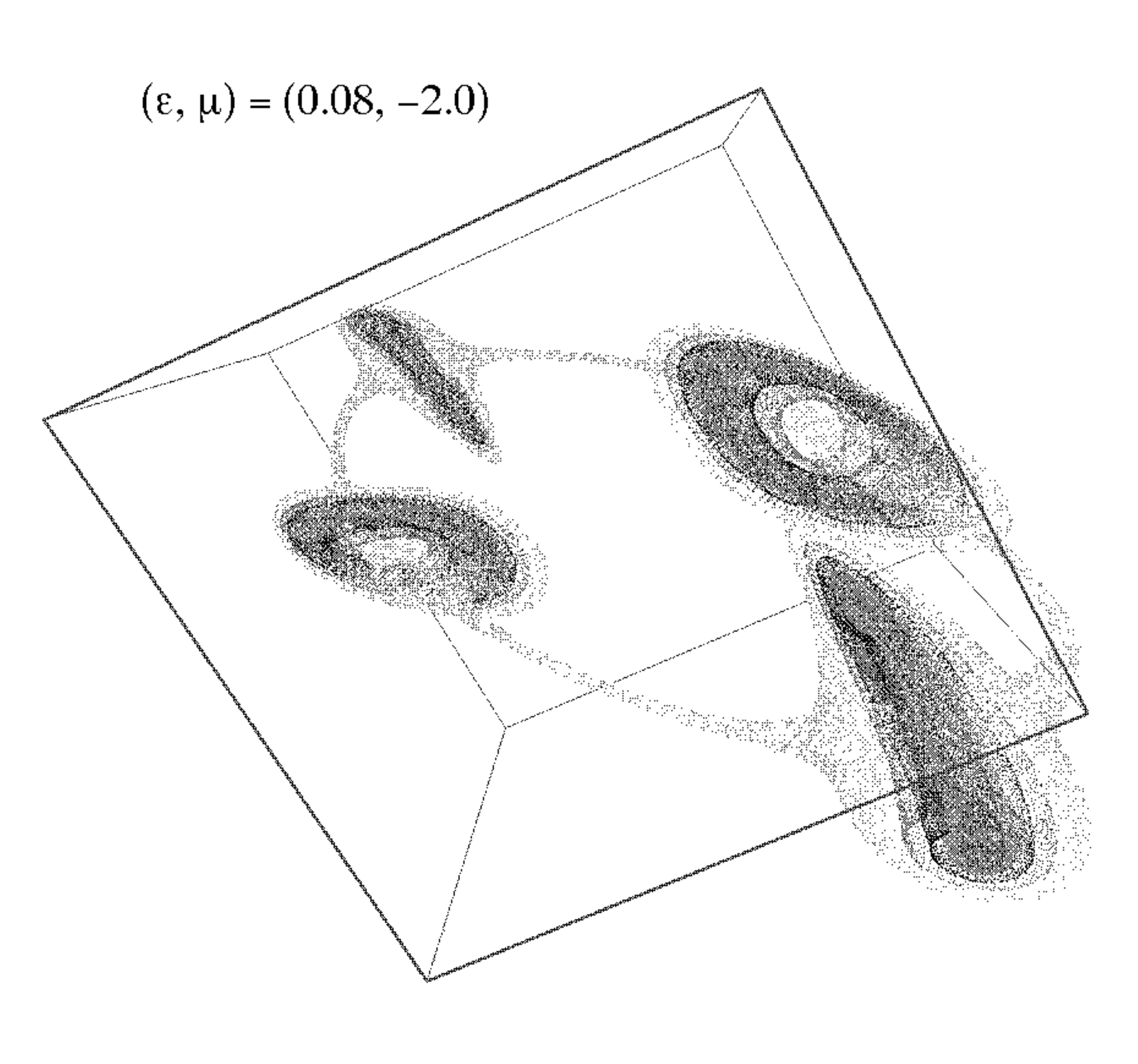}{Phase space near the $\omega_L = \frac14$ string of Pearls bifurcation. The red cube is centered at the origin and has sides of length $\frac{10}{3}\sqrt{\eps}$.}{PeriodFourPearls}{3in}

This structure can be expected generically. Suppose that as $\omega_L \to \frac{n}{m_1}$ the invariant circle $\cC$ remains elliptic and that $f$ can be approximately described by the SCNS normal form \eqref{eq:SNHFinal}. When $\omega_L$ is rational every point on the invariant circle of the normal form becomes $m_1$-periodic with multipliers $(1, \lambda^{m_1}, \lambda^{-m_1})$, where $\lambda = e^{2\pi i \omega_T}$. In the resonant case, there are additional terms that must be kept in the normal form, but if $|m_1| >3$ and $\omega_T$ is irrational, then the quadratic terms in $f$ do not change. Consequently $f^{m_1}$ corresponds to the critical case on the saddle-center-Hopf line of \Fig{Stability} and, when any parameter is varied, it will generically undergo a SCNS bifurcation. If this bifurcation is supercritical, the singular fixed point of $f^{m_1}$ will be replaced by a pair of saddle-focus fixed points and a new elliptic invariant circle will be created. For the original map this corresponds to the creation of a pair of period-$m_1$ orbits and a period-$m_1$ family of invariant circles linked with the original circle. Recall from \App{SNHNormalForm} that when the unfolding of this bifurcation is constructed, the coordinate transformations include a shift along the direction of the unit eigenvector: this shift selects the particular points along $\cC$ that will correspond to the fixed points.

\InsertFig{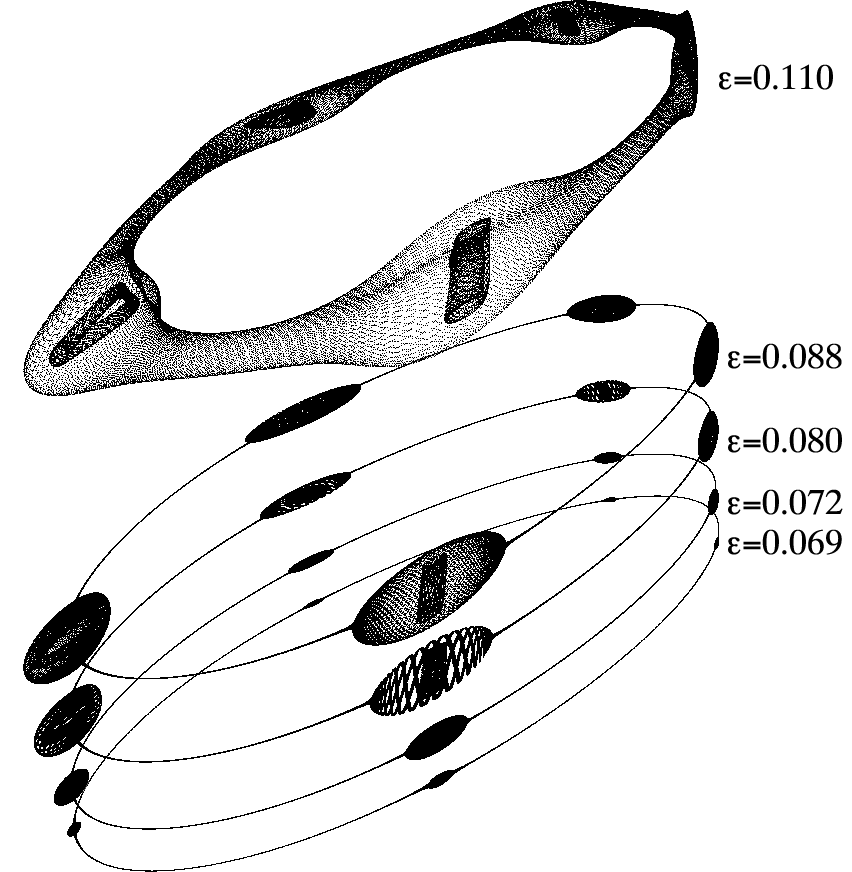}{String of Pearls bifurcation for $\mu = -1.383$ and the values of $\eps$ shown. The circles are shifted vertically and scaled by $\eps^{-\frac12}$ so that their size appears to be roughly fixed}{1-5Pearls}{4in}

Another example of this bifurcation is shown in \Fig{1-5Pearls} for $\omega_L = \frac15$. When $\eps = 0$, this longitudinal rotation number occurs at $\mu \approx -1.3819$ according to \eqref{eq:fixedPtRotNum}. Fixing $\mu = -1.383$ and increasing $\eps$ from zero, the $(5,0,1)$ resonance occurs at $\eps \approx 0.069$ where the ``circle" develops five small bulges formed from the nearly-coincident, two-dimensional manifolds of two new, period-five saddle-foci. As $\eps$ increases the points on the new periodic orbits move apart and the corresponding \textit{pearls} grow in size. The one-dimensional manifolds of the saddle-foci also nearly coincide, forming an approximate circle up to $\eps \approx 0.09$. The string of pearls is still enclosed by a family of two-tori; one  of these is shown in the uppermost portrait at $\eps = 0.11$. These tori and the new period-five tori apparently vanish near $\eps = 0.12$ and when $\eps$ reaches $0.18$ the computation of \Sec{BoundedOrbits} reports no bounded orbits.  A similar scenario occurs at $(\eps,\mu) = (0.076,-2.6)$ where $\omega_L = \frac{3}{10}$, see \Fig{3-10-3-8Pearls}. In this case the pearls exist only for a very small window in $\eps$, and the circle reforms as $\eps$ is increased.
Note that the ellipse algorithm is typically able to continue the circle through the string-of-pearls bifurcation when the slice is not too close to the periodic orbits. 

\InsertFig{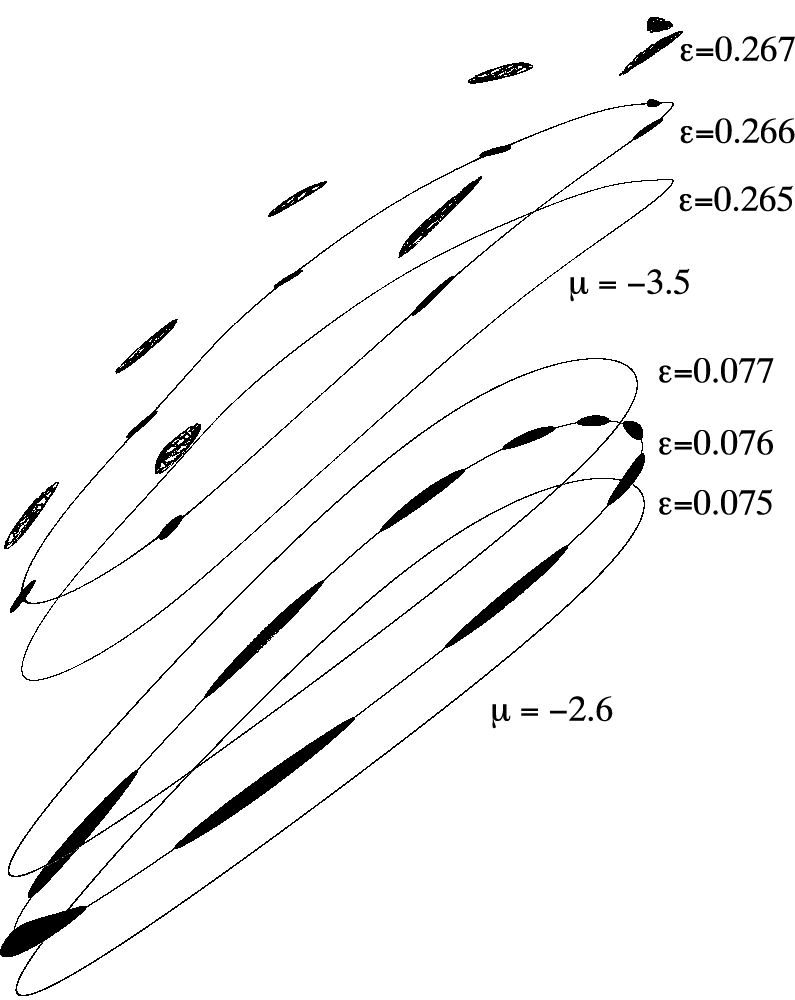}{String of pearls bifurcations for $\mu = -2.6$ (lower three portraits) and $\mu = -3.5$ (upper three portraits), where $\omega_L = \frac{3}{10}$ and $\frac{3}{8}$, respectively.}{3-10-3-8Pearls}{3in}

In some cases an $(m_1,0,n)$ resonance is responsible for the final destruction of the invariant circle. This happens for $\mu = -3.5$ where the circle is undergoes an $(8,0,3)$ bifurcation at $\eps = 0.266$, see \Fig{OmegaLTmu-35}. A sequence of three portraits for this case is also shown in \Fig{3-10-3-8Pearls}. As for the $\frac15$ and $\frac{3}{10}$ cases, an approximate, frequency-locked invariant circle appears to persist for a small parameter interval beyond the bifurcation (at least up to $\eps = 0.2686$); however, in this case the circle does not reconstitute as $\eps$ increases.  Destruction of the invariant circle also occurs for $\mu = -2.4$ when $\omega_L = \frac27$ at $\eps = 0.32$. This bifurcation was shown in the frequency map \Fig{OmegaLTmu-24}. The $(7,0,2)$ bifurcation appears to be subcritical as we can find no stable orbits beyond $\eps = 0.32$. This is suggested by the rightmost pane of \Fig{OrbitsMu-24} where the first iterates of seven unbounded orbits are plotted. These orbits appear to be near the two-dimensional unstable manifolds of a period-seven saddle and these manifolds do not appear to intersect as would be predicted by the subcritical case in \Fig{Vp_sn}.

\InsertFig{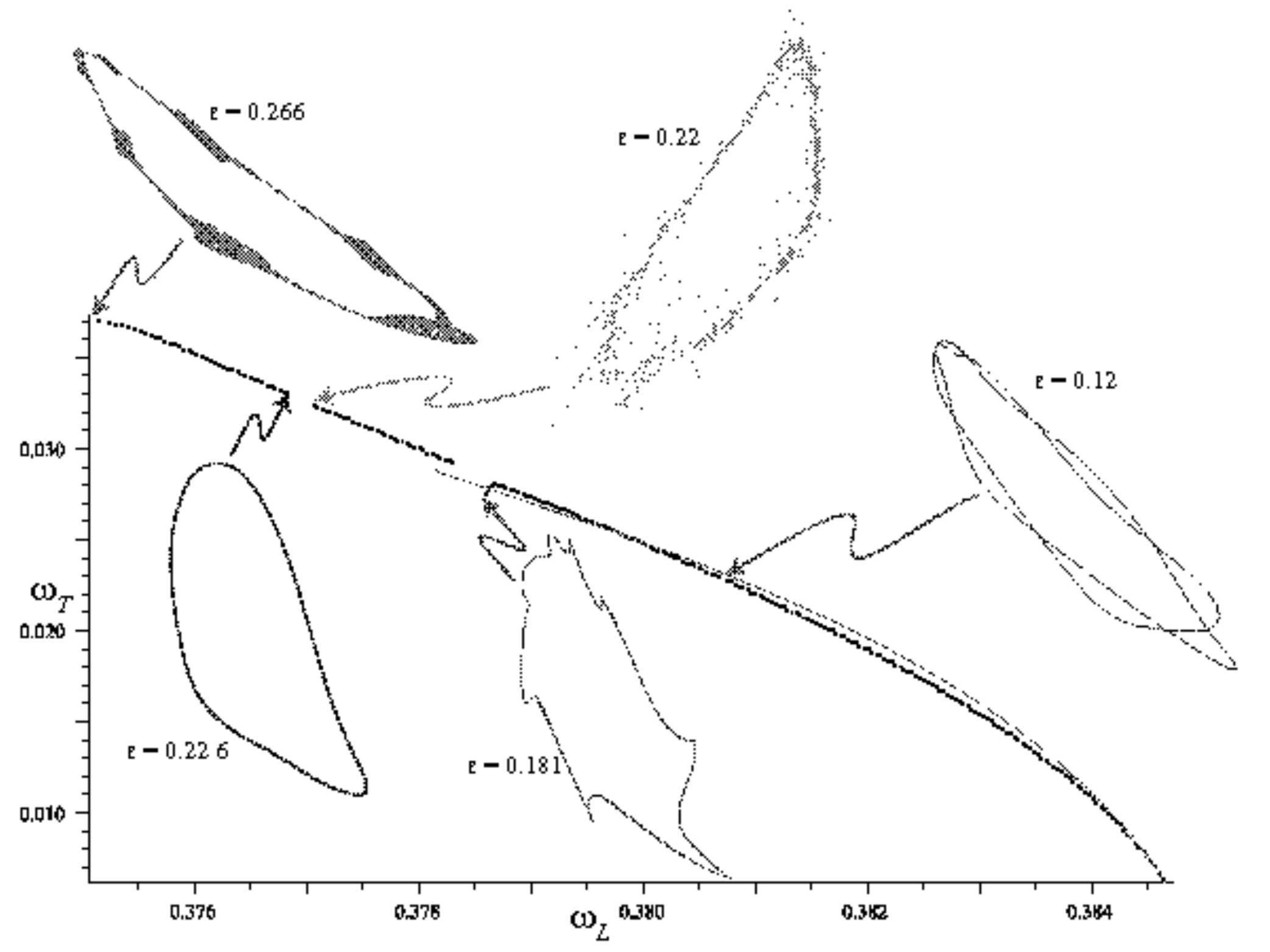}{Frequency map for $\mu = -3.5$. Inserts show the $(8,-2,3)$ resonance at $\eps = 0.12$ resulting in a doubled circle, $(7,12,3)$ at $\eps = 0.181$, $(8,2,3)$ at 
$\eps =0.22$ and the final disintegration of the invariant circle in the $(8,0,3)$ resonance at $\eps = 0.266$ leading to the formation of a string of pearls.}{OmegaLTmu-35}{5in}

\subsection{Pairs of Saddle-center bifurcations of invariant circles}\label{sec:SaddleNode}

As we have seen, resonances on an elliptic invariant circle can lead to a local bifurcation that creates new invariant circles linked with the original. Invariant circles can also be created by resonances on two-tori and more generally by saddle-center bifurcations that are not associated with any regular structure. Sometimes saddle-center bifurcations are related to other resonances. For example, near the tripling resonance with $m_2 = \pm 3$ (if $\gamma \neq 0$) a pair of new invariant circles is born in a saddle-center bifurcation,
as seen in \Fig{SectionsMu-24}. In the following we discuss a new saddle-center bifurcation that is caused by a resonance with $m_2 = \pm 1$.

We observe that these circle-saddle-center bifurcations come in pairs, as can be seen in the frequency maps of \Fig{OmegaLTmu-15-19}, for $m_2 = \pm 1$. In these cases, the circle $\cC$ created at $\eps = 0$ approaches a resonance line of the form $(m_1,\pm1, n)$ and its frequency map curves dramatically to ``avoid" crossing the resonance. As this happens, an elliptic/hyperbolic pair of circles are born in a circle-saddle-center bifurcation far from the central circle. This may happen inside a chaotic region or region of unbounded motion or within the family of elliptic two-tori surrounding the invariant circle. In the latter case a torus bifurcates into a figure eight crossed with a circle. 
According to our observations, the rotation numbers of the elliptic member of the new pair lie on the opposite side of the resonance line. When $\mu = -1.5$, this is the $(5,-1,1)$ resonance line shown in \Fig{OmegaLTmu-15-19}. A sequence of Poincar\'e sections for this case is shown in \Fig{SaddleNode5-1-1}. The new circles appear near $\eps = 0.097$; they form a $(5,-1)$ torus knot. Note, however, that the new elliptic circle (red in the figures) is relatively smooth and it is more appropriate to think of the highly deformed original circle (blue) as a $(5, -1)$ torus knot wrapping about the new one. As $\eps$ grows the stable region around the original circle shrinks and that around the newly created elliptic circle grows until it takes over as the primary elliptic circle and persists up to $\eps = 0.175$.  
The old orbit (blue) eventually dies in a circle-saddle-center bifurcation. 
So overall there is a pair of saddle-center bifurcations of invariant circles, first the creation of a new elliptic circle, and later the destruction of the old elliptic circle.
This bifurcation also occurs at $(\eps, \mu) = (0.049, -1.1)$ corresponding to the $(6,-1,1)$ resonance. When $\mu = -1.9$ a similar bifurcation appears to occur near $\eps = 0.07$ for the resonance $(4,1,1)$, however, there are also many other resonances near these parameter values that complicate this case.


\InsertFig{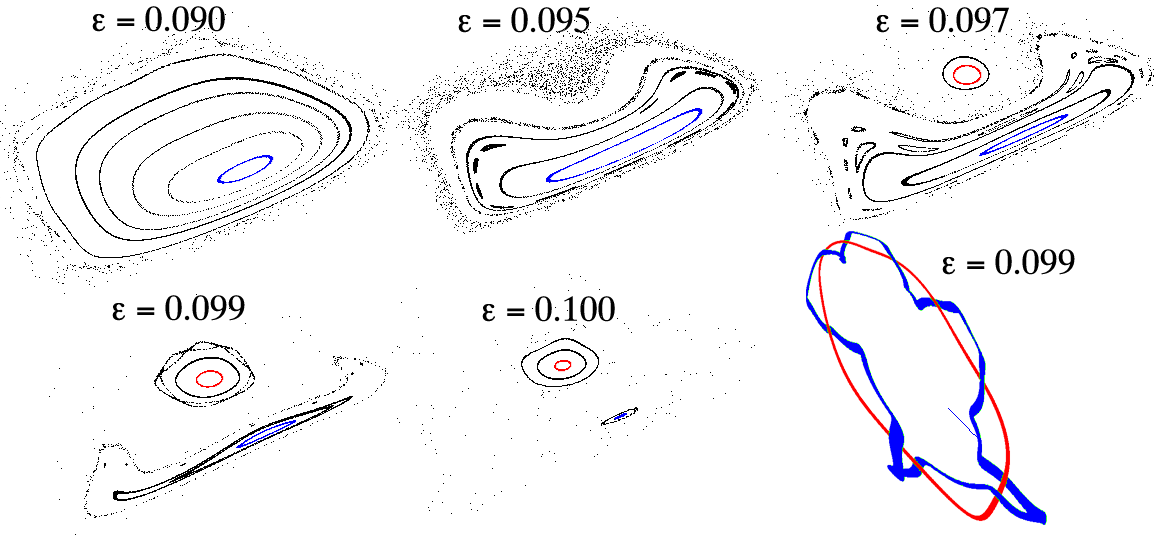}{Circle-saddle-center bifurcation for $\mu = -1.5$. Five Poincar\'e slices show the creation of a new elliptic circle with $(5,-1)$ helicity near $\eps = 0.096$. The stable region around the new circle grows while that around the original circle shrinks, until near $\eps = 0.101$ the original circle looses stability. Tori near the new original circle are shown in green and near the new circle are shown in red at $\eps = 0.099$ in the last inset.}{SaddleNode5-1-1}{5in}

\section{Conclusion}

The three-dimensional analogue of the two-dimensional, area-preserving H\'enon map can be found  by normal form expansion and unfolding near a triple-one multiplier \cite{Dullin08a}; we believe this quadratic map, \eqref{eq:StdMap}, should serve as the prototype for volume-preserving dynamics in $\R^3$.

We computed the volume of bounded orbits for \eqref{eq:StdMap} as a function of its two essential parameters, $(\epsilon, \mu)$, obtaining an intricate structure reminiscent of Arnold tongues for circle maps. In the H\'enon map bounded motion appears near the elliptic fixed point. By contrast, in the quadratic volume-preserving map bounded motion appears 
near elliptic invariant circles. These invariant circles in turn are created in 
saddle-center-Neimark-Sacker bifurcations. These occur on a codimension-one line in parameter space that emanates from the triple-one-multiplier, codimension-two point. 
The SCNS bifurcation also creates two fixed points that are at the poles of the vortex bubble containing the bounded orbits. 

Near the SCNS bifurcation, the map can be decoupled in cylindrical coordinates into a semi-direct product of an area-preserving map independent of the angle and a map of the angle depending on all coordinates, \eqref{eq:SNHFinal}. A consequence is that close to the bifurcation, the two fixed points have one-dimensional manifolds that nearly coincide along the bubble's polar axis and two-dimensional manifolds that form the bubble's outer boundary. The curves of transversal intersections of the latter manifolds undergo bifurcations that were described in \cite{Lomeli98a}; these give rise to bounded chaotic motion near the boundary of the vortex bubble. 

When the SCNC bifurcation is supercritical, it creates an invariant circle $\cC$ in the vortex bubble that corresponds to the third, elliptic, fixed point of the area-preserving, decoupled map. We computed the location and frequencies of $\cC$ for the normal form and obtained analytical expansions for the longitudinal, $\omega_L$, and transverse, $\omega_T$, frequencies of the invariant circle (or in fact for any torus nearby) for the map \eqref{eq:StdMap}.

The frequency map, $\Omega$ is an essential object of KAM theory. In a Hamiltonian setting $\Omega$ maps $n$ actions into $n$ frequencies for tori of maximal dimension. In the volume-preserving case the dimension of the action and angle spaces are typically not equal. In $\R^3$, the tori in the neighborhood of the circle $\cC$ have effectively one action, and $\Omega$ defines a curve in the two-frequency space that depends upon that single action. As is done in the study of lower-dimensional tori in Hamiltonian systems, the frequency map can be extended to a local diffeomorphism by including parameters in the domain. In this way we study the frequency map $\Omega: (\eps, \mu) \to (\omega_L, \omega_T)$ of the invariant circle of the quadratic map. This mapping organises the bifurcations of invariant circles. Our analytic approximation for $\Omega$ compares very well with the numerical computations. 

In order to numerically compute $\Omega$, we developed a simple method to find the location of elliptic invariant circles of maps. The idea is to consider a slice of the mapping that is transverse to the invariant circle and fit a thin ellipsoid to iterated points returning into the slice. This method converges very fast and appears to be a robust and accurate way to continue the invariant circles. Having found a good approximation of $\cC$, the computation of the longitudinal rotation number can be made quadratically convergent by considering its continued fraction expansion. For the transverse rotation number we have a linearly convergent numerical method.

Equipped with the continuation method and the frequency map an astounding variety of resonant bifurcations of invariant circles can be found whenever $m_1 \omega_L + m_2 \omega_T = n$. 
We classified these into three cases: string-of-pearls bifurcations ($m_2 = 0$), paired circle-saddle-center bifurcations ($m_2 = 1$), and torus link bifurcations ($m_2 \ge 2$). More precisely a torus knot of type $(m_1, m_2)$ is found when $k = \gcd(m_1, m_2) = 1$, while a torus link of type $(m_1, m_2)$ consisting of $k$ circles is found when $k > 1$. For example, there are two types of circle-doublings: either the new circle is a double cover of the original one, or there are two new invariant circles, each covering the original circle once. 

The most spectacular bifurcation is the string of pearls that occurs when the longitudinal frequency becomes rational. Here the invariant circle breaks into a pair of saddle-focus periodic orbits, and each pair recapitulates the original vortex bubble on a smaller scale. In this way, we expect that is is possible to find self-similarity in the SCNS bifurcation: the invariant circle bifurcates to create new bubbles that contain invariant circles linked to the original one and that will themselves undergo string-of-pearls bifurcations under parameter variation, etc.  It seems likely that the there exist parameter values with infinite sequence of linked pearls around pearls around pearls \ldots. 

Similar bifurcations also occur in dissipative $3D$ maps. For example circle doublings can occur, but unlike the infinite sequences of doublings of periodic orbits, these appear to be typically finite in one-parameter families \cite{Argoul84}.
Moreover, a bifurcation similar to the string of pearls also occurs in the dissipative case near the codimension-two a saddle-center point. The three-parameter unfolding of this bifurcation is discussed by Broer et al \cite{Broer08}; they also studied a model of the dynamics near $\omega_L = \frac15$. It is interesting that they observed that there can be an attracting invariant two torus containing a repelling period-five bubble.  We hope to apply similar methods to study the volume-preserving, resonant cases in a future paper. These are both simpler (since there is one fewer parameter) and more complex (since there are no attractors) than the dissipative case.

\appendix

\section{Appendix: Parameter Reduction}\label{app:generic}
Though the map \eqref{eq:111NF} nominally depends upon the three unfolding parameters ($\eps, \mu_1, \mu_2)$, one of these can generically be eliminated by an affine shift in $x$. If we set $x \to x_0 + \tilde{x}$ then the new map has the same form \eqref{eq:111NF} with $p(x,y,\eps,\mu)$ replaced by $p(x_0+\tilde{x},\eps,\mu)$. If there is an $x_0$ so that $\partial_x p(x_0,0,\eps,\mu) =0$, then the linear term $\mu_1 x$ can be eliminated. Near the origin, this equation has a solution providing $\partial^2_x p(0,0;0,0) \neq 0$.\footnote
{
	If this is not the case, but $\partial^2_yp(0,0,0,0) \neq 0$, then we 
	can choose $x_0$ so that $\partial_y p(x_0,0,\eps,\mu) = 0$ 
	eliminating the linear term in $y$.
}
In this paper, we assume this is the case, so that without loss of generality we can set $\mu_1 = 0$. Moreover, even though $p$ is nominally a function of $(\eps, \mu)$ we will take the coefficients of this function to be independent parameters thus suppress this dependence. The normal form \eqref{eq:111NF} then becomes \eqref{eq:StdMap}.

For the quadratic case \eqref{eq:StdMap}, one of the parameters in $P$ could be eliminated by scaling the variables; for example, under the scaling $\xi \to \sqrt{|a|} \xi$, the parameter $a$ is effectively replaced by $\pm 1$. Thus for this case \eqref{eq:StdMap} reduces to a four parameter family.

\section{Appendix: Basic Properties}\label{app:properties}
In this section we study some of the simple dynamical properties of \eqref{eq:StdMap}, especially the fixed points and low period orbits. At first we let $P$ be an arbitrary, smooth function that is $\cO(2)$ in the variables $(x,y)$, and such that
\beq{adefine}
	a \equiv \frac12 P_{xx}(0,0) > 0 \;,
\eeq
which we used to eliminate the parameter $\mu_1$, see \App{generic}.

When $P$ is smooth, the map \eqref{eq:StdMap} is a diffeomorphism; indeed, its inverse can be easily computed
\beq{quadInv}
	f^{-1}(x,y,z)=
	\begin{pmatrix} 
		x-y+z \\  y-z\\ z+\eps -\mu(y-z) -P(x-y+z,y-z)
	\end{pmatrix} \;. 
\eeq

The fixed points of \eqref{eq:StdMap} have the form $(x^*,0,0)$, where
\[
	P(x^*,0) = \eps \;.
\]
By \eqref{eq:adefine}, this equation has two solutions,
\[
	x^*_\pm \approx \pm \sqrt{\frac{\eps}{a}}\left(1 + \cO(\eps)\right) \;,
\]
that emerge from the origin for $\eps > 0$. This corresponds to the saddle-center bifurcation in the normal form. Indeed, the Jacobian of $f$ at these points has characteristic polynomial \eqref{eq:charPoly} with
\[
    \tau_\pm = 3 + \mu +P_y(x^*_\pm,0) \;,
    \quad \sigma_\pm = 3 +\mu + P_y(x^*_\pm,0)-P_x(x^*_\pm,0) \;.
\]
The saddle-center line, corresponding to a multiplier $\lambda_1 = 1$, occurs on the line $\tau = \sigma$ in \Fig{Stability}. Since $a \ne 0$, this corresponds to the line $\eps = 0$. 

Recall that a diffeomorphism is \emph{reversible} if it is conjugate to its inverse, $r \circ f = f^{-1} \circ r$.  The normal form \eqref{eq:StdMap} is only reversible if $P$ satisfies special conditions.  For example, if
\[
	P(x,y) = P(-x-y,y) \;,
\]
then there is an involutory reversor
\beq{reversor}
	r(x,y,z) = (-x, y-z,-z) \;.
\eeq
This condition on $P$ is essentially necessary for reversibility as well.
Recall that a reversor maps each period-$n$ orbit into another such orbit with reciprocal multipliers. Orbits that lie on the fixed set of $r$, in this case the $y$-axis, are mapped into themselves by $r$ and are called \emph{symmetric}.  Orbits that are related by the reversor have reciprocal multipliers, and a symmetric orbit must have self-reciprocal multipliers and therefore one unit multiplier. When $\eps \neq 0$, the fixed points do not lie on the saddle-center line and consequently cannot be symmetric. If there are only two such points, then they must be mapped into each other by the reversor and must have reciprocal multipliers. This means that $\tau_\pm = \sigma_\mp$, or equivalently
\[
    P_y(x^*_+,0) -P_y(x^*_-,0)  = P_x(x^*_+,0) = -P_x(x^*_-,0) \;.
\]
For the quadratic case, this implies that $P$ has the form
\[
P = a(x^2 + xy) +cy^2 \;.
\]
and that the reversor is \eqref{eq:reversor}. Consequently, the map \eqref{eq:StdMap} is only rarely reversible.

For the case that $P$ is a positive definite, quadratic form, that is $a > 0$ and $D \equiv ac-b^2/4 > 0$, it  was shown in \cite{Lomeli98a}  that every bounded orbit of the quadratic map is contained in the cube $|\xi|_\infty \le \kappa_B$ where
\[
	\kappa_B = \alpha\left(|3+\mu|+1 + \sqrt{(|3+\mu|+1)^2+|\eps|/\alpha} \right) \;,
\]
where $\alpha = \frac{1}{D}\max(c,a-b+c)$.\footnote
{
	This corrects an error in \cite{Lomeli98a} in the formula on page 572 for $\kappa$.
}

\section{Quadratic Map Fixed Points}\label{app:fixedPoints}
The stability properties of the fixed points \eqref{eq:fixedPoints} of the quadratic map\eqref{eq:StdMap} are easily determined from the characteristic polynomial \eqref{eq:charPoly}. For the fixed points the trace and second trace become
\begin{align*}
	\tau_\pm &= 3+\mu + b x_\pm \;, \\
	\sigma_\pm &= 3+\mu +(b-2a) x_\pm \;.
\end{align*}
Since $a > 0$, the stability diagram \Fig{Stability} maps diffeomorphically onto $(\mu,x_\pm)$ plane, the half-planes $ \tau -\sigma \lessgtr 0$ corresponding to $x_\pm$, respectively. As for the case of a general nonlinear function $P$, the saddle-center bifurcation curve corresponds to $\eps = 0$. Along this curve, when $-1 < \tau = \sigma < 3$ the other two multipliers have the form $e^{\pm 2\pi i \omega}$, which implies that $-4 < \mu < 0$ and 
$\omega = \omega_0(\mu)$, \eqref{eq:fixedPtRotNum}. Thus we call this the \emph{saddle-center-Hopf} segment.

Note that for most parameter values, one fixed point is type $(2,1)$ (two-dimensional stable manifold) and the other is type $(1,2)$ (two-dimensional unstable manifold). The exception corresponds to a parabolic region near $\mu = -4$ that happens to coincide with the existence of a period-two orbit. Indeed, the period-doubling bifurcation in \Fig{Stability} corresponds to $\tau+\sigma = -2$. For the quadratic map, this implies $\mu = (a-b)x_\pm-4$,
which is equivalent to the parabola
\beq{pd}
	\Delta \equiv (a-b)^2\eps -a(\mu+4)^2 = 0 \;,
\eeq
as shown by the dashed lines in \Fig{QuadStability}. It is easy to see that the quadratic map has at most one period-two orbit, which has the form $(x,y,z) \to (x+y, -y, -z) \to (x,y,z)$ where
\[
	(x,y,z) = \frac{1}{a-b}\left(\mu+4 \mp \rho, \pm 2 \rho, \pm 4 \rho \right) \;.
\] 
Consequently, this orbit exists only when $a \neq b$ and
\[
   \rho^2 = \frac{\Delta}{a-2b+4c} > 0 \;.
\]
For the case that $a-2b+4c > 0$, the orbit exists in the interior of the period-doubling parabola (the dashed curves in \Fig{QuadStability}); otherwise, it exists in the exterior of this parabola. 
The characteristic polynomial for the period two orbit has the form \eqref{eq:charPoly} with
\[
	\tau-\sigma = \frac{4\Delta}{a-b} \;.
\]
As expected, the period-two orbit is born on the parabola $\Delta = 0$ with a multiplier $1$. Thus, for fixed $(a,b,c)$ only four of the eight stability regimes of \Fig{Stability} are reached; an example for $a > b$ and $a+4c > 2b$ is shown in \Fig{Period2stability}. In this case, the period-two orbit is created on the upper half of the doubling parabola $\Delta = 0$ by being emitted from the fixed point $(x_+,0,0)$ and is destroyed on the lower half of the parabola by being absorbed into the second fixed point $(x_-,0,0)$.

\InsertFig{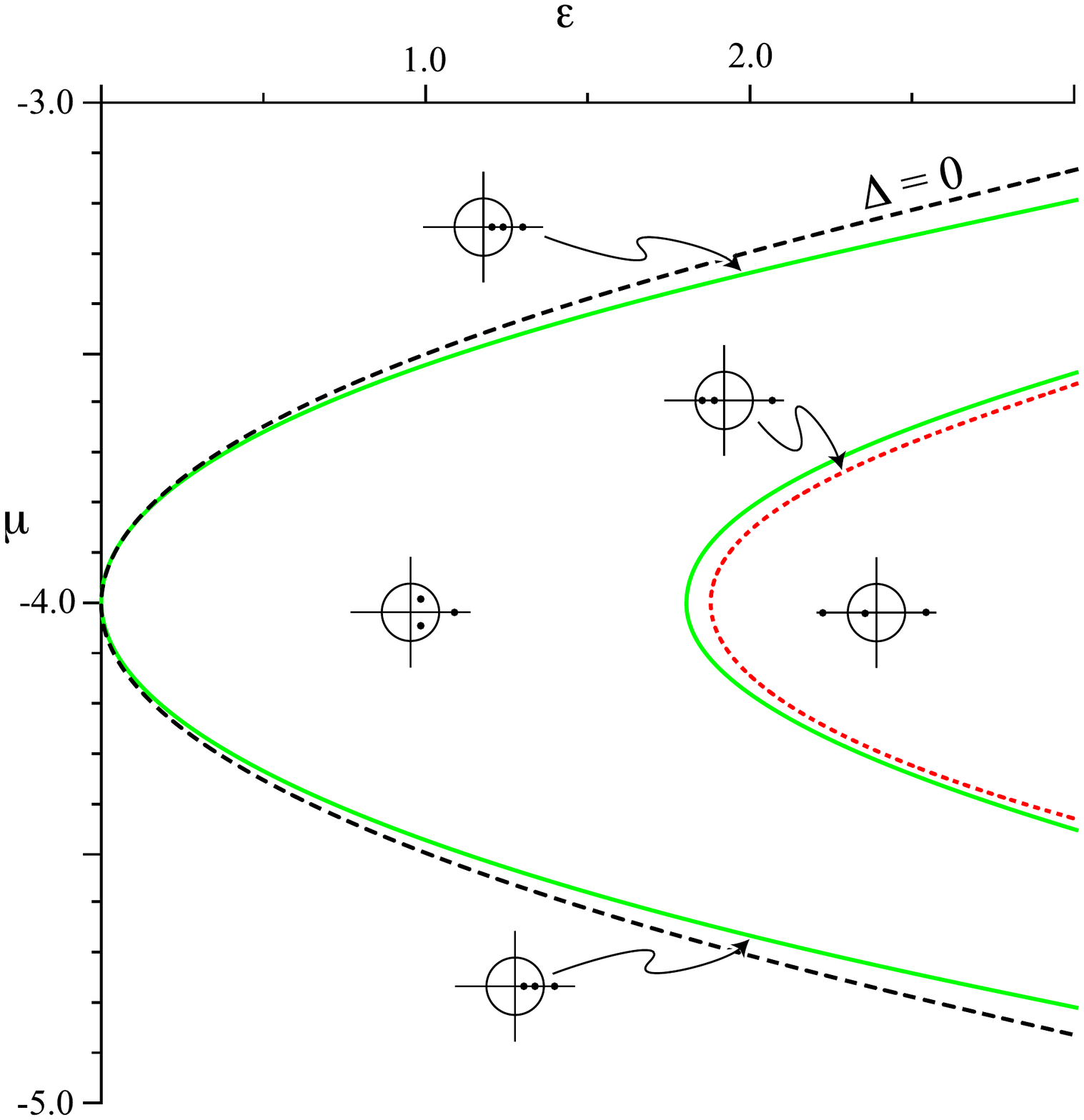}{Stability diagram for the period-two orbit of \eqref{eq:StdMap} with the polynomial \eqref{eq:quadForm} and $a = 1$, $b=c=0.5$.The dashed (black) parabola is the curve $\Delta = 0$ where one multiplier of the period-two orbit is $1$. This is the same as the period-doubling curve in \Fig{QuadStability}. The period-two orbit exists for $\Delta>0$, to the right of the $\Delta = 0$ parabola.  The solid (green) curves correspond to a double eigenvalue, and the dotted (red) curve to a multiplier $-1$.}{Period2stability}{4in}

The two curves of double eigenvalues in \Fig{Stability} correspond to $\lambda_1 = \lambda_2 = r$ giving $\tau = 2r+1/r^2$ and $\sigma = r^2+2/r $. For the fixed point these become
\begin{align*}
     2a x_\pm &=  -r^2+2r-2/r+1/r^2  \;,\\
     \mu      & = b x_\pm -\left(2r+\frac{1}{r^2} \right) -3 \;.
\end{align*}
These are shown as solid curves in \Fig{QuadStability}.

Finally, fixed points with multipliers of the form $\lambda = (r e^{2\pi i \omega}, r e^{-2\pi i \omega}, \frac{1}{r^2})$ have $\tau = 2r \cos(2\pi\omega) + 1/r^2$ and $\sigma = r^2 + 2\cos(2\pi\omega)/r$, which implies
\bsplit{bifcurves}
     2a x_\pm &=  \left( r - \frac{1}{r}\right)\left( 2\cos(2\pi\omega) 
        - r-\frac{1}{r}\right)  \;,\\
     \mu      & = 2r \cos(2\pi\omega) + \frac{1}{r^2}-3-b x_\pm \;.
\end{split}\eeq
Several curves of this type for fixed, rational values of $\omega$ are shown in \Fig{QuadStability}.

The quadratic map \eqref{eq:StdMap} has at most one pair of period-three orbits; this follows
because the set of equations $\xi = f^3(\xi)$ can be reduced to a degree-eight polynomial in one variable. This polynomial has a quadratic factor with zeros at the fixed points; the remaining sixth degree equation has zeros on the pair of period three orbits. They
are created in a saddle-center bifurcation along the curve
\[
  \eps = K(\mu + 3)^2 \;,
\]
where
\[
   K = \frac {a (2a-3b+6c)((a-b)^2+ 3c(a-b+c))}
   {(a^2- 3ab + 3ac + 3b^2 - 9bc + 9c^2)(a^3- 3a^2b + 3a^2c + ac^2 + 2b^2c - 5abc  + 3ab^2 - b^3) } \;.
\]
For example, when $a=1$, $b=c=\frac12$, $K = 4$. Note that this curve is not the same as the tripling curve of the fixed point.

One of the period-three orbits is type-$(2,1)$ and the other type-$(1,2)$. 
When $\mu < -3$, the period-three orbits are born  with real eigenvalues, corresponding to $\tau = \sigma > 3$, and (for $a=1, b=c=\frac12)$ when $-3 < \mu < -2.24$ they are spiral-saddles (saddle-center-Hopf bifurcation). At $\mu \approx -2.24$ the period three orbits sit at the point $\tau = \sigma = -1$, the $(-1,-1,1)$ point. $\eps$ just to the right of this parabola, these orbits have real eigenvalues. They then cross a curves of double multipliers, and become spiral saddles. 

Overall it should be noted that the importance of fixed pointed for three-dimensional, volume-preserving maps is far less than for two-dimensional, area-preserving maps, since they are generically unstable. The important object in the 3D volume-preserving case are elliptic invariant circles.


\section{SCNS Normal Form}\label{app:SNHNormalForm}

In this section we summarize the computation of the normal form for the saddle-center-Neimark-Sacker (SCNS) bifurcation of a volume-preserving map that is discussed in \Sec{SNHMap}, and thus the technical part of the proof of \Th{SCNS}. In the complex coordinates $(u,v,z)$ the map takes the form
\beq{SNHstart}
	\zeta' = f(\zeta) = M(\zeta + b(\zeta))\;, \quad 
	M = \diag(e^{2\pi i\omega}, e^{-2\pi i \omega}, 1) \;,
\eeq
where $b = \cO(2)$ represents the nonlinear terms.

The formal, normal form for $f$ is obtained by constructing a near identity transformation 
$\eta = \psi(\zeta) = \zeta + h(\zeta)$ such that the conjugacy, $g \circ \psi = \psi \circ f$ gives a new map 
\[
     g(\eta) = M(\eta+c(\eta))
\]
that is ``simpler" than $f$---for example contains fewer nonlinear terms. Assuming that $f$ is in normal form through terms of degree $d-1$, this means that we must find a degree-$d$ vector function $h$ such that the conjugacy holds to degree $d$. To do this
we must solve the homological equation,
$ \Lmap_M h(\zeta)  = c(\zeta) - b(\zeta)$,  at degree $d$ for $h$ where
\beq{homoMap}
	\Lmap_M h(\zeta)  \equiv M^{-1} h( M\zeta)  - h(\zeta) \;.
\eeq
The terms in $b$ that are in the range of $\Lmap_M$ can be eliminated by the transformation. Thus the nonlinear terms $c$ in the normal form correspond to the parts of $b$ in a 
complement to the range of $\Lmap_M$.

The normal form is formal in the sense that the map and transformation are expanded in formal power series in the variables $(u,v,z)$. A basis for the space of vectors of polynomials of degree-$d$ are the ``vector monomials,''
\beq{monomialBasis}
   p_{i,m} =  \zeta^m \hat{e}_i \;,\quad m \in \N^3 \;,
\eeq
where $\hat{e}_i\;,  i = 1, 2, 3$ are the unit vectors $\hat{e}_{i,j} = \delta_{i,j}$, $\zeta^m = u^{m_1}v^{m_2}z^{m_3}$ and  
$|m| \equiv \Sigma_{i=1}^3 m_i = d$. When $M$ is diagonal, the homological operator \eqref{eq:homoMap} has a diagonal representation in the monomial basis \eqref{eq:monomialBasis}:
\[
    \Lmap_M(p_{i,m}) = \left( \frac{\lambda^m}{\lambda_i} - 1\right) p_{i,m} \;.
\]
Consequently $\ker(\Lmap_M)$ is spanned by those monomials that satisfy the ``resonance conditions"
\beq{resonanceCondition}
   \lambda^m = \lambda_i \;,
\eeq
and we can choose the normal form to lie in the kernel of $\Lmap_M$. The elements of the kernel correspond to eigenvectors with zero eigenvalue; if we assume that $\omega$ in \eqref{eq:HopfEigen} is irrational, then the resonance condition is satisfied for arbitrary integers $m_3$ and when
\[
   m_1-m_2 = \left\{ \begin{array}{rc} 
   					1, & i=1 \\
   					-1, & i=2 \\
   					0, & i=3 \\
			\end{array} \right. \;.
\]
Thus to arbitrary degree, the normal form can be written
\beq{SNHform}\begin{split}
	u'  &= \lambda u (1+c_1(|u|^2, z)) \;,\\
	z'  &= z + c_3(|u|^2,z)  \;,
\end{split}\eeq
where $c_i$ are arbitrary polynomials subject to the requirements that $c_i(0,0) = 0$ and $D_z c_3(0,0) = 0$. 

Since $f$ is volume-preserving, $h$ can be selected to be volume-preserving as well \cite{Dullin08a}; to implement this efficiently, it is better to use Lie series instead of power series. We compute the power series by truncating the exponentials.
Through cubic degree \eqref{eq:SNHform} becomes
\beq{SNH2}\begin{split}
	u'  &= \lambda u ( 1 + A z + B z^2 + C |u|^2 ) + \cO(4)\;, \\
	z'  &=  z + \alpha z^2 + \beta|u|^2 +\gamma |u|^2 z + \kappa z^3 + \cO(4) \;,
\end{split}\eeq
where $A,B,C \in \C$ and $\alpha, \beta,\gamma, \kappa \in \R$. This map is volume-preserving to $\cO(3)$ when 
\beq{vpCoeffs}\begin{split}
   A_r &= -\alpha \;,\\
   2 B_r &= 3( \alpha^2-\kappa) - A_i^2 \;,\\
   4 C_r &= -\gamma -2 \alpha\beta \;,
\end{split}\eeq
where we denote $A = A_r + i A_i$, etc.\footnote{
	Alternatively, we could eliminate $\alpha = -A_r$, 
	$\gamma = -4C_r + 2A_r \beta$, and $\kappa =\alpha^2-\frac23 B_r- \frac13 A_r^2$
}

The normal form \eqref{eq:SNHform} can be unfolded by assuming that the map $f$ depends upon a set of parameters $p$. Following Elphick et al \cite{Elphick87}, formally expand $f, h$, and $g$ in double power series in $p$ and $\zeta$; then, for each degree in $p$, the coefficients of $h$ must still satisfy the homological equation. To zeroth order in $p$ and all orders in $\zeta$, the normal form is given by \eqref{eq:SNHform}. The new feature occurs at first order in $p$: there are now terms in $f$ that are constant and terms that are linear in the variables $\zeta$. For the constant terms, the homological equation reduces to
\beq{constantHom}
	(M-I)h_0 = c_0 - b_0 \;.
\eeq
Thus any terms in $b$ not in the null space of $M-I$ can be removed. This implies that there is generically an affine term $(0,0,\delta(p))$ in the normal form at this order. Since this is the sole term that occurs at this order, we can use $\delta$ as a primary parameter; it corresponds to unfolding the saddle-center bifurcation.

The terms of first degree in $\zeta$ can be represented by matrices, e.g. $h_1(\zeta) = H \zeta$, etc., and the homological equation becomes the matrix equation
\beq{linearHom}
	M^{-1}HM - H = C-B \;.
\eeq
Since $M$ is diagonal, this operator is diagonal in the components of $H$ and its kernel consists of the matrices that commute with $M$: the diagonal matrices. Thus the normal form must include an added diagonal, linear perturbation. For the volume-preserving case when $\omega \neq 0$ or $\frac12$, this means that 
\[
	M \to \diag(\Lambda e^{2\pi i \tilde{\omega}}, 
	            \Lambda e^{-2\pi i \tilde{\omega}}, \frac{1}{\Lambda^2}) \;.
\]
for a new frequency $\tilde{\omega} = \omega + O(p)$ and expansion/contraction factor $\Lambda = 1 + O(p)$. The three parameters $(\delta, \Lambda, \tilde{\omega})$ unfold the saddle-center Hopf-bifurcation; however, as we will see below one of them can generically be eliminated.

For all terms of higher degree in $\zeta$, the homological equations are the same as before; consequently, the form of the terms in the normal form is the same as in \eqref{eq:SNH2}, except that the coefficients are now allowed to depend upon the parameters.

To quadratic degree, the unfolded normal form is thus
\beq{SNH3}\begin{split}
	u'  &=  \Lambda e^{2\pi i \omega} u (1+ A z) +\cO(3) \;, \\
	z'  &=  -\delta + z/\Lambda^2 + \alpha z^2 + \beta|u|^2 +\cO(3) \;.
\end{split} \eeq
Note that when $\alpha$ is nonzero, the implicit function theorem implies the existence of an affine shift $\tilde{z} = z + \Delta$ such that the new map has unit multiplier at the new origin.
Hence this shift can be used eliminate $\Lambda$, and the final normal form is identical to \eqref{eq:SNH2} with the addition of an affine term $-\delta$ in the $z'$ equation. This coordinate transformation modifies the rotation number $\omega$ as well.

Transforming \eqref{eq:SNH2} to symplectic, cylindrical coordinates gives the final result \eqref{eq:SNHFinal}.

\section{Appendix: Resonance tables}\label{app:tables}

Table~\ref{tbl:pearls} shows some parameter values for different string-of-pearls bifurcations.
Table~\ref{tbl:resonances} shows lower resonances encountered in the family with $\mu = -2.4$.

\begin{table}
\centering
\begin{tabular}{|l |l |c c c|}
$\mu$ & $\eps$ & $m_1$ & $m_2$ & $n$\\
\hline
-1.95&	0.23		&	4	&	0	&	1 \\
-2.4&	0.318	&	7	&	0	&	2 \\
-2.6	&	0.0760	&	10	&	0	&	3 \\
-2.6	&	0.095	&	283	&	0	&	85 \\
-2.7&   0.027	&	13	&	0	&	4 \\  
-2.8&   0.0354	&	41	&	0	&	13 \\ 
-3.3&   0.1164	&	14	&	0	&	5	\\ 
-3.5	&	0.265	&	8	&	0	&	3 \\
-3.8	&	0.24		&	12	&	0	&	5 \\  
-3.85&	0.152	&	7	&	0	&	3 \\  
-3.85&	0.305	&	19	&	0	&	8 \\  
-3.9&	0.053	&	9	&	0	&	4	\\ 
\end{tabular}
\caption{Parameters for some of the string-of-pearls bifurcations}
\label{tbl:pearls}
\end{table}

\begin{table}
\centering
\begin{tabular}{||l |c c c||l |c c c|}
$\eps$ range & $m_1$ & $m_2$ & $n$ & $\eps$ range & $m_1$ & $m_2$ & $n$ \\
\hline
0.016-0.017 	 & 7	 	 &   2	 &   2 		& 	 0.146-0.148	 &   3	 &   4	 &   1 	 \\ 
0.019       	 & 4	 	 &   -10	 &   1 		& 	 0.157-0.159	 &   8	 &   -7	 &   2 	 \\ 
0.023    	 & 4	 	 &   -9	 &   1		& 	 0.168-0.17	 &   10	 &   4	 &   3 	 \\ 
0.026	 	 &   3	 &   10	 &   1		& 	 0.171-0.173	 &   -1	 &   7	 &   0 	 \\ 
0.032		 &   3	 &   9	 &   1 		& 	 0.183	 	 &   5	 &   -10	 &   1 	 \\ 
0.039		 &   4	 &   -7	 &   1 		& 	 0.184-0.185	 &   6	 &   7	 &   2 	\\ 
0.041		 &   3	 &   8	 &   1 		& 	 0.192		 &   2	 &   10	 &   1 	\\ 
0.042		 &   10	 &   9	 &   3 		& 	 0.200		 &   9	 &   10	 &   3	\\ 
0.048-0.052 	 &  7	 &   1	 &   2 		& 	 0.210-0.215	 &   4	 &   -3	 &   1 	\\ 
0.053		 &   4	 &   -6	 &   1 		& 	 0.221		 &   5	 &   -9	 &   1 	\\ 
0.054		 &   7	 &   1	 &   2 		& 	 0.224-0.226	 &   -1	 &   6	 &   0	\\ 
0.066-0.067 	 &  10	 &   7	 &   3 		& 	 0.228-0.229	 &   2	 &   9	 &   1   \\ 
0.07-0.071 	 & 3	 	 &   6	 &   1		& 	 0.233-0.237	 &   3	 &   3	 &   1   \\ 
0.07632-0.07702 	 & 4	 	 &   -5	 &   1 		& 	 0.250-0.254	 &   10	 &   3	 &   3  	\\ 
0.087		 &   10	 &   6	 &   3		& 	 0.267		 &   5	 &   -8	 &   1 	\\ 
0.088		 &   -1	 &   10	 &   0 		& 	 0.272-0.273	 &   2	 &   8	 &   1	\\ 
0.095		 &   8	 &   -9	 &   2		& 	 0.277		 &   9	 &   8	 &   3 	\\ 
0.098-0.099 	 & 3	 	 &   5	 &   1		& 	 0.287-0.288	 &   8	 &   -5	 &   2 	\\ 
0.107-0.0108 & -1	 &   9	 &   0		& 	 0.291		 &   9	 &   -10	 &   2 	\\ 
0.118-0.12 	 & 10	 &   5	 &   3  		& 	 0.293-0.294	 &   -1	 &   5	 &   0 	\\ 
0.121-0.122	 &   4	 &   -4	 &   1		& 	 0.296		 &   5	 &   10	 &   2 	\\ 
0.134-0.135	 &   -1	 &   8	 &   0		& 	 0.298-0.299	 &   6	 &   5	 &   2	\\ \end{tabular}
\caption{Approximate resonances, $|m \cdot \omega -n| < 0.001$,  with $|m_i| \leq 10$ for invariant circles with $\mu = -2.4$, see table~\ref{tbl:circles}, page~\pageref{tbl:circles}, and \Fig{OmegaLTmu-24}.}
\label{tbl:resonances}
\end{table}

\bibliographystyle{alpha}
\bibliography{VP111}

\end{document}